\documentclass[floatfix,aps,preprint,showpacs,nofootinbib,superscriptaddress]{revtex4}
\usepackage{epsfig}
\usepackage{amsmath}
\def\lsi{\raise0.3ex\hbox{$<$\kern-0.75em\raise-1.1ex\hbox{$\sim$}}}
\def\gsi{\raise0.3ex\hbox{$>$\kern-0.75em\raise-1.1ex\hbox{$\sim$}}}
\newcommand{\lsim}{\mathop{\lsi}}
\newcommand{\gsim}{\mathop{\gsi}}
\newcommand{\wt}{\widetilde}

\def\be{\begin{equation}}
\def\ee{\end{equation}}
\def\ba{\begin{eqnarray}}
\def\ea{\end{eqnarray}}

\def\l{\left(}
\def\r{\right)}
\def\la{\langle}
\def\ra{\rangle}
\def\e{{\rm e}}
\def\Br{{\rm Br}}

\begin{document}


\title{How to find neutral leptons of the $\nu$MSM?}

\author{Dmitry Gorbunov} 
\affiliation{
Institute for Nuclear Research of the Russian Academy of Sciences,
60th October Anniversary prospect 7a, 
Moscow 117312, Russia}
\author{Mikhail Shaposhnikov}
\affiliation{
Institut de Th\'eorie des Ph\'enom\`enes Physiques,
Ecole Polytechnique F\'ed\'erale de Lausanne,
CH-1015 Lausanne, Switzerland}


\begin{abstract}

An extension of the Standard Model by three singlet fermions with
masses smaller than the electroweak scale allows to explain
simultaneously neutrino oscillations, dark matter and baryon
asymmetry of the Universe. We discuss the properties of neutral
leptons in this model and the ways they can be searched for in
particle physics experiments. We establish, in particular, a lower
and an upper bound on the strength of interaction of neutral leptons
coming from cosmological considerations and from the data on neutrino
oscillations. We analyse the production of neutral leptons in the
decays of different mesons and in $pp$ collisions. We study in detail
decays of neutral leptons and establish a lower bound on their mass
coming from existing experimental data and Big Bang Nucleosynthesis.
We argue that the search for a specific missing energy signal in kaon 
decays would allow to strengthen considerably the bounds on neutral
fermion couplings and to find or definitely exclude them below the
kaon threshold. To enter into cosmologically interesting parameter
range for masses above kaon mass the dedicated searches similar to CERN
PS191 experiment would be needed with the use of intensive proton
beams. We argue that the use of CNGS, NuMI, T2K or NuTeV beams could
allow to search for singlet leptons below charm in a large portion of
the parameter space of the $\nu$MSM. The search of singlet fermions
in the mass interval  $2-5$ GeV would require  a considerable
increase of the intensity of proton accelerators or the detailed
analysis of kinematics of more than $10^{10}$ B-meson decays.

\end{abstract}


\maketitle

\section{Introduction}
In a search for physics beyond the Standard Model (SM) one can use
different types of guidelines. A possible strategy is to attempt to
explain the phenomena that cannot be fit to the SM by minimal means,
that is by introducing the smallest possible number of new particles
without adding any new physical principles (such as supersymmetry or
extra dimensions) or new energy scales (like the Grand Unified
scale). An example of such a theory is the renormalizable extension
of the SM, the $\nu$MSM (neutrino Minimal Standard Model)
\cite{Asaka:2005an,Asaka:2005pn}, where three {\em light} singlet
right-handed fermions (we will be using also the names neutral
fermions, or heavy leptons, or sterile neutrinos interchangeably) 
are introduced. The leptonic sector of the theory has the same
structure as the quark sector, i.e. every left-handed fermion has its
right-handed counterpart. This model is consistent with the data on
neutrino oscillations, provides a candidate for dark matter particle
-- the lightest singlet fermion (sterile neutrino), and can explain
the baryon asymmetry of the Universe \cite{Asaka:2005pn}. A further
extension of this model by a light singlet scalar field allows to
have inflation in the Early Universe \cite{Shaposhnikov:2006xi}.

A crucial feature of this theory is the relatively small mass scale
of the new neutral leptonic states, which opens a possibility for a 
direct search of these particles. Let us review shortly the physical
applications of the $\nu$MSM. 

{\em 1. Neutrino masses and oscillations.} The $\nu$MSM contains 18
new parameters in comparison with SM. They are: 3 Majorana masses for
singlet fermions, 3 Dirac masses associated with the mixing between
left-handed and right-handed neutrinos, 6 mixing angles and 6
CP-violating phases. These parameters can describe any pattern (and
in particular the observed one) of masses and mixings of active
neutrinos, which is characterized by 9 parameters only (3 active
neutrino masses, 3 mixing angles, and 3 CP-violating phases). Inspite
of this freedom, the {\em absolute} scale of active neutrino masses
can be established in the $\nu$MSM from cosmology and astrophysics of
dark matter particles
\cite{Asaka:2005an,Boyarsky:2006jm,Boyarsky:2006fg,Asaka:2006rw,Asaka:2006nq}:
one of the active neutrinos must have a mass smaller than ${\cal
O}(10^{-5})$ eV. The choice of the small mass scale for singlet
fermions leads to the small values of the Yukawa coupling constants,
at the level $10^{-6}-10^{-12}$, which is crucial for explanation of
dark matter and baryon asymmetry of the Universe. 

{\em 2. Dark matter.} Though the $\nu$MSM does not have any extra stable
particle in comparison with the SM, the lightest singlet fermion,
$N_1$,  may have a life-time $\tau_{N_1}$ greatly exceeding the age
of the Universe and thus play a role of a dark matter particle
\cite{Dodelson:1993je,Shi:1998km,Dolgov:2000ew,Abazajian:2001nj}. Dark
matter sterile neutrino  is likely to have a mass in the ${\cal
O} (10)$ keV region. The arguments leading to the keV mass for dark
matter neutrino are related to structure formation and to the problems
of missing satellites and cuspy
profiles in the Cold Dark Matter cosmological models
\cite{Moore:1999nt,Bode:2000gq,Goerdt:2006rw,Gilmore:2007fy};
the keV scale is also favoured by the cosmological considerations of
the production of dark matter due to transitions between active and
sterile neutrinos \cite{Dodelson:1993je,Shi:1998km}; warm DM may help
to solve  the problem of galactic angular
momentum~\cite{Sommer-Larsen:1999jx}. However, no upper
limit on the mass of sterile neutrino exists
\cite{Asaka:2006ek,Shaposhnikov:2006xi} as this particle can be
produced in interactions beyond the $\nu$MSM. The radiative decays of
$N_1$ can be potentially observed in different X-ray observations
\cite{Dolgov:2000ew,Abazajian:2001vt}, and the stringent limits on the
strength of their interaction with active neutrinos
\cite{Boyarsky:2005us,Boyarsky:2006zi,Boyarsky:2006fg,
Riemer-Sorensen:2006fh,Watson:2006qb,Boyarsky:2006kc,Boyarsky:2006ag,
Riemer-Sorensen:2006pi,Abazajian:2006jc,Boyarsky:2006hr} and their
free streaming length at the onset of cosmological structure formation
\cite{Hansen:2001zv,Viel:2005qj,Seljak:2006qw,Viel:2006kd} already
exist. An astrophysical lower bound on their mass is $0.3$ keV,
following from the analysis of the rotational curves of dwarf
spheroidal galaxies
\cite{Tremaine:1979we,Lin:1983vq,Dalcanton:2000hn}. The dark matter
sterile neutrino can be searched for in particle physics experiments
by detailed analysis of the kinematics of $\beta$ decays of different
isotopes \cite{Bezrukov:2006cy} and may also have interesting
astrophysical applications \cite{astro}.

{\em 3. Baryon asymmetry of the Universe.} The  baryon (B) and lepton
(L)  numbers are not conserved in the $\nu$MSM. The lepton number is
violated by the Majorana neutrino masses, while  $B+L$ is broken by
the electroweak anomaly. As a result, the sphaleron processes with
baryon number non-conservation \cite{Kuzmin:1985mm}  are in thermal
equilibrium  for  temperatures $100$ GeV $< T < 10^{12}$ GeV. As for
CP-breaking, the $\nu$MSM contains  $6$ CP-violating phases in the
lepton sector and a Kobayashi-Maskawa phase in the quark sector. This
makes two of the Sakharov conditions \cite{Sakharov:1967dj} for
baryogenesis satisfied. Similarly to the SM, this theory does not
have an electroweak phase transition with allowed values for the
Higgs mass \cite{Kajantie:1996mn}, making impossible the electroweak
baryogenesis, associated with the non-equilibrium bubble expansion.
However, the $\nu$MSM contains extra degrees of freedom - sterile
neutrinos - which may be out of thermal equilibrium exactly because
their Yukawa couplings to ordinary fermions are very small. The
latter fact is a key point for the baryogenesis in the $\nu$MSM,
ensuring the validity of the third Sakharov condition. 

In \cite{Akhmedov:1998qx} it was proposed that the baryon asymmetry
can be generated through CP-violating sterile neutrino oscillations.
For small Majorana masses the total lepton number of the system,
defined as the lepton number of active neutrinos plus the total
helicity of sterile neutrinos, is conserved and equal to zero during
the Universe's evolution. However, because of oscillations the lepton
number of active neutrinos becomes different from zero and gets
transferred to the baryon number due to rapid sphaleron transitions.
Roughly speaking, the resulting baryon asymmetry is equal to the
lepton asymmetry at the sphaleron freeze-out. 

The kinetics of sterile neutrino oscillations and of the transfers
of  leptonic number between active and sterile neutrino sectors has
been worked out in \cite{Asaka:2005pn}.  The effects to be taken into
account include oscillations, creation and destruction of sterile and
active neutrinos, coherence in sterile neutrino sector and its lost
due to interaction with the medium, dynamical asymmetries in active
neutrinos and charged leptons. For masses of sterile neutrinos
exceeding $\sim 20$ GeV the mechanism does not work as the sterile
neutrinos equilibrate. The temperature of baryogenesis is right above
the electroweak scale. 

In \cite{Asaka:2005pn} it was shown that the $\nu$MSM can provide
simultaneous solution to the problem of neutrino oscillations, dark
matter and baryon asymmetry of the Universe. 

{\em 4. Inflation.} In \cite{Shaposhnikov:2006xi} it was proposed that 
the $\nu$MSM may be extended by a {\em light} inflaton in order to
accommodate inflation. To reduce the number of parameters and to have
a common source for the Higgs and sterile neutrino masses the 
inflaton-$\nu$MSM couplings can be taken to be scale invariant at the
classical level and the Higgs mass parameter can be set to
zero. The mass of the inflaton can be as small as few
hundreds MeV, and the coupling of the lightest sterile neutrino to it
may serve as an efficient mechanism for the dark matter production. 

{\em 5. Fine-tunings in the $\nu$MSM.} The phenomenological
success of the $\nu$MSM requires a number of fine tunings. In
particular, one of the singlet fermion masses should be in the ${\cal
O} (10)$ keV region to provide a candidate for the dark-matter
particle, while two other masses must be much larger but almost
degenerate \cite{Asaka:2005pn,Akhmedov:1998qx} to enhance the
CP-violating effects in the sterile neutrino oscillations leading to
the baryon asymmetry. In addition, the Yukawa coupling of the dark
matter sterile neutrino must be much smaller than the Yukawa
couplings of the heavier singlet fermions, to satisfy cosmological
and astrophysical constrains \cite{Asaka:2005an}. These fine-tunings
are ``natural'' in a sense that they are stable against radiative
corrections. Moreover, in \cite{Shaposhnikov:2006nn} was shown that a
specific mass-coupling pattern for the singlet fermions, described
above, can be a consequence of a lepton number symmetry, slightly
broken by the Majorana mass terms and Yukawa coupling constants. At
the same time not all 18 new parameters are fixed: the allowed region
in parameter space is quite large to yield variety of signatures to
be tested with different experiments and methods.

To summarize, none of the {\em experimental facts}, which are
sometimes invoked as the arguments for the existence of the large
$\sim 10^{10}-10^{15}$ GeV intermediate energy scale between the
$W$-boson mass and the Planck mass, really requires it. The smallness
of the active neutrino masses may find its explanation in small
Yukawa couplings rather than in large energy scale. The dark matter
particle, associated usually with some stable SUSY partner of the
mass ${\cal O}(100)$ GeV or with an axion, can well be a much lighter
sterile neutrino, practically stable on the cosmological scales. The
thermal leptogenesis \cite{Fukugita:1986hr}, working well only at
large masses of Majorana fermions, can be replaced by the
baryogenesis through light singlet fermion oscillations. The
inflation can be associated with the light inflaton field rather than
with that with the mass $\sim 10^{13}$ GeV, with the perturbation
power spectrum coming from inflaton self-coupling rather than from
its mass.

Putting all the physics beyond the Standard Model below the
electroweak scale is not harmless, as it can be confronted with
experiment at {\em low} energies  (see e.g. \cite{Bezrukov:2005mx}
for a discussion of neutrinoless double beta-decay in the framework
of the $\nu$MSM).  The aim of this paper is to analyse the
possibilities to search for singlet fermions responsible for baryon
asymmetry of the Universe in the $\nu$MSM.  Finding these particles
and studying their properties in detail (in particular, CP-violating
amplitudes) would allow to compute the {\em sign} and the {\em
magnitude} of the baryon asymmetry of the Universe theoretically
along the lines of \cite{Asaka:2005pn} and confront this prediction
with observations. The existence of the U(1) lepton symmetry provides
an argument in favour of ${\cal O} (1)$ GeV mass of these singlet
leptons \cite{Shaposhnikov:2006nn}. In addition, the structure of
their couplings to the particles of the SM is almost fixed by the
data on neutrino oscillations. It is interesting to know, therefore,
what would be the experimental signatures of the neutral singlet
fermions in this mass range and in what kind of experiments they
could be found. To answer this question, in this paper we will
consider the variant of the $\nu$MSM without addition of the
inflaton; we will discuss what kind of differences one can expect if
the light inflaton is included elsewhere.

Naturally, several distinct strategies can be used for the
experimental search of these particles. The first one is related to
their production. The singlet fermions  participate in all reactions
the ordinary neutrinos do with a probability suppressed roughly by a
factor $(M_D/M_M)^2$, where $M_D$ and $M_M$ are the Dirac and
Majorana masses correspondingly. Since they are massive, the
kinematics of, say, two body decays $K^\pm \rightarrow \mu^\pm N$,
$K^\pm \rightarrow e^\pm N$ or three-body decays  $K_{L,S}\rightarrow
\pi^\pm + e^\mp + N$ changes when $N$ is replaced by
an ordinary neutrino. Therefore, the study of  {\em kinematics} of
rare meson decays can constrain the strength of the coupling of heavy
leptons. This strategy has been used in a number of experiments for
the search of neutral leptons in the past
\cite{Yamazaki:1984sj,Daum:2000ac}, where the spectra of electrons
or muons originating in decays of $\pi$- and $K$-mesons have been studied.
The second strategy is to look for the  decays of neutral leptons
inside a detector 
\cite{Bernardi:1985ny,Bernardi:1987ek,Vaitaitis:1999wq,Astier:2001ck}
(``nothing" $\rightarrow$ leptons and hadrons).  Finally, these two
strategies can be unified, so that the production and the decay
occurs inside the same detector \cite{Achard:2001qw}.

Clearly, to find the best way to search for neutral leptons, their
decay modes have to be identified and branching ratios must be
estimated. A lot of work in this direction has been already done in
Refs. 
\cite{Shrock:1980ct,Shrock:1981wq,Gronau:1984ct,Johnson:1997cj} for
the general case; we add new general results for three body meson
decays.  To analyze the corresponding quantities in the
$\nu$MSM we will constrain ourselves by the singlet fermion masses
below the mass of the beauty mesons, $M_N \lsim 5$ GeV, considering
this mass range as the most plausible because of the reasons
presented above. We will use the specific $\nu$MSM predictions for
the branching ratios. 

We arrived at the following conclusions.\\ 

(i) The singlet fermions with the masses smaller than $M_\pi$ are
already disfavoured on the basis of existing experimental data of
\cite{Bernardi:1985ny,Bernardi:1987ek} and from the requirement that
these particles do not spoil the predictions of the Big Bang
Nucleosynthesis (BBN) \cite{Dolgov:2000jw,Dolgov:2000pj}
(s.f. \cite{Kusenko:2004qc}).

(ii) The mass interval $M_\pi < M_N < M_K$ is perfectly allowed from
the cosmological and experimental points of view. Moreover, it is not
excluded that further constraints on the couplings of singlet
fermions can be derived from the reanalysis of the {\em already
existing but never considered from this point of view} experimental
data of KLOE collaboration and of the E949 experiment\footnote{We
thank Gino Isidori and Yury Kudenko for discussion of these points.}.
In addition, the NA48/3 (P326) experiment at CERN would allow to find
or to exclude completely singlet fermions with the mass below that of
the kaon\footnote{We thank Augusto Ceccucci for discussion of this
point.}. The search for the missing energy signal, specific for the
experiments mentioned above, can be complemented by the search of
decays of neutral fermions, as was done in CERN PS191 experiment
\cite{Bernardi:1985ny,Bernardi:1987ek}. To this end quite a number of
already existing or planned neutrino facilities (related, e.g. to
CNGS, MiniBoone, MINOS or T2K),  complemented by a near (dedicated) 
detector (like the one of CERN PS191) can be used\footnote{We thank
Francois Vannucci for discussion of this point.}. 
At the same time, the existing setups of the MiniBooNE or MINOS
experiments  would unlikely allow to probe the cosmologically
interesting parameter space of the $\nu$MSM for  $M_N <  450$ MeV,
where strong bounds on the parameters coming from CERN PS191
experiment already exist.  However, MiniBooNE and MINOS can possibly 
improve the existing limits or find neutral fermions in the mass
region  $450~{\rm MeV}< M_N <  M_K$, where current bounds are
weak (s.f. \cite{Kusenko:2004qc}). The record intensity of the neutrino
beam at CNGS and T2K experiment are quite promising for heavy neutrino
searches and calls for a detailed study of the possibility
of neutral fermions detection at (possible) near detectors.

(iii) For $M_K < M_N < M_D$ the search for the missing energy signal,
potentially possible at beauty, charm and $\tau$ factories, is
unlikely to gain the necessary statistics and is very difficult if
not impossible at hadronic machines like LHC\footnote{We thank Tasuya
Nakada for discussion of this point.}. So, the search for decays of
neutral fermions is the most effective opportunity. In short, an
intensive beam of protons hitting the fixed target, creates,
depending on its energy, pions, strange, charmed and beauty mesons
that decay and produce heavy neutral leptons. A part of these leptons
then decay inside a detector, situated some distance away from the
collision point. The  dedicated experiments on the basis of the
proton beams NuMI or NuTeV at FNAL, CNGS at CERN, or JPARC  can touch
a very interesting parameter range for $M_N \lsim 1.8$ GeV.

(iv) Going above $D$-meson but still below $B$-meson thresholds is
very hard if not impossible with present or planned proton machines or
B-factories. To enter into cosmologically interesting parameter space
would require the increase of the present intensity of, say, CNGS
beam by two orders of magnitude or producing and studying the
kinematics of more than $10^{10}$ B-mesons.  

The paper is organized as follows. In Section~\ref{Sec:Lagrangian} we
discuss the relevant part of the $\nu$MSM Lagrangian and specify the
predictions for the couplings of these particles coming from the data
on neutrino oscillations and cosmological considerations. In
Section~\ref{Sec:exp-limits} we analyze the present experimental and
cosmological limits on the properties of these particles. In
Section~\ref{decays} we analyze the decay modes of singlet
fermions. In Section~\ref{production} we consider the production of
heavy neutral leptons in decays of $K$-, $D$- and $B$-mesons and of
$\tau$-lepton.  In Section 6 we analyze the possibilities of their
detection in existing and future experiments. We conclude in Section
7.

\section{The Lagrangian and parameters of the $\nu$MSM} 
\label{Sec:Lagrangian} 
For our aim it is more convenient to use the Lagrangian of the
$\nu$MSM\footnote{ Of course, this Lagrangian is not new and is
usually used for the explanation of the small values of neutrino
masses via the see-saw mechanism \cite{Seesaw}. The see-saw scenario
assumes that the Yukawa coupling constants of the singlet fermions
are of the order of the similar couplings of the charged leptons or
quarks and that the Majorana masses of singlet fermions are of the
order of the Grand Unified scale. The theory with this choice of
parameters can also explain the baryon asymmetry of the Universe but
does not give a candidate for a dark matter particle. Another
suggestion is to fix the Majorana masses of sterile neutrinos in
$1-10$ eV energy range (eV see-saw) \cite{deGouvea:2005er} to
accommodate the LSND anomaly \cite{Aguilar:2001ty}. This type of
theory, however, cannot explain dark matter and baryon asymmetry of
the universe. Also, the MiniBooNE experiment
\cite{Aguilar-Arevalo:2007it} did not confirm the LSND result. The 
$\nu$MSM paradigm is to determine the Lagrangian parameters  from
available observations, i.e. from  requirement that it should explain
neutrino oscillations, dark matter and baryon asymmetry of the
universe in a unified way. This leads to the singlet fermion Majorana
masses  {\em smaller} than the electroweak scale, in the contrast
with the see-saw choice of \cite{Seesaw}, but much larger than few
eV, as in the eV see-saw of \cite{deGouvea:2005er}.} in
parameterization of Ref. \cite{Shaposhnikov:2006nn}:
\begin{eqnarray}
  {\cal L}_{\nu\rm{MSM}}
  = {\cal L}_{\rm{MSM}} + \bar {\wt N_I} i \partial_\mu \gamma^\mu
  {\wt N_I}
  - F_{\alpha I} \,  \bar L_\alpha {\wt N_I} \tilde \Phi
  - M \bar {\wt N_2}^c {\wt N_3} 
  - \frac{\Delta M_{IJ}}{2} \; \bar {\wt N_I}^c {\wt N_J} + \rm{h.c.} \,, 
   \label{lagr}
\end{eqnarray}  
where $\wt N_I$ are the right-handed singlet leptons (we will keep
the notation without tilde for mass eigenstates),
$\tilde\Phi_i=\epsilon_{ij}\Phi_j^*$, $\Phi$ and $L_\alpha$
($\alpha=e,\mu,\tau$) are  the Higgs and lepton doublets, 
respectively, $F$ is a matrix of Yukawa coupling constants, $M$ is
the common mass of two heavy neutral fermions, $\Delta M_{IJ}$ are  
related to the mass of the lightest sterile neutrino $N_1$ 
responsible for dark matter and produce the small splitting of the
masses of $N_2$ and $N_3$, $\Delta M_{IJ}\ll M$.  The Yukawa coupling
constants of the dark  matter neutrino $|F_{\alpha 1}| \lsim
10^{-12}$ are strongly bounded by cosmological considerations
\cite{Asaka:2005an} and by the X-ray observations
\cite{Boyarsky:2006fg} and can be safely neglected for the present
discussion and the sterile neutrino $N_1$ field can be omitted from
the Lagrangian. 

In the limit $\Delta M_{IJ} \to 0$, $F_{\alpha 2}  \to 0$ the
Lagrangian (\ref{lagr}) has a global U(1) lepton symmetry
\cite{Shaposhnikov:2006nn}. In this paper we will assume that the
breaking of this symmetry is small not only in the mass sector (which
is required for baryogenesis and explanation of dark matter), but
also in the Yukawa sector, $|F_{\alpha 3}|  \ll |F_{\beta 2}|$.  For
the case when $|F_{\alpha 3}|  \sim |F_{\beta 2}|$ our general
conclusions remain the same, but the branching ratios for different
reactions can change. In this work we also neglect all CP-violating
effects, which go away if the lepton number symmetry is exact. 

To characterize the measure of the $U(1)_L$ symmetry breaking, we
introduce a small parameter $\epsilon =F_3/F \ll 1$, where $F_i^2 =
[F^\dagger F]_{ii}$, and $F_2\equiv F$. As was shown in 
\cite{Shaposhnikov:2006nn}, there is a lower bound on $\epsilon$
coming from the baryon asymmetry of the Universe, $\epsilon \gsim 
0.6\cdot\kappa\cdot 10^{-4}(M/{\rm GeV})$, where $\kappa=1(2)$ for
the case of normal(inverted) hierarchy in active neutrino sector. 

The mass eigenstates ($N_{2,3}$ without tilde) are related to $\wt
N_{2,3}$ by the unitary transformation,
\be
\wt N = U_R N\;,
\ee
where the $2 \times 2$ matrix $U_R$ has the form 
\be
U_R \simeq \frac{e^{i\phi_0}}{\sqrt{2}}
\left(
   \begin{array}{c c }
      e^{i\phi_1} &  e^{i\phi_2}\\
      - e^{-i\phi_2}&  e^{-i\phi_1}
   \end{array}
  \right)~,
  \label{UR}
\ee
where the phases $\phi_k$ can be expressed through the elements of
$\Delta M_{IJ}$, the explicit form of which is irrelevant for us.

As a result, for $\epsilon \ll 1$ the interaction of the mass
eigenstates $N_2$ and $N_3$ has a particular simple form, 
\be 
L_N
\simeq - \frac{1}{\sqrt{2}} f_\alpha \bar L_\alpha (N_2+N_3) \tilde\Phi -
\frac{M_2}{2}\bar {N_2{}^c} N_2 - \frac{M_3}{2}\bar {N_3{}^c} N_3+
\rm{h.c.} \,,  \label{simpl} 
\ee 
where $f_\alpha =|F_{\alpha 2}|$. The masses $M_2$ and $M_3$ must be
almost the same (baryogenesis constraint), $\Delta M^2=|M_2^2-M_3^2|
\lsim 10^{-5} M^2$
\cite{Asaka:2005pn,Akhmedov:1998qx,Shaposhnikov:2006nn}. The baryon
asymmetry generation occurs most effectively if  $\Delta M^2 \simeq
({2~\rm keV})^2$, but smaller and larger degeneracy works well
also. 

The fact that two heavy fermions are almost degenerate in mass may be
important for analysis of the experimental constraints. In decays of
different mesons or $\tau$-lepton a coherent combination $(N_2+N_3)$
will be created, while in a detector of size $l$ situated on the
distance $L$ from the creation point an admixture of the $(N_2-N_3)$
state with the probability (in the relativistic limit) $P \sim
\sin^2\phi$ will appear ($E$ is the energy of the neutral fermion,
$\phi=\Delta M^2 L/(4E)$). For $\phi l/L \gg 1$ coherence effects are
not essential and the description of the process in terms of $N_2$
and $N_3$ is completely adequate, while if  $\phi l/L \sim 1$ the
coherence effects are important, and order $\epsilon$ terms
describing the interactions of $(N_2-N_3)$ with the particles of the
SM must be included. Numerically, if $\Delta M^2 \gsim ({2~ \rm
keV})^2$, $l\sim 10$ m, and $E < 100$ GeV, then  $\phi l/L \gsim
10^3$, and $N_2 \leftrightarrow N_3$ oscillations can be safely
neglected. Only this case will be considered in what follows.

As it was demonstrated in \cite{Shaposhnikov:2006nn}, the coupling
constants $f_\alpha$ can be expressed through the elements of the
active neutrino mass matrix $M_\nu$. To present the corresponding
relations, we parameterize $M_\nu$ following Ref.~\cite{Strumia:2005tc}:
\be
M_\nu= V^*\cdot {\rm diag}(m_1,m_2e^{2i\delta_1},m_3e^{2i\delta_2})
\cdot V^\dagger~,
\label{act}
\ee
with $V = R(\theta_{23}) \rm{diag}(1,e^{i\delta_3},1) R(\theta_{13})
R(\theta_{12})$ the active neutrino mixing matrix \cite{numix}, and
choose for normal hierarchy $m_1<m_2<m_3$ and for inverted hierarchy
$m_3<m_1<m_2$. All active neutrino masses are taken to be positive.
As was shown in \cite{Asaka:2005an,Boyarsky:2006jm,Boyarsky:2006fg,Asaka:2006nq}, the one of the
active neutrino masses must be much smaller than the solar mass
difference, $m_{\rm{sol}}=\sqrt{\Delta m^2_{\rm{sol}}}\simeq 0.01$
eV, so that other active neutrino masses are simply equal to 
$m_{\rm{atm}}=\sqrt{\Delta m^2_{\rm{atm}}}\simeq 0.05$ eV and to
$m_{\rm{sol}}$ for the case of normal hierarchy and to $m_{\rm{atm}}$
with a mass splitting $\delta m = m_{\rm{sol}}^2/2 m_{\rm{atm}}$ for
the case of inverted hierarchy.

The coupling $F$ is given by  \cite{Shaposhnikov:2006nn}:
\be
F^2  \simeq \kappa \frac{m_{\rm{atm}} M}{2v^2 \epsilon}~,
\label{Ffix}
\ee
where  $v=174$ GeV is the vev of the Higgs field and  $\kappa \simeq
1 (2)$ for the case of normal (inverted) hierarchy.

The ratios of the Yukawa couplings $f_\alpha$ can be expressed
through the elements of the active neutrino mixing matrix
\cite{Shaposhnikov:2006nn}. A  simple expression can be derived for
the case $\theta_{13}=0,~\theta_{23}=\pi/4$, which is in agreement
with the experimental data. For normal hierarchy there are
possibilities:
\be
f_e^2:f_\mu^2:f_\tau^2 \approx
\frac{m_2}{m_3}\sin^2\theta_{12}|1\pm
x|^2:\frac{1}{2}|1-x^2|^2:\frac{1}{2}|1\pm x|^4~,
\label{Yukawanormal}
\ee
where  $x=i
e^{i(\delta_1-\delta_2-\delta_3)}\sqrt{\frac{m_2}{m_3}}\cos\theta_{12}$,
and all combinations of signs are admitted.  For a numerical estimate
one can take \cite{Strumia:2005tc} $\sin^2\theta_{12}\simeq 0.3$,
leading to $x\simeq 0.35 i e^{i(\delta_1-\delta_2-\delta_3)}$ and to
$f_e^2/(f_\mu^2+f_\tau^2)\sim 0.05$. In other words, the coupling of
the singlet fermion to the leptons of the first generation is
suppressed, whereas the couplings to the second and third generations
are close to each other.  

For the case of inverted hierarchy two out of four solutions are
almost degenerate and one has \cite{Shaposhnikov:2006nn}:
\be
f_e^2:f_\mu^2:f_\tau^2 \approx
\frac{1+p}{1-p}:\frac{1}{2}:\frac{1}{2}~,
\label{Yukawainverted}
\ee
where $p=\pm \sin\delta_1\sin(2\theta_{12})$. Taking the same value
of $\theta_{12}$ as before, we arrive at 
$f_e^2/(f_\mu^2+f_\tau^2)\sim (0.04-25)$, depending on the value of
unknown CP-violating phase $\delta_1$. The couplings of $N_{2,3}$ to
$\mu$ and $\tau$ generations are almost identical, but the coupling
to electron and its neutrino can be enhanced or suppressed
considerably. The corrections to relations
(\ref{Yukawanormal},\ref{Yukawainverted}) are of the order of ${\cal
O}(\epsilon)$ and for $\epsilon \sim 1$ the ratios of the coupling
constant can be quite different from those in eqns.
(\ref{Yukawanormal},\ref{Yukawainverted}).

The relations (\ref{Ffix},\ref{Yukawanormal},\ref{Yukawainverted})
form a basis for our analysis of experimental signatures of heavy
neutral leptons. In most of the works the strength of the coupling of
a neutral lepton $X$ to charged or neutral currents of flavour
$\alpha$ is characterized by quantities $U_{\alpha X}$ and $V_{\alpha
X}$. In the case of the $\nu$MSM there are two neutral leptons with
almost identical couplings (if $\epsilon \ll 1$), so that
\be
|U_{\alpha 1}|=|V_{\alpha 1}|=|U_{\alpha 2}|=|V_{\alpha 2}|\equiv
|U_\alpha|~.
\ee
The overall strength of the coupling is given by
\be
U^2 \equiv \sum_\alpha |U_{\alpha}|^2= \frac{F^2 v^2}{2M^2}~,
\ee
whereas the relations between different flavours follow from 
(\ref{Yukawanormal},\ref{Yukawainverted}). 
 
As it was found in \cite{Shaposhnikov:2006nn}(see also
\cite{Akhmedov:1998qx,Asaka:2005pn}), for successful baryogenesis the
constant $F$ must be small enough, $F \lsim 1.2\times 10^{-6}$,
otherwise $N_2$ and $N_3$ come to thermal equilibrium above the
electroweak scale and the baryon asymmetry is erased. This leads to
the upper bound
\be
U^2 < 2\kappa \times 10^{-8} \left(\frac{\rm GeV}{M}\right)^2~.
\label{upper}
\ee

It is the smallness of the required strength of coupling which makes the
search for neutral leptons of the $\nu$MSM be a very challenging
problem, especially for large $M$.  

In the framework of the $\nu$MSM, a lower bound on $U$ can be derived
as well. The maximal value of the parameter $\epsilon$,
characterizing the breaking of the U(1) leptonic symmetry is
$\epsilon = 1$. This results in  
\be 
U^2 > 1.3\kappa \times 10^{-11} \left(\frac{\rm GeV}{M}\right)~.
\label{lower}
\ee

Further cosmological constraints on the couplings of heavy sterile
neutrinos are coming from BBN.  The cosmological production rate of
these particles peaks roughly at the temperature 
\cite{Dolgov:2000jw}  $T_{peak} \sim 10 \left(M/{\rm
GeV}\right)^{1/3}~{\rm GeV}$ and for  $U^2 > 2 \times 10^{-13}
\left({\rm GeV}/M\right)~$ they were in thermal equilibrium in some
region of temperatures around $T_{peak}$. This is always true, since
in the $\nu$MSM the constraint \eqref{lower} is required to be valid.
We will see below that the BBN constraints are in fact stronger than
those of \eqref{lower} for relatively small fermion masses $M_N < 1$
GeV.  On the basis of inequalities (\ref{upper},\ref{lower}) and
limits from BBN, the $\nu$MSM can be probed  (either confirmed or
ruled out) in  particle physics experiments. 

The relations (\ref{Yukawanormal},\ref{Yukawainverted}) still allow a
lot of freedom in relations between Yukawa couplings to different
leptonic flavours, since the Majorana CP-violating phases in the
active neutrino mass matrix are not known. Therefore,   to present
quantitative predictions we will consider  three sets of Yukawa
couplings corresponding to  three ``extreme hierarchies'', when value
of Yukawa constants $f_\alpha$, $f_\beta$ are taken to be as small 
as possible compared to another one $f_\gamma$,
$\alpha\neq\beta\neq\gamma$, which thus mostly determines the overall
strength of mixing $U^2$. In what follows we will refer to these sets
as benchmark models I, II  and III   with ratios of coupling
constants which can be read off from eqs.~\eqref{Yukawanormal},
\eqref{Yukawainverted}: 
\begin{align*}
{\rm model~I}\;:& \hskip 0.5cm f_e^2:f_\mu^2:f_\tau^2\approx 52:1:1\;,
~~~\kappa=1\;,\\
{\rm model~II}\;:& \hskip 0.5cm f_e^2:f_\mu^2:f_\tau^2\approx 1:16:3.8\;,
~~~\kappa=2\;,\\
{\rm model~III}\;:& \hskip 0.5cm f_e^2:f_\mu^2:f_\tau^2\approx 0.061:1:4.3\;,
~~~\kappa=2\;.
\end{align*}

Let us explain how these numbers were obtained. For the  model I we
simply increase  in a maximal way the value of the coupling constant to 
electron, choosing the appropriate combination of signs in eq.
(\ref{Yukawainverted}). In case of model III the coupling of $N$ to
the third generation of leptons is stronger than to the others. This
could only happen if the hierarchy of active neutrino masses is
normal, see eq. (\ref{Yukawanormal}). Choosing real and positive $x$
one can see that the maximum value of the ratio $|f_\tau/f_\mu|^2$ is
given by
\be
\frac{|f_\tau|^2}{|f_\mu|^2}
\simeq\left(\frac{1+x}{1-x}\right)^2\;.
\label{max}
\ee
As reference point we choose the central values of parameters of
neutrino mixing (see, e.g. \cite{Strumia:2005tc}), that gives
$x\approx 0.35$.  This means that the ratio \eqref{max} can be as
large as $4.3$ (varying the parameters of the active neutrino mixing
matrix within their error bars one arrives at a bit larger number). 
By the same type of reasoning the maximal values of the ratio 
$|f_\tau/f_e|^2$ is given by 
\be
\frac{|f_\tau|^2}{|f_e|^2}
\simeq
\left(\frac{m_2}{2m_3}\sin^2{\theta_{12}}\cdot
\left(\frac{1-x}{|1+x|^2}\right)^2\right)^{-1}
\simeq 71\;.
\label{maxe}
\ee
Similar considerations provide values of Yukawa couplings in
model II.  

These benchmark models are choosen to show the variety of
quantitative predictions within originally 18-dimensional parameter
space of $\nu$MSM, constrainted already by cosmology, astrophysics,
and observations of neutrino oscillatuions. For a given process, they
should be confined between numbers given for benchmark models for
$\epsilon\ll 1$. A special study should be undertaken to outline the
actual range of $\nu$MSM predictions  in case of $\epsilon\sim 1$,
when relations \eqref{Yukawanormal} and  \eqref{Yukawainverted} become
invalid.

\section{Laboratory and BBN constraints on the properties of heavy
leptons} 
\label{Sec:exp-limits}
The aim of this section is to discuss whether the past experiments
devoted to the search for neutral leptons have entered into
cosmologically interesting parameter range defined by eqns.
(\ref{upper},\ref{lower}). In addition, we will consider the Big Bang
Nucleosynthesis constraints on the properties of heavy leptons in the
$\nu$MSM. 

The analysis of the published works of different collaborations
reveals that for the mass of the neutral lepton $M > 450$ MeV none of
the past or existing experiments enter into interesting for $\nu$MSM
region defined by eq.~(\ref{upper}). The NuTeV  upper limit on the
mixing is at most $10^{-7}$ in the region $M\simeq 2$ GeV
\cite{Vaitaitis:1999wq}, whereas the NOMAD \cite{Astier:2001ck} and
L3 LEP experiment \cite{Achard:2001qw} give much weaker constraints.
Note that the eqns.~(\ref{upper},\ref{lower}) give at $M=2$ GeV:
$6\cdot 10^{-12} < U^2/\kappa < 5\cdot 10^{-9}$. 

The best constraints in the small mass region, $M < 450$ MeV are
coming from the CERN PS191 experiment
\cite{Bernardi:1985ny,Bernardi:1987ek}, 
giving\footnote{{\it The most recent 
published results} of CERN SPS experiment 
\cite{Bernardi:1987ek}
contain the exclusion plots up to 400 MeV. In a previous publication, 
\cite{Bernardi:1985ny}, the limit on $U_e^2$, though not as strong as
in \cite{Bernardi:1987ek}, was presented up to 450 MeV. We 
became aware of PhD Thesis of J.-M. Levy \cite{Levy} 
(we thank F. Vannucci for providing us a
copy of this manuscript) which contains the experimental exclusion plots for
$U_e^2$ and $|U_e U_\mu|$ up to $450$ MeV.  We use these 
unpublished results in our work. If the results of \cite{Levy} are ignored, our
plots should be modified accordingly in the region $400$ MeV $< M_N <450$
MeV, and phenomenologically viable region expands.}
roughly $|U_{e,\mu}|^2
\lsim 10^{-9}$ in the region $250$ MeV $< M < 450$ MeV (the NuTeV
limit in this mass range is some two orders of magnitude
weaker). These numbers are already in the region (\ref{upper}) and
thus provide non-trivial limits on the parameters of the
$\nu$MSM. Moreover, as it will be seen immediately, the considerations
coming from BBN allow to establish a number of {\em lower} bounds on
the couplings of neutral leptons which decrease considerably the
admitted window for the couplings and masses of the neutral leptons.

The successful predictions of the BBN are not spoiled provided the
life-time of sterile neutrinos is short enough. Then neutrinos decay
before the onset  of the BBN and the products of their decays
thermalize.  This question has been studied in \cite{Dolgov:2000jw}
and we will use the results of their general analysis for the case of
Models I-III described in Section \ref{Sec:Lagrangian}.

First, we note that \cite{Dolgov:2000jw} considered the case of one
sterile neutrino of Dirac type, whereas we have two Majorana sterile
neutrinos\footnote{The concentration of the dark matter sterile
neutrinos is well below the equilibrium one so that its existence may
be safely neglected at this time.}. This  means that we have exactly
the same number of degrees of freedom and that the constraints of
\cite{Dolgov:2000jw}, expressed in terms of  {\it lifetime} of
sterile neutrino are applicable to our case.

Ref. \cite{Dolgov:2000jw} studied in detail only the mass range $10$
MeV $ < M_N < 140$ MeV, for higher masses these authors argued that
the life-time $\tau_N$ of the heavy lepton must be smaller than $0.1$
s  to definitely avoid any situation when heavy lepton decay products
could change the standard BBN pattern of  light element abundances. 
We note in passing that it would be extremely interesting to repeat
the computation of \cite{Dolgov:2000jw} for $M_N > 140$ MeV in order
to have a robust BBN constraints in this mass range; meanwhile we
will just require  (conservatively)   that $\tau_N < 0.1$ s for
neutral fermions heavier than $\pi$-meson.

For the masses in the interval $10$ MeV $ < M < 140$ MeV the
constraint on the mixing angle, based on a fit to numerical BBN
computations \cite{Dolgov:2000jw}, reads 
\be 
U_{I\beta}^2 >
\frac{1}{2}\left(s_{1,\beta} \left(M/{\rm MeV}\right)^\alpha_\beta
+s_{2,\beta}\right)
\label{bbn1}
\ee 
with $s_{1,e}=140.4$, $s_{1,\mu}=s_{1,\tau}=568.4$,
$s_{2,e}=-1.05\cdot 10^{-5}$,
$s_{2,\mu}=s_{2,\tau}=-5.17\cdot10^{-6}$, $\alpha_e=-3.070$ and
$\alpha_\mu=\alpha_\tau =-3.549$ (we took a conservative bound
equivalent to adding one extra neutrino species, as explained in
\cite{Dolgov:2000jw}); the limits \eqref{bbn1} are valid in the
models where sterile neutrino mix predominantly with only one active
flavor.  Here we took into account that in Ref.~\cite{Dolgov:2000jw}
neutrinos of Dirac type have been considered, while we discuss
neutrino of Majorana type, hence the total width contains an extra factor
$2$ in comparison with the Dirac case and the constraint of $U^2$ is
in fact $2$ times weaker than that of \cite{Dolgov:2000jw}. The
limits \eqref{bbn1} can be converted into limits on  the mixing $U^2$
for the models I-III. 

To consider higher masses we computed the life-time of heavy leptons
(the details of computation can be found in Section~\ref{decays}) and
required  that it exceeds $0.1$ s, to make a conservative exclusion
plot. The
most important decay channels for $M_N < M_K$ are the two-body
semileptonic ones $N\rightarrow \pi^0\nu,~~N\rightarrow \pi^\pm
e^\mp,~~ N\rightarrow \pi^\pm \mu^\mp$.

\begin{figure}[!htb]
\centerline{
\includegraphics[width=0.33\textwidth]{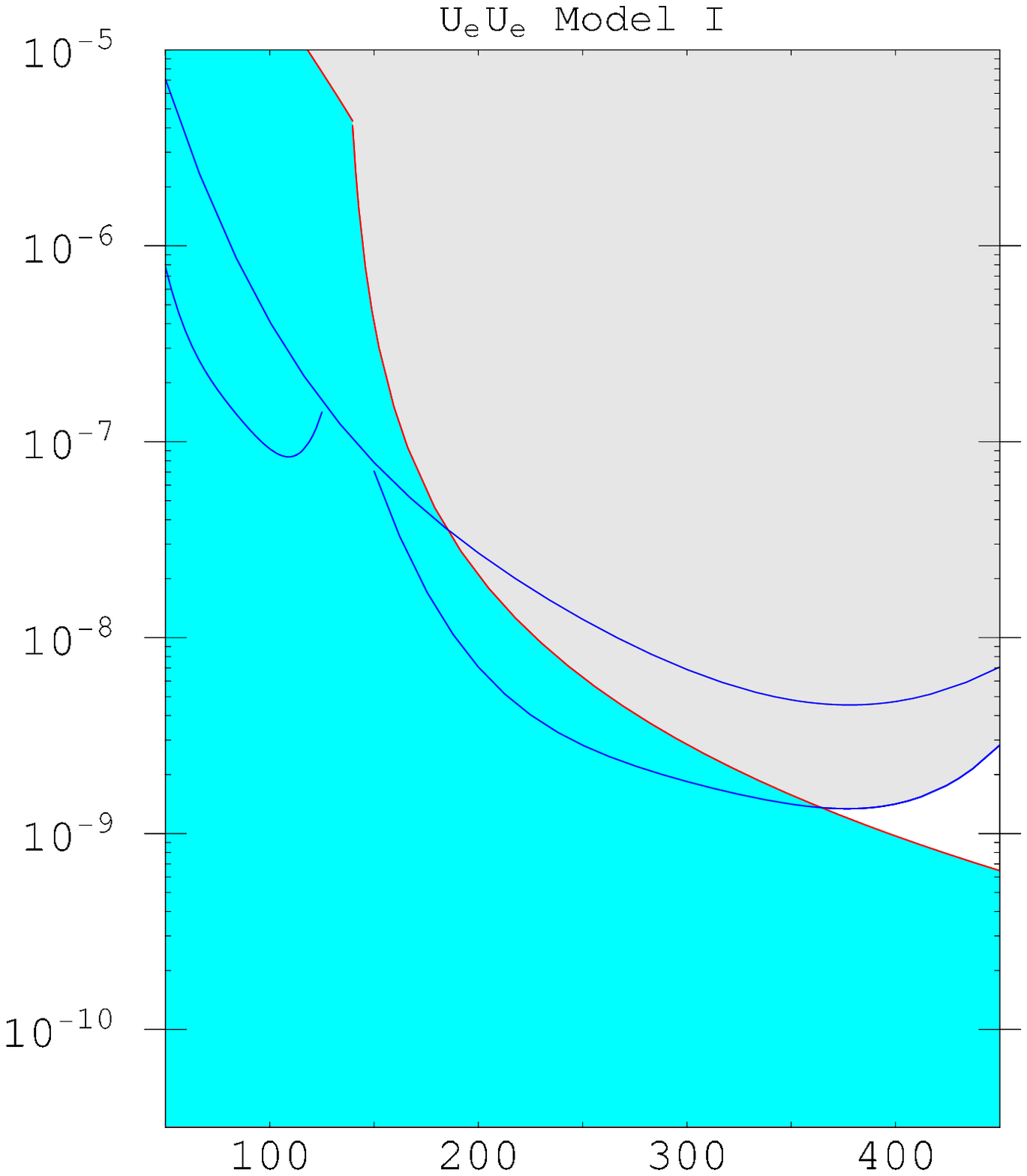}
\includegraphics[width=0.33\textwidth]{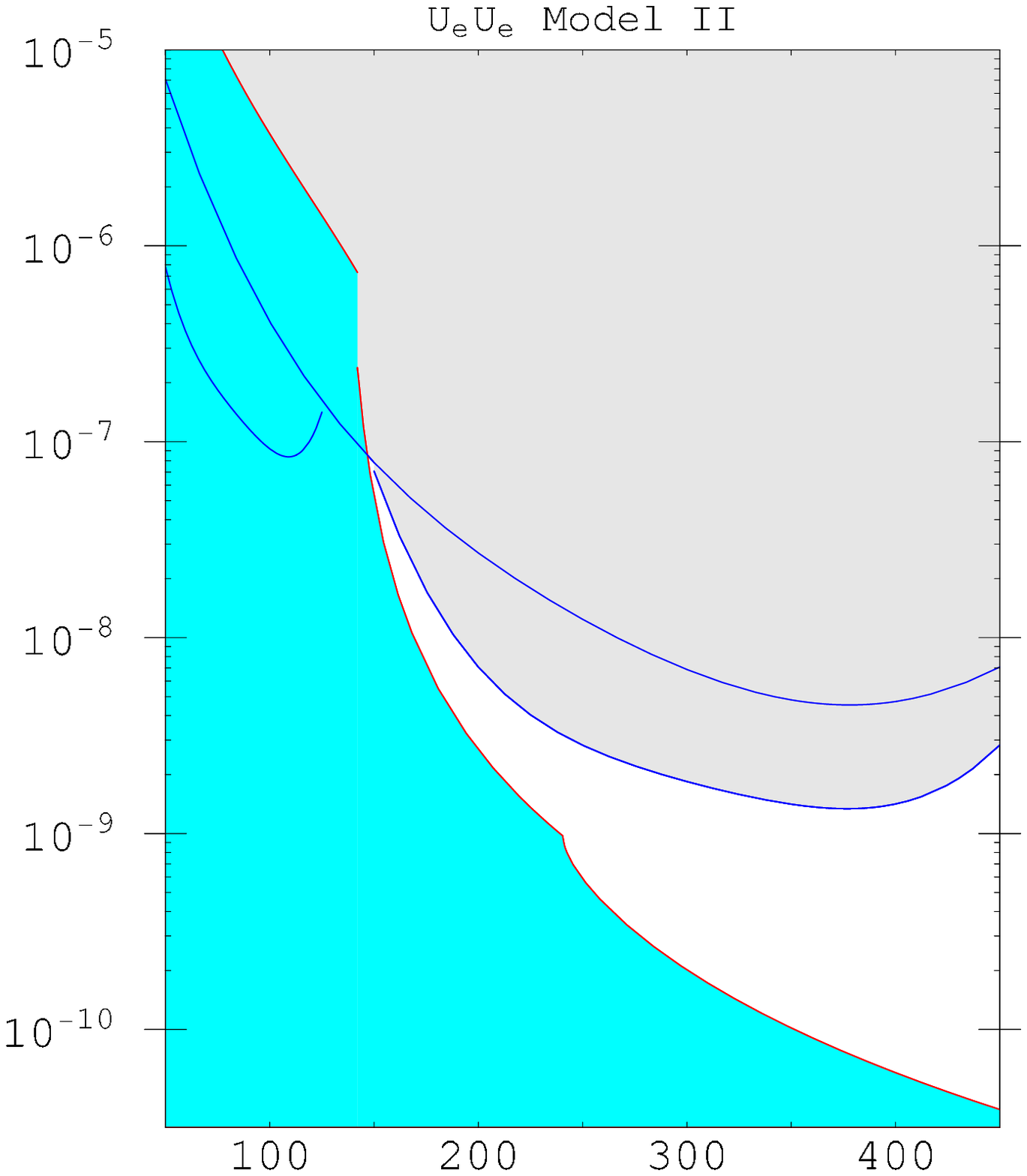}
\includegraphics[width=0.33\textwidth]{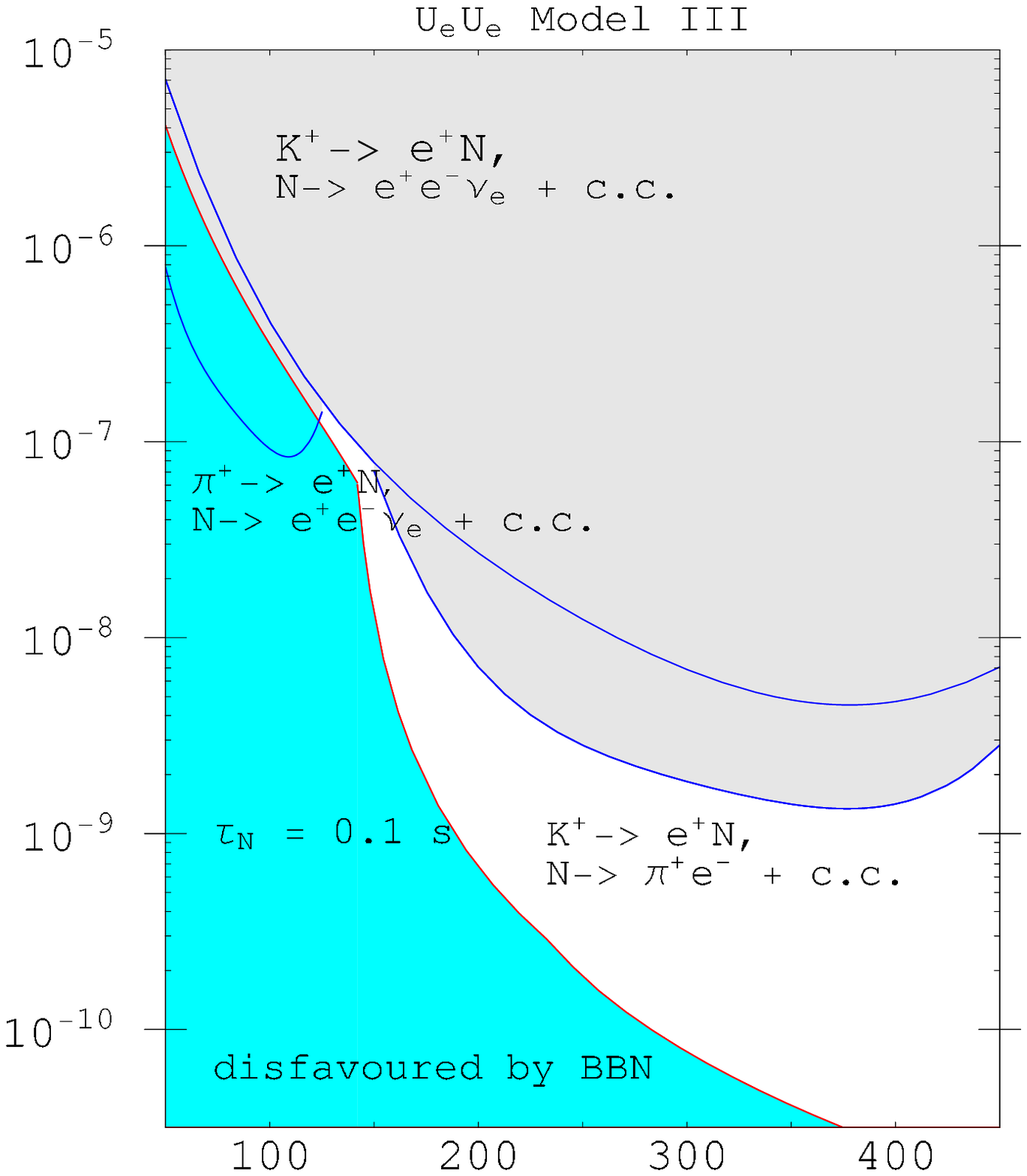}
}
\centerline{
\includegraphics[width=0.33\textwidth]{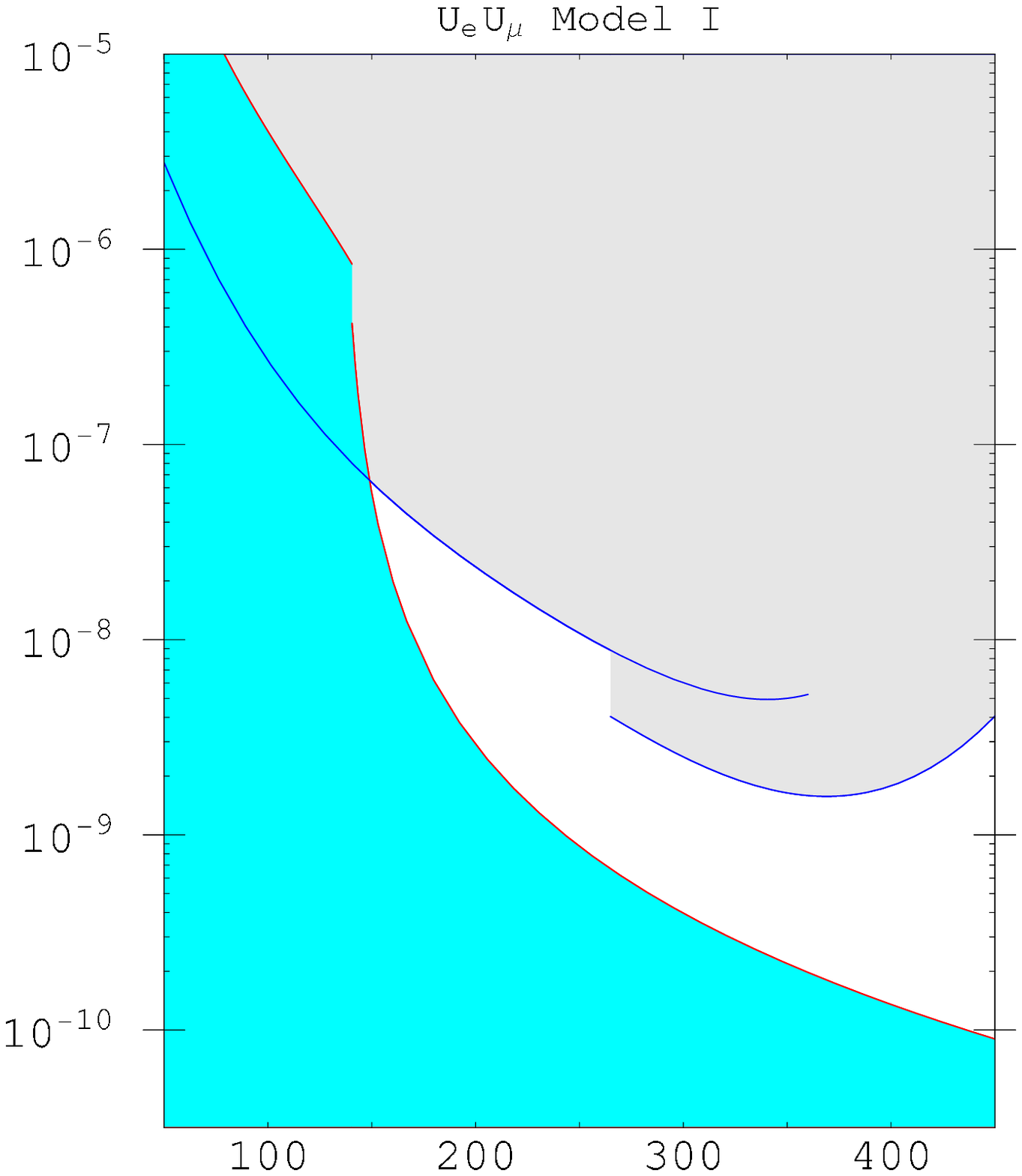}
\includegraphics[width=0.33\textwidth]{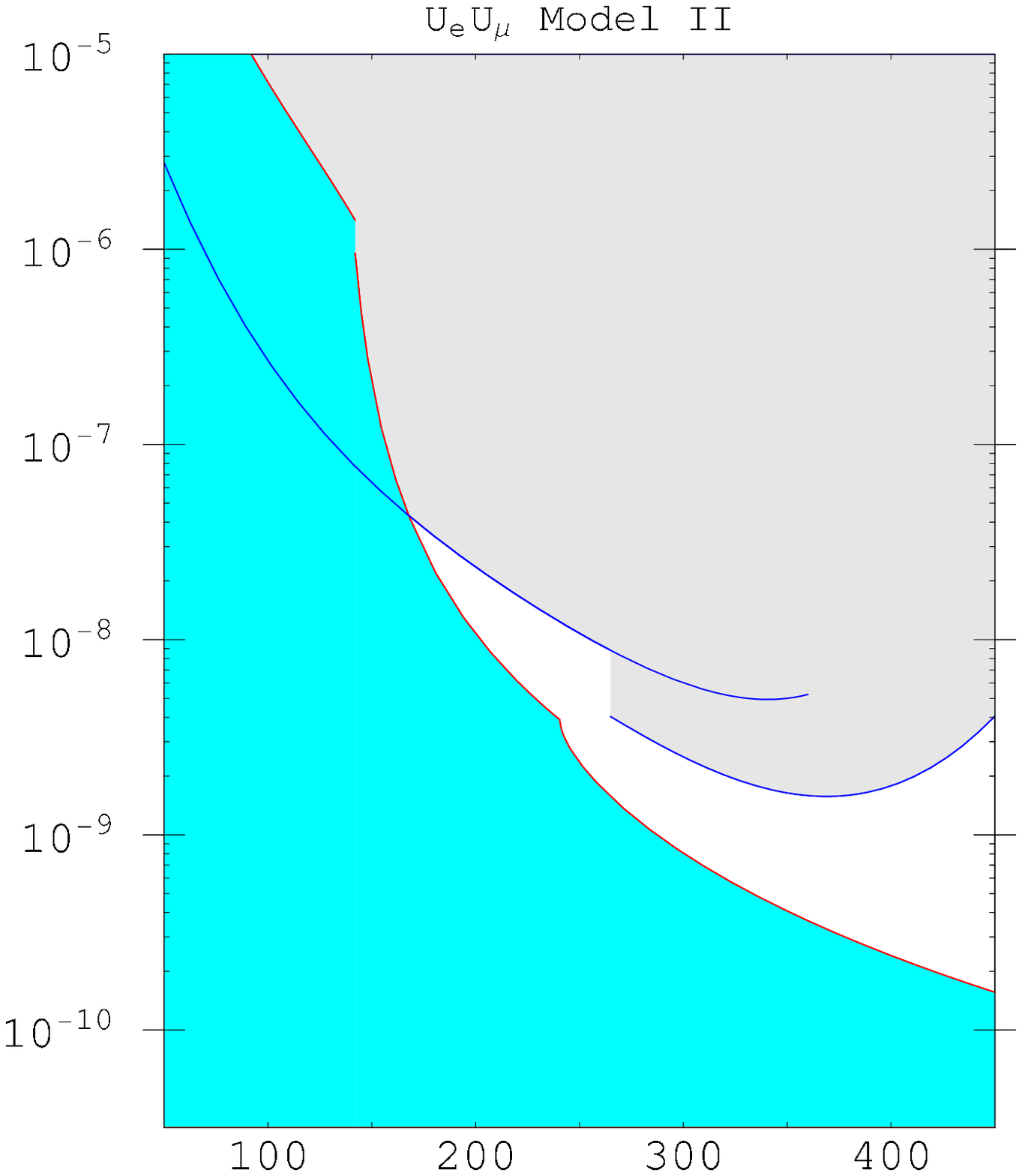}
\includegraphics[width=0.33\textwidth]{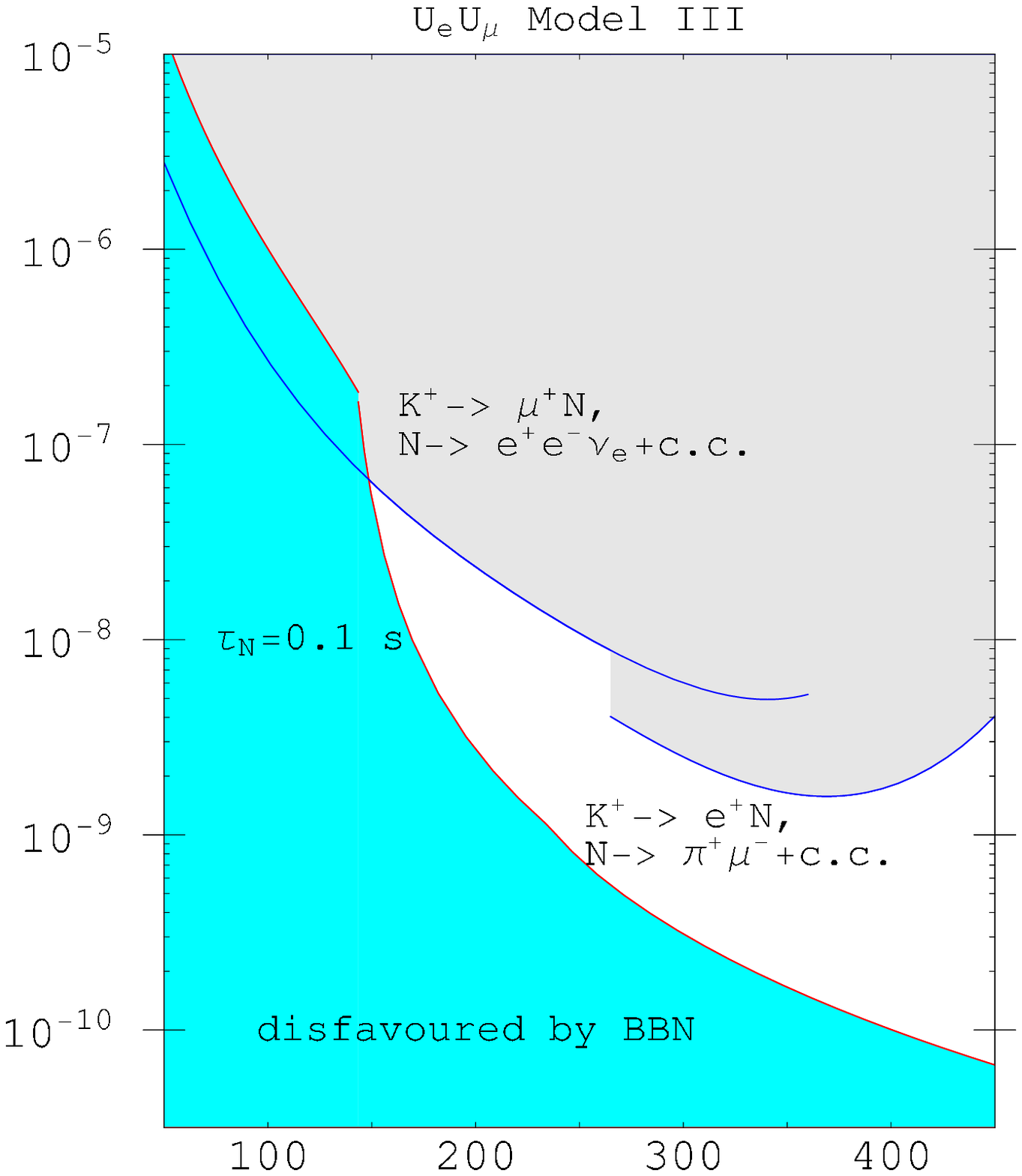}
}
\centerline{
\includegraphics[width=0.33\textwidth]{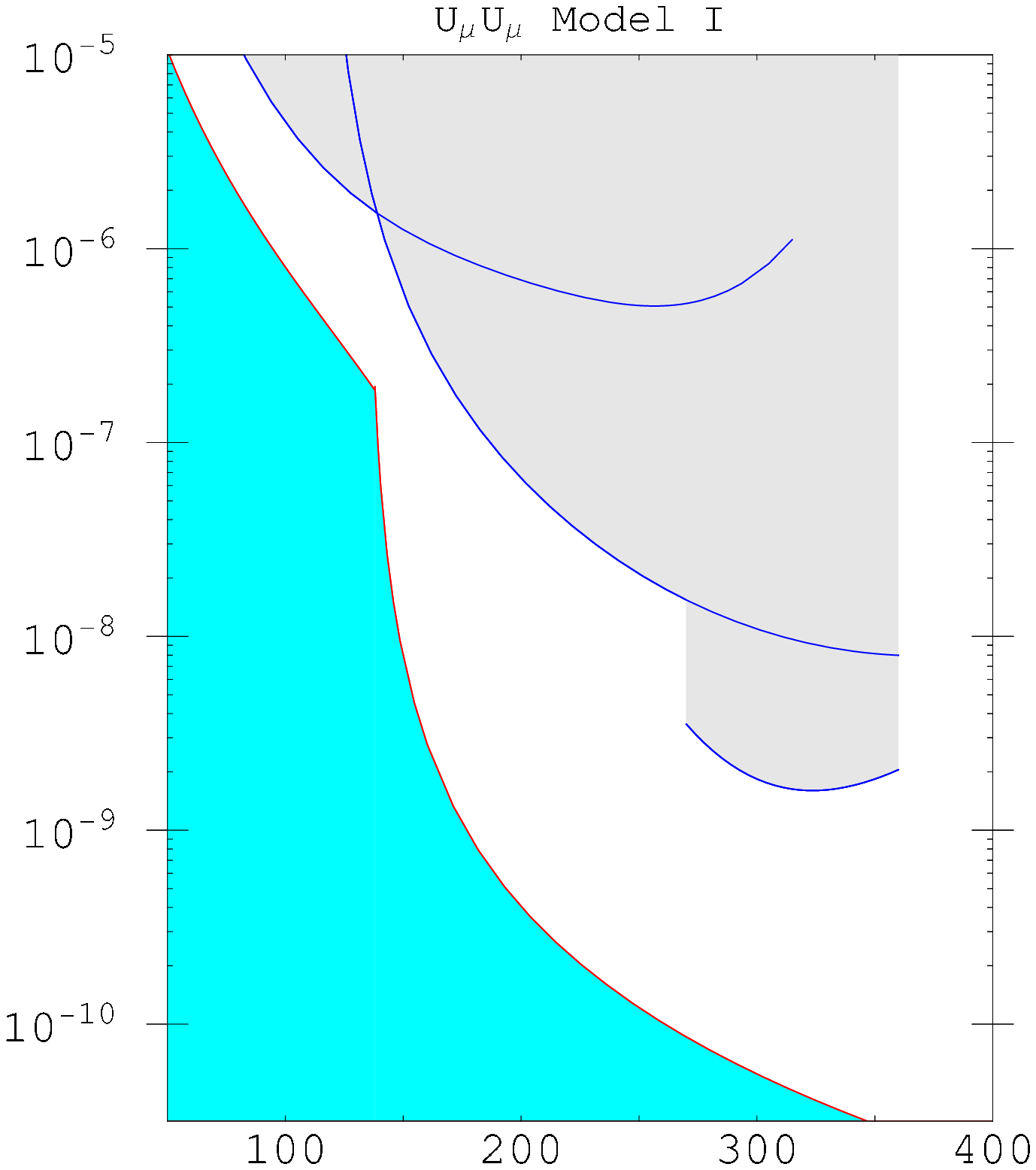}
\includegraphics[width=0.33\textwidth]{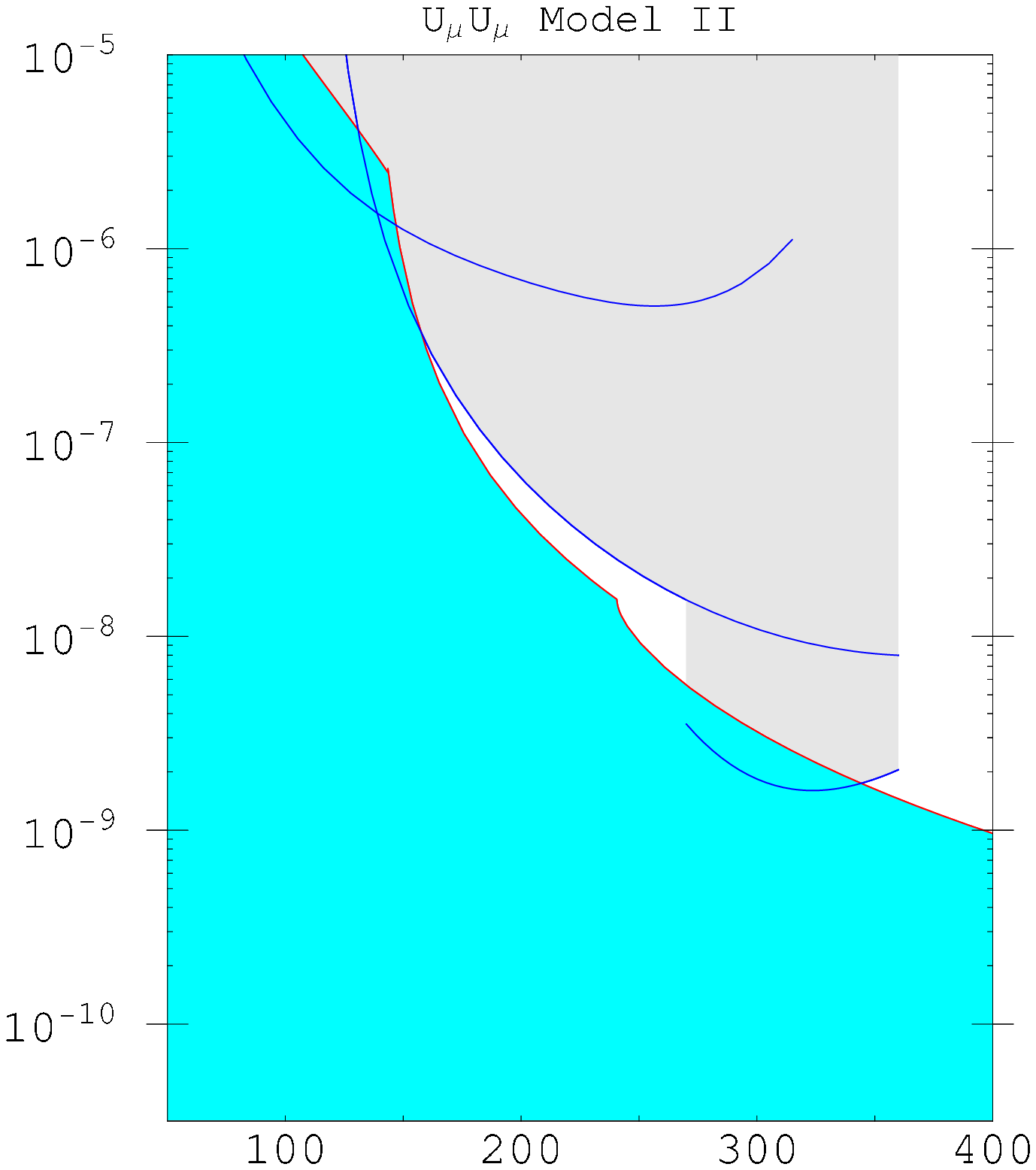}
\includegraphics[width=0.33\textwidth]{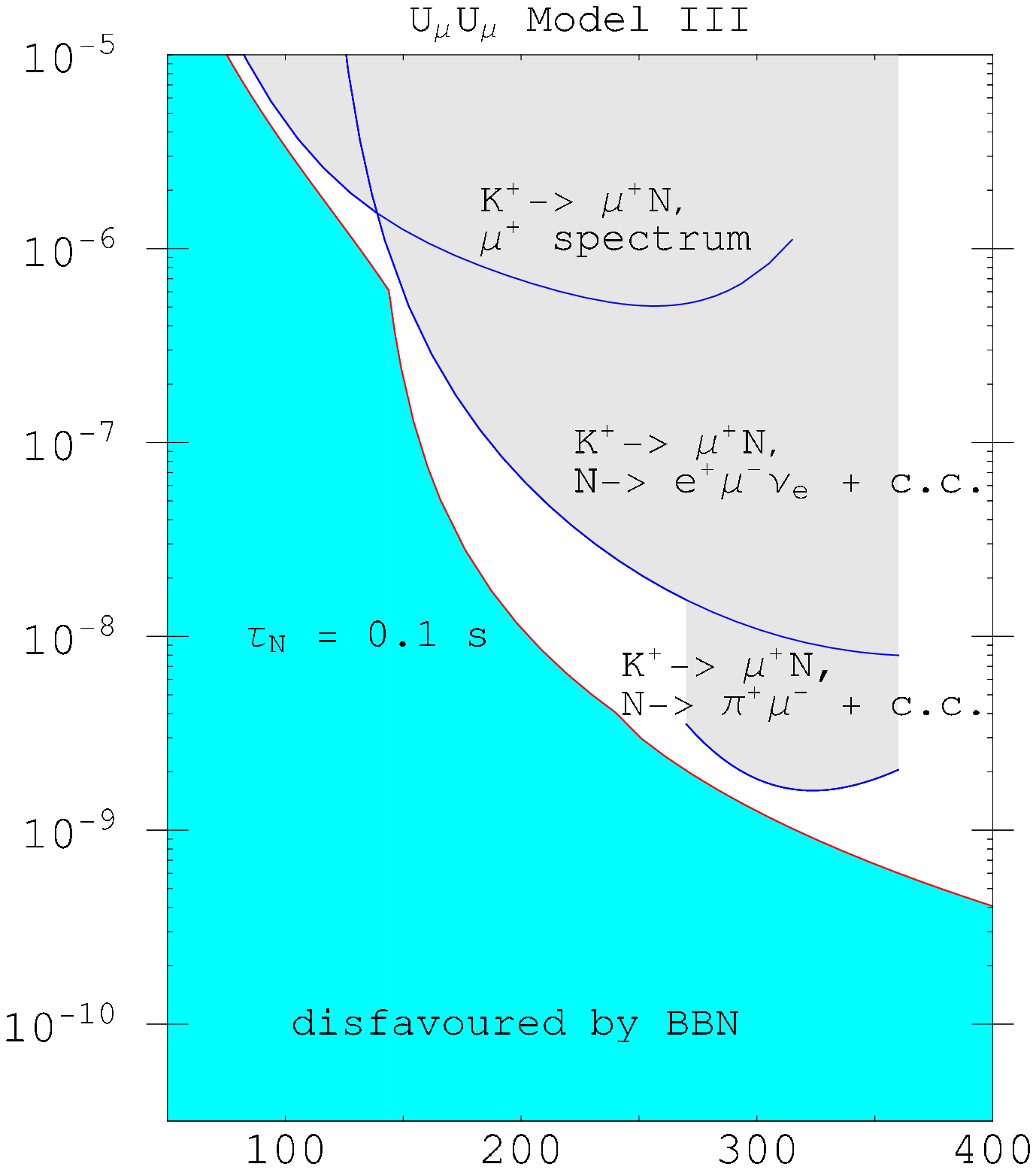}
}
\begin{picture}(0,0)(0,0)
\end{picture}
\vskip -1cm
\caption{Limits on $|U_e|^2$, $|U_e||U_\mu|$ and $|U_\mu|^2$ for three
  benchmark models (I-III from left to right) 
from BBN (lower bound) and from direct searches 
in the CERN PS191 experiment (upper bound). Blank regions 
are phenomenologically allowed.    
\label{direct-limits}
}
\end{figure}

For various patterns of neutrino mixing we present the experimental
and BBN constraints in Fig.~\ref{direct-limits}. Note that in
extracting the limits on mixing from
\cite{Bernardi:1985ny,Bernardi:1987ek} (this experiment presented 90\%
confidence level exclusion plot) we also take into account that there
are two degenerate neutrinos in the $\nu$MSM, and that the constraints
in \cite{Bernardi:1985ny,Bernardi:1987ek} are given for Dirac type
sterile neutrinos. For the same value of the mixing angles, the same
number of sterile neutrino helicity states are created in both Dirac
and Majorana cases, but in the former case only half of states
contribute to each decay channel.  Hence, the constraints on
$|U_e|^2$, $|U_e||U_\mu|$ and $|U_\mu|^2$ are in fact by a factor $2$
stronger, since the number of decay events is proportional to $|U|^4$.

One can see that depending on the type of the neutrino mass hierarchy
and specific branching ratios in the benchmark models I-III the
phenomenologically allowed region of parameter space can be reduced
or enlarged. Moreover, the masses below the $\pi$ meson mass are
excluded in most cases\footnote{For the $\nu$MSM with light inflaton
BBN bounds are weaker and masses below pion are certainly allowed
\cite{Shaposhnikov:2006nn}.} but still there are models where small
regions of the parameter-space above the pion mass are perfectly
allowed\footnote{Note that our exclusion plot is different from that
of Ref.~\cite{Kusenko:2004qc}, where the coupling of
sterile neutrino to $\tau$ generation was not considered. Moreover,
eq. (3.1) of this paper contains a factor $4$ error. In addition, the
formula (21) of \cite{Dolgov:2000pj}  for the probability of
$N\rightarrow \pi^0\nu$ decay is not correct, see discussion in 
Section \ref{decays}.}. We would also like to stress that the
branching ratios for $\epsilon \sim 1$ can be quite different from  
(\ref{Yukawanormal},\ref{Yukawainverted}) leading to extra
uncertainties. 

Above pion mass, the BBN limits are down to two order of magnitude
below the direct limits form CERN PS191 experiment, thus one-two
orders of magnitude improvement is required to either confirm or
disprove the $\nu$MSM with sterile neutrinos lighter than $450$~MeV.
For the three benchmark models we transfered these limits to the upper
limits on overall mixing $U^2$ and neutrino lifetime and plotted them
in Fig.~\ref{Nu-lifetime}. 

The improvement required to test the $\nu$MSM with sterile neutrinos
lighter than 450~MeV can be done with either new kaon experiments,
such as one planned in JPARC, or special analysis of the available
data on kaon decays collected in Brookhaven and Frascati. In
particular, E787/E949 Collaboration reported limit on $K^+\to\pi^+X$ decay
with $X$ being hypothetical long-lived neutral
particle~\cite{Adler:2004hp}.  With statistics of thousand of
billions charged kaons, available in this experiment, one can expect
to either prove or completely rule out $\nu$MSM with sterile
neutrinos lighter than $450$~MeV. The same conclusion is true for the
third stage of CERN NA48 experiment.

In the next two Sections we discuss the decays and production of
neutral fermions for a mass range up to $5$ GeV, to understand the
requirements to possible future experiments that could allow to enter
into interesting parameter space for neutral fermion masses above
$400$ MeV.


\section{Decays of heavy neutral leptons}
\label{decays}

Heavy neutral leptons we consider ($M_N\gtrsim 10$~MeV) are unstable,
since decay channels to light active leptons, $N\to \bar\nu_\alpha
\nu_\alpha \nu_\beta$, $N\to e^+e^- \nu_\alpha$ are open;  the modes
like $N_{2,3}\to N_1+\dots$ are strongly suppressed. Hereafter charge
conjugated modes are also accounted resulting in double rates for
Majorana neutrinos as compared to Dirac case. For heavier leptons
more decay modes are relevant,
\[
N\to\mu e \nu,\;\pi^0\nu,\;\pi e,\;\mu^+ \mu^- \nu,\;\pi \mu ,\; K e 
,\;
K \mu \;, \eta\nu,\rho\nu,\dots 
\]

Decays of sterile neutrinos have been exhaustedly studied in
literature. For convenience we present explicit formulae for relevant
decay rates in Appendix~\ref{Appendix-neutrino-decays}. Most of them
(but not all) can be be obtained straightforwardly by making use of
the formulae for Dirac neutrinos 
presented in Ref.~\cite{Johnson:1997cj}, which we found to be
correct.

\begin{figure}[!htb]
\centerline{
\includegraphics[width=0.33\textwidth]{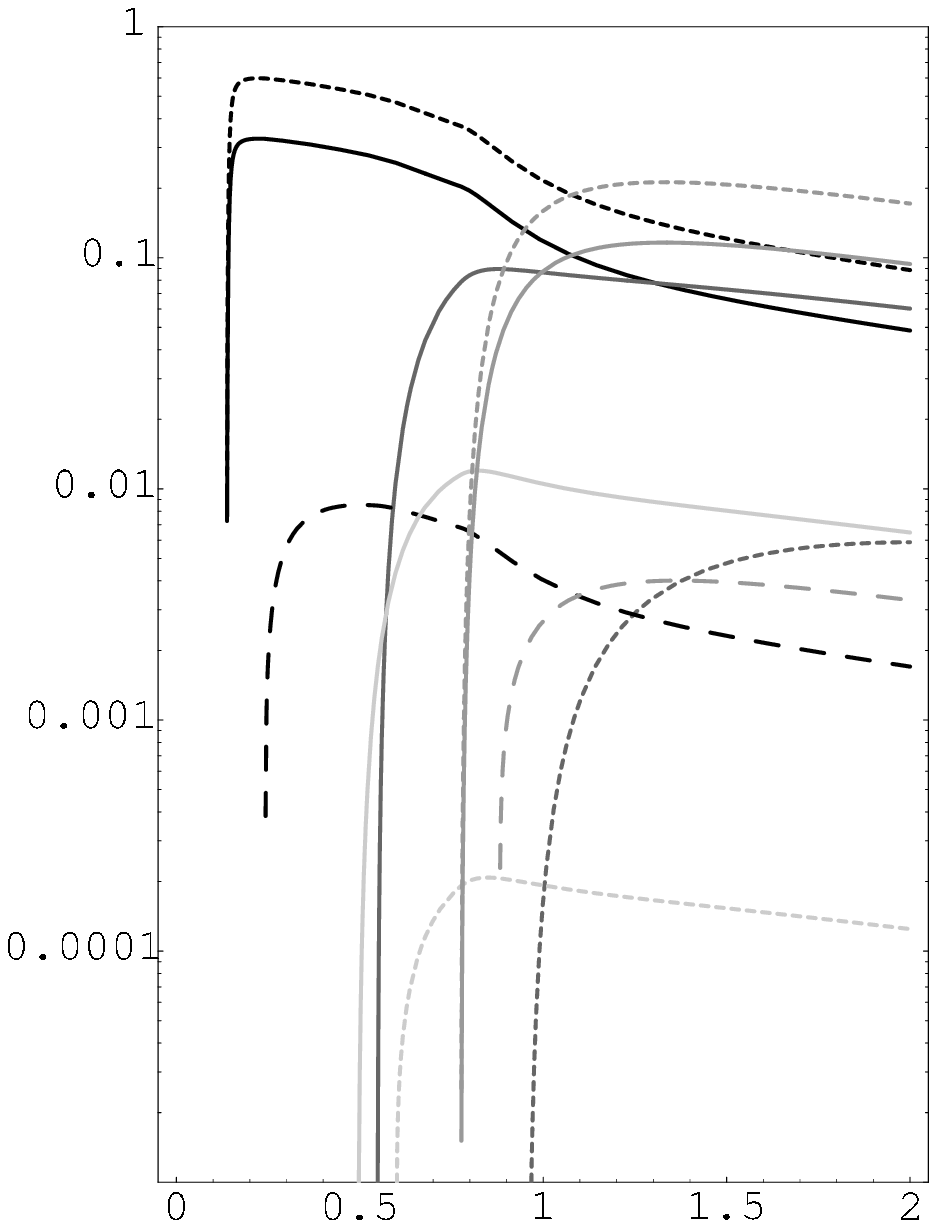}
\includegraphics[width=0.33\textwidth]{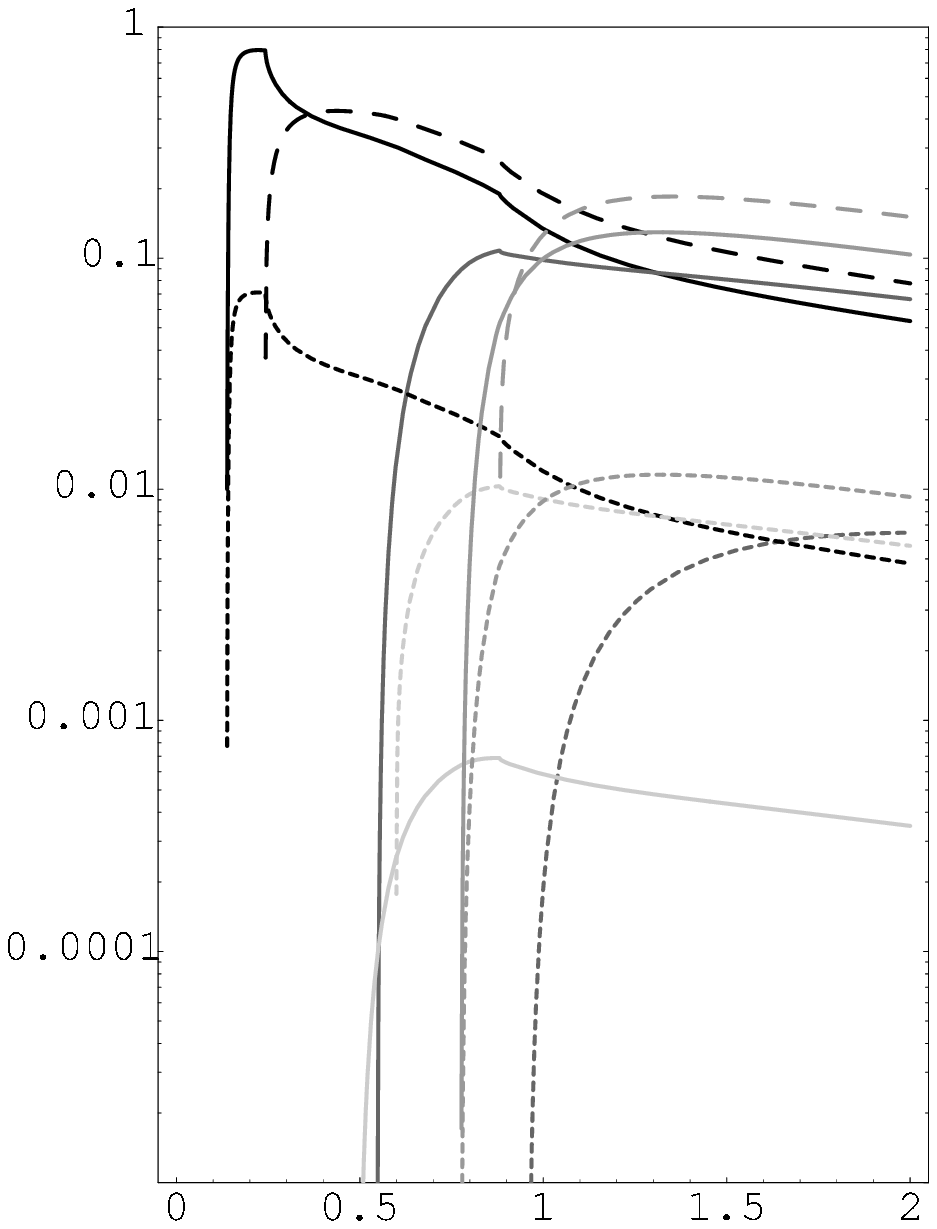}
\includegraphics[width=0.33\textwidth]{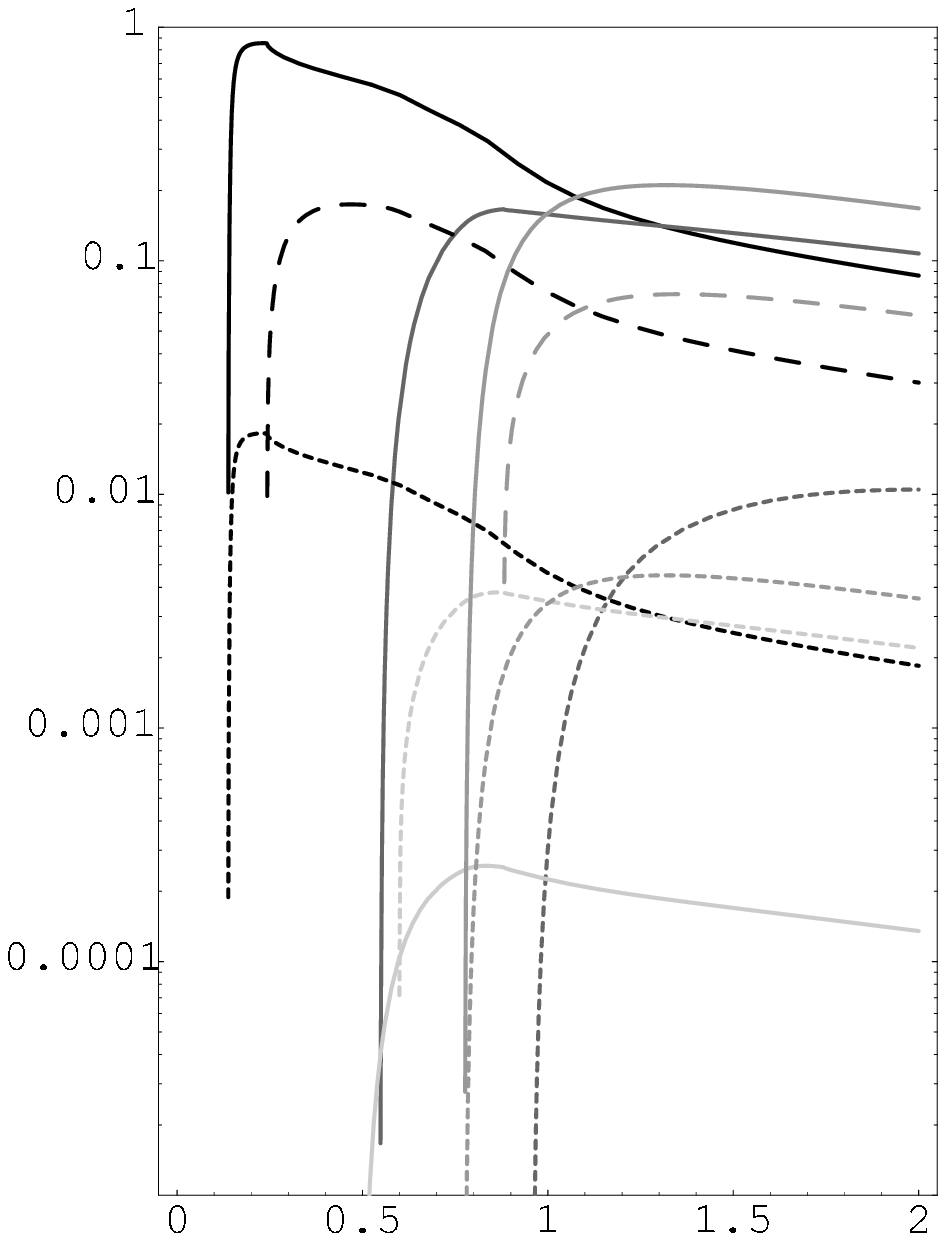}}
\begin{picture}(0,0)(0,0)
\put(-165,22){{\small$M_N$, GeV}}
\put(-5,22){{\small$M_N$, GeV}}
\put(150,22){{\small$M_N$, GeV}}
\put(-205,50){{\small a)}}
\put(-45,50){{\small b)}}
\put(110,50){{\small c)}}
\end{picture}
\vskip -1cm
\caption{Branching ratios of neutrino two-body decays $N_I\to XY$ as
functions of neutrino mass $M_N$ for models with the same hierarchy in
mixing as in models: a) I, b) II, c) III; different lines 
correspond to different modes: $\pi\nu$ (solid
black), $\pi e$ (short-dashed black), $\pi\mu$ (long-dashed black),
$Ke$ (solid light gray), $\eta\nu$ (solid dark gray), $\eta'\nu$
(short-dashed dark gray), $K\mu$ (dashed light gray), $\rho \nu$
(solid gray), $\rho e$ (short-dashed gray), $\rho\mu$ (long-dashed
gray).
\label{neutrino-2-body-branchings}
}
\end{figure}

\begin{figure}[htb]
\centerline{
\includegraphics[width=0.33\textwidth]{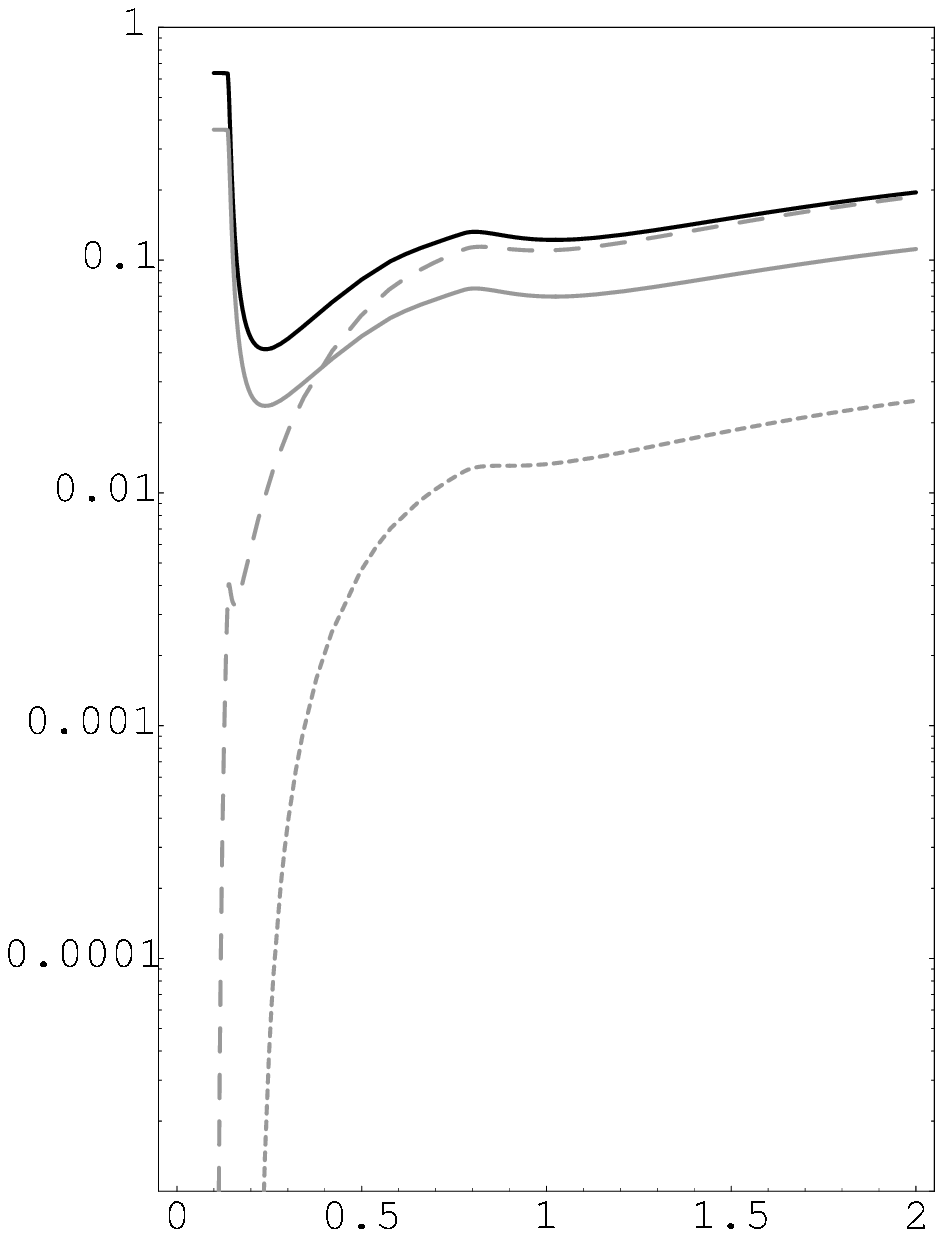}
\includegraphics[width=0.33\textwidth]{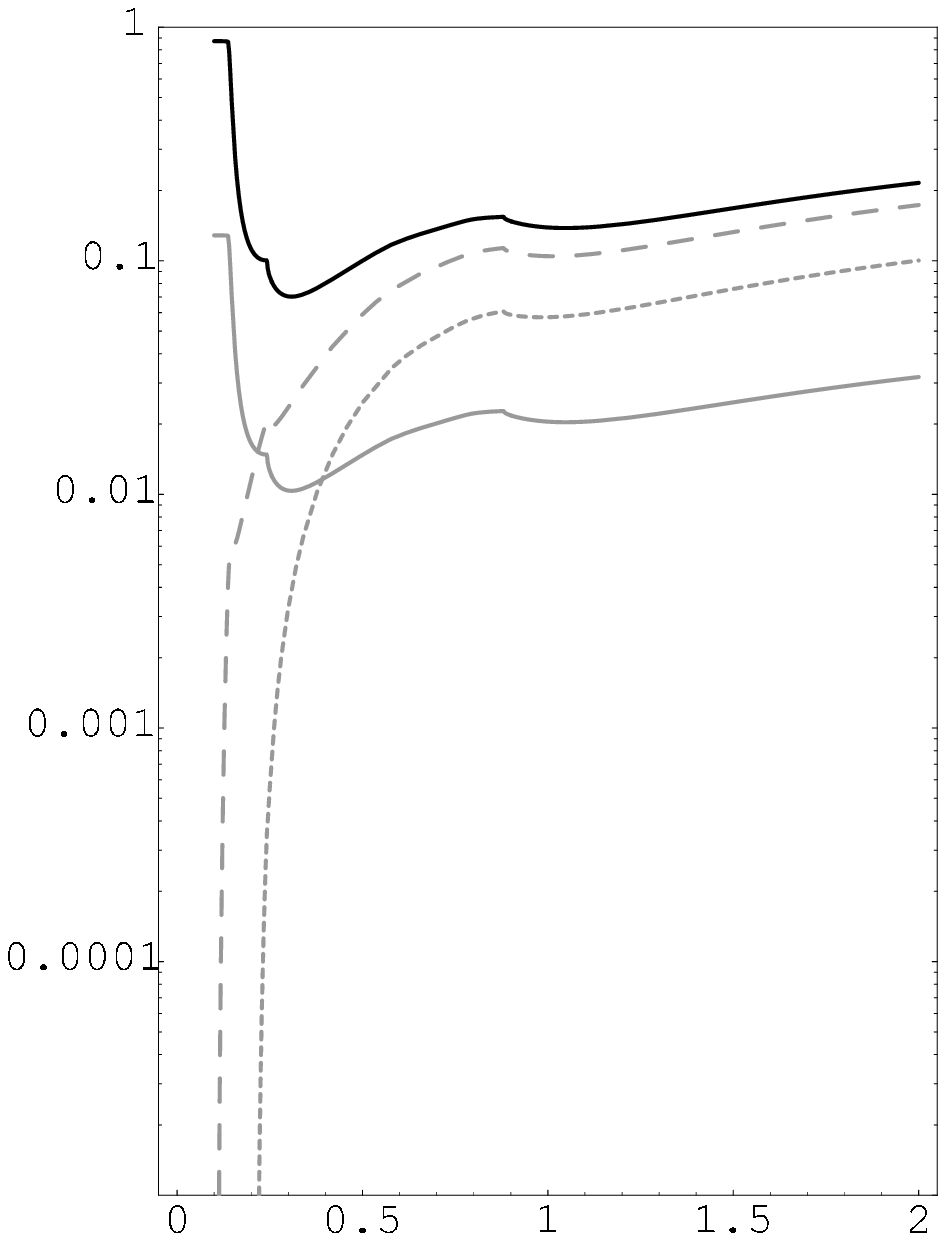}
\includegraphics[width=0.33\textwidth]{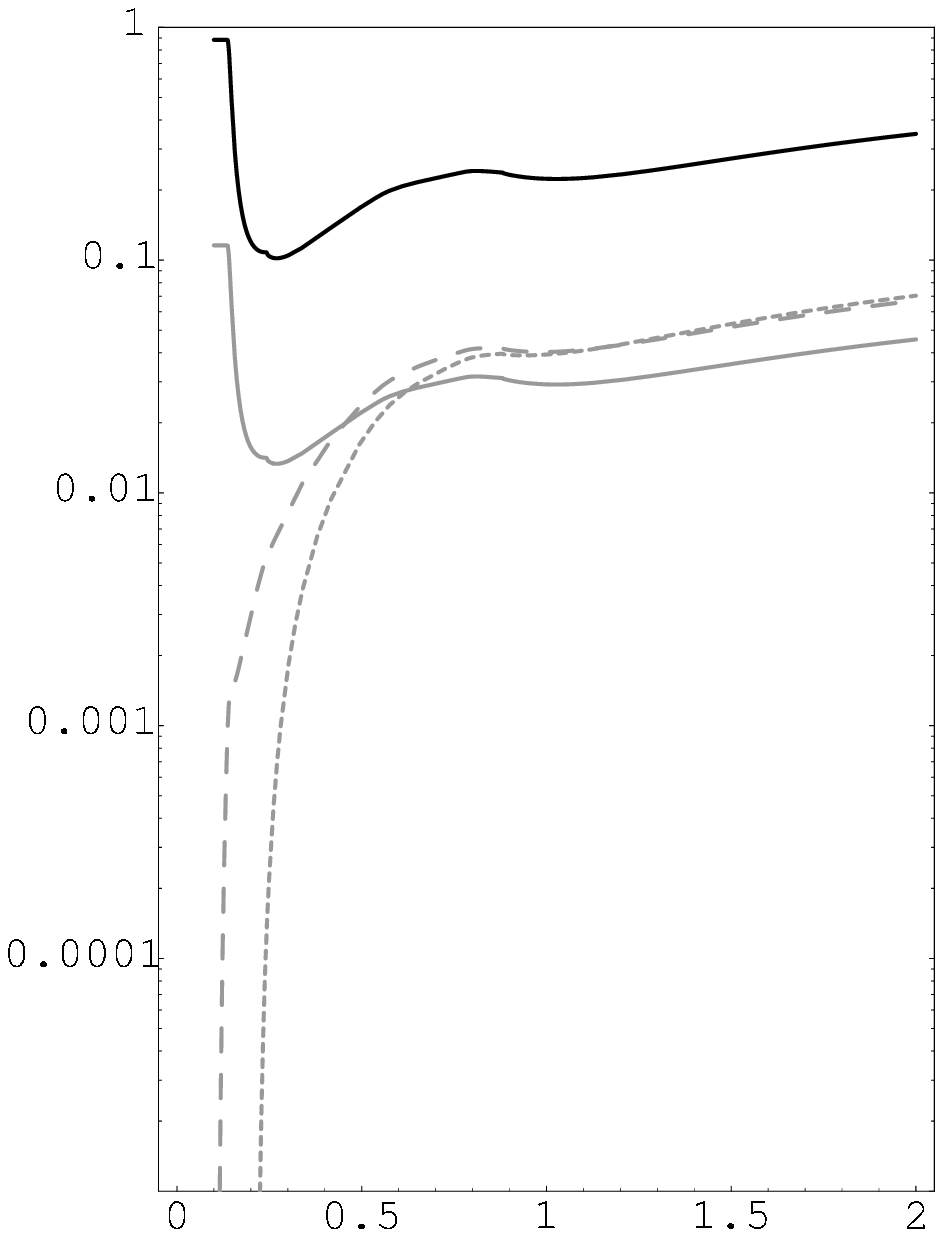}}
\begin{picture}(0,0)(0,0)
\put(-165,22){{\small$M_N$, GeV}}
\put(-5,22){{\small$M_N$, GeV}}
\put(150,22){{\small$M_N$, GeV}}
\put(-110,50){{\small a)}}
\put(50,50){{\small b)}}
\put(210,50){{\small c)}}
\end{picture}
\vskip -1cm
\caption{Branching ratios of neutrino three-body decays $N_I\to ABC$ as
functions of neutrino mass $M_N$ for models with the same hierarchy in mixing 
as in models: a) I, b) II, c) III; different lines 
correspond to different modes: $\bar\nu \nu\nu$ (sum over all
invisible modes, solid black), $\nu e^+e^-$ (solid gray), $\nu e\mu$ (sum over
two modes, long-dashed gray), $\nu \mu^+\mu^-$ (short-dashed gray).
\label{neutrino-3-body-branchings}
}
\end{figure}

Neutrino decays branching ratios for 
benchmark models I-III and $M_\pi<M_N<2$~GeV are plotted in
Figs.~\ref{neutrino-2-body-branchings},
\ref{neutrino-3-body-branchings}.   
For heavier neutrino many-hadron final
states become important, and one can use spectator quarks to calculate
the corresponding branching ratios. Below $2$~GeV the contribution of
these modes to total neutrino width is less than 10\%. 
 Neutrino lifetime is constrained by limits~ \eqref{upper},
\eqref{lower} on overall strength of mixing.  The results for models
I, II and III are presented in Fig.~\ref{Nu-lifetime}a:
\begin{figure}[htb]
\centerline{
\includegraphics[width=0.5\textwidth]{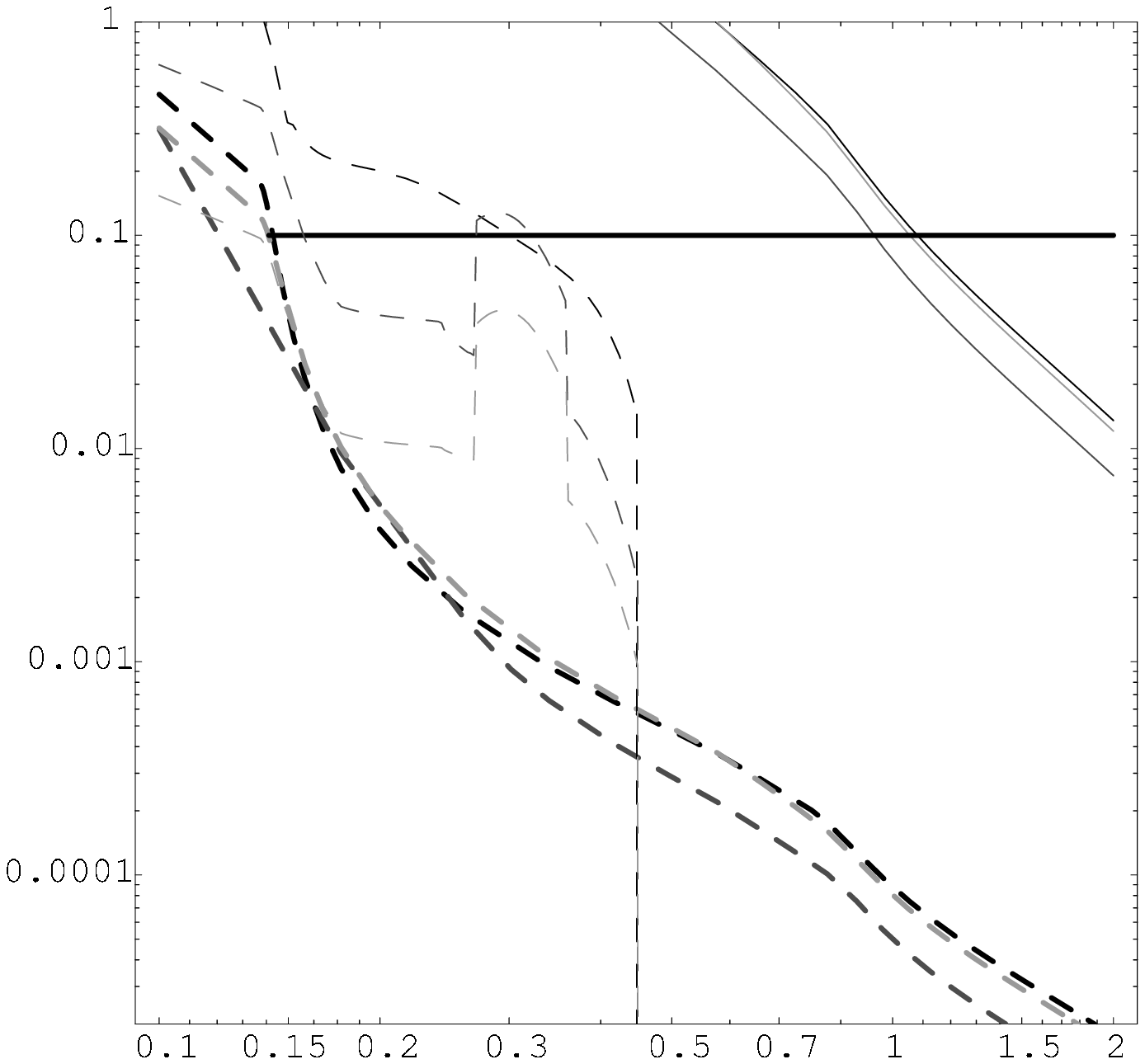}
\includegraphics[width=0.5\textwidth]{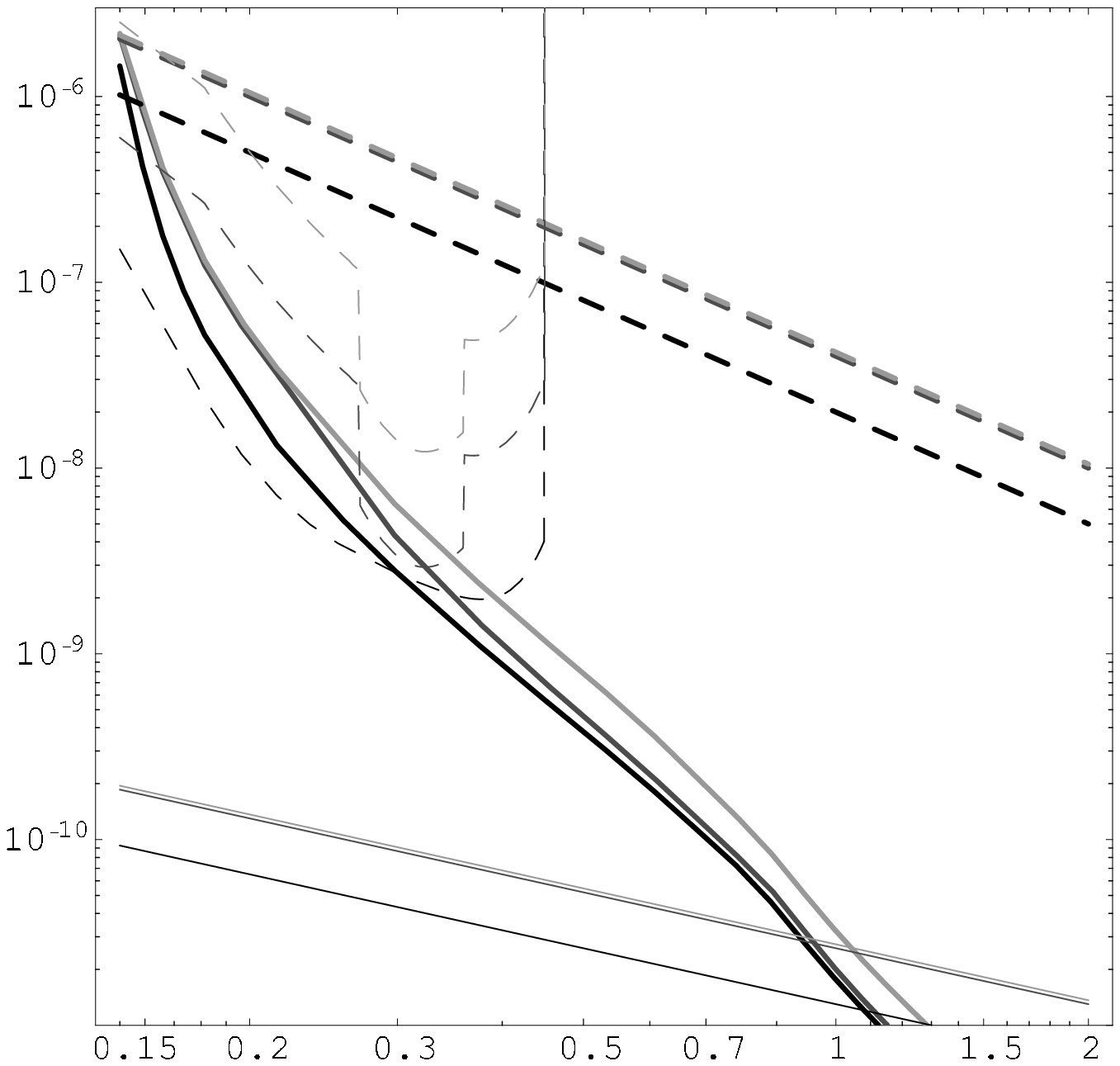}}
\begin{picture}(0,0)(0,0)
\put(-120,20){{\small$M_N$, GeV}}
\put(110,20){{\small$M_N$, GeV}}
\put(-190,50){{\small a)}}
\put(60,50){{\small b)}}
\put(-230,255){$\tau_N$, s}
\put(20,255){$U^2$}
\end{picture}
\vskip -1.cm
\caption{a) Upper (solid lines) and lower limits (dashed lines) on
neutrino lifetime in models I (black), II (dark gray) and III (light
gray); a horizontal thick solid black 
line indicates the upper limit from BBN, $\tau_N<0.1$~c, suggested in
Ref.\cite{Dolgov:2000jw}, thin solid lines are limits from
eq.~\eqref{lower}, thick dashed lines refer to eq.~\eqref{upper}, thin
dashed lines correspond to limits from direct searches for sterile
neutrinos discussed in Section~\ref{Sec:exp-limits}; charge-conjugated
modes are accounted. b) Lower (solid
lines) and upper limits (dashed lines) on overall mixing $U^2$ in
models I (black), II (dark gray) and III (light gray).  Thin solid lines and
thick dashed lines depict limits from eqs.~\eqref{lower} and
\eqref{upper}, respectively.  Thick solid lines indicate lower limits
from order-of-magnitude BBN bound on neutrino lifetime, $\tau_N<0.1$~c
for $M_N>140$~MeV, thin dashed lines refer to limits from direct
searches for sterile neutrinos discussed in Section~\ref{Sec:exp-limits}.
\label{Nu-lifetime}
}
\end{figure}
in phenomenologically viable models neutrino lifetime is confined
by corresponding solid (upper limits) and dashed (lower limits)
lines. The horizontal solid line indicates 
the order-of-magnitude upper limit on
neutrino life time, $\tau_N<0.1$~c, which guarantees that the results
of standard BBN remain intact~\cite{Dolgov:2000jw} for
$M_N\gtrsim140$~MeV. In a given model the range of neutrino mass,
where the corresponding solid line(s) is(are) above the corresponding
dashed one(s) is disfavoured.  

These limits imply limits on overall mixing $U^2$ plotted in
Fig.\ref{Nu-lifetime}b: in phenomenologically viable models mixing
$U^2$ is confined by corresponding solid and dashed lines. One can
see that the constraint from BBN is stronger than the see-saw
constraint \eqref{lower} for $M \lsim 1$~GeV. However, it is worth
noting that the limit $\tau_N<0.1$~s may happen to be too
conservative and can presumably be relaxed to some extent provided
careful study of processes in primordial plasma in BBN epoch. In what
follows, for the three benchmark models we give upper and lower
limits on various neutrino rates.  For a given neutrino mass these
limits are saturated respectively by the  tightest among  upper
limits and  tightest among  lower limits on neutrino mixing,
presented in Fig.~\ref{Nu-lifetime}b.  Only these tightest limits are
used below. 

 Note in passing that as we already mentioned the $\nu$MSM
  predictions beyond benchmark models could deviate to some extent
  from a naive interplay between benchmark numbers. At the same time
  for any set of parameters the presence of both upper and lower
  bounds on neutrino rates is a general feature of $\nu$MSM, which
  allows it to be falsified. 

\section{Production of heavy neutral leptons}%
\label{production}

In high-energy experiments the most powerful sources of heavy neutral
leptons are the kinematically allowed weak decays of mesons (and
baryons) created in beam-beam and beam-target collisions. Obviously,
the relevant hadrons are those which are stable with respect to
strong and electromagnetic decays.

The spectrum of outgoing heavy neutral leptons $N$ in a given experiment is
determined mostly by the spectrum of produced hadrons $H$ subsequently
decaying into heavy leptons. Since relevant hadrons contain one heavy quark
$Q$, differential cross section of their direct production $d\sigma^{dir}_H$
can be estimated by use of the factorization theorem
\begin{equation}
\label{Sec3:1+}
\frac{d\sigma^{dir}_H}{dp_{H,L}dp^2_{H,T}}= 
\int_0^1 dz\cdot\delta\l p_Q-zp_H\r \cdot 
D_{H,Q}(z)\cdot\frac{d\sigma^{dir}_Q}{dp_{Q,L}dp^2_{Q,T}}\;,
\end{equation}
where $d\sigma^{dir}_Q$ is differential cross section of direct
$Q$-quark production\footnote{We assume non-polarized beam(s) and
target and hence
axial symmetry.},
$p_{H,L}$, $p_{H,T}$ and  $p_{Q,L}$, $p_{Q,T}$ are longitudinal and transverse
spatial momenta of hadron $H$ and heavy quark $Q$, respectively; $zp_H$ is a
part of hadron momentum carried by heavy quark and a fragmentation function  
$D_{H,Q}(z)$ describes the details of hadronization. The differential cross section
entering the integrand in eq.\eqref{Sec3:1+} can be calculated within
perturbative QCD, while function $D_{H,Q}(z)$ comprises non-perturbative
information. There are several approximations to $D_{H,Q}(z)$ in literature,
e.g. commonly used in high energy physics generator PYTHIA adopts modified
Lund fragmentation function~\cite{Fragmentation-in-PYTHIA} 
\[
D(z)\propto\frac{\l 1-z\r^a}{z^{1+b\cdot m_Q^2}}\cdot
\e^{-\frac{b}{z}\cdot\l M_H^2+p^2_{H,T}\r}
\]
with default parameters $a=0.3$ and $b=0.58$~GeV$^{-2}$. 

The rate of hadron production depends on the intensity of
collisions. The distribution of total number of directly produced
hadrons $dN^{dir}_H$ reads
\[
\frac{dN^{dir}_H}{dp_{H,L}dp^2_{H,T}}=
\frac{d\sigma^{dir}_H}{dp_{H,L}dp^2_{H,T}} \cdot {\cal L}_{acc}\;,
\] 
where ${\cal L}_{acc}$ is an integrated luminosity of a given experiment and
we neglect tiny imprints of real bunch structure on outgoing hadronic spectra.
Note that we are interested in hadrons stable with respect to strong and
electromagnetic decays, thus apart of direct production they emerge due to
strong and electromagnetic decays of other hadrons, which give indirect
contribution $dN^{ind}_H$. The distribution of the total number of produced
hadrons $dN_H$ is a sum of both contributions,
\[
\frac{dN_H}{dp_{H,L}dp^2_{H,T}}=\frac{dN^{dir}_H}{dp_{H,L}dp^2_{H,T}}+ 
\frac{dN^{ind}_H}{dp_{H,L}dp^2_{H,T}}\;.
\]

Produced hadrons stable with respect to strong and electromagnetic decays
travel distances of about $\beta_H\cdot\tau_H\cdot\gamma_H$
($\beta_H$, $\tau_H$ and
$\gamma_H$ are speed, lifetime and boost factor of a given hadron) and then decay
weakly, producing some amount of heavy neutral leptons. In the hadron 
rest frame the spatial momentum of heavy lepton $p_N$ can be correlated
with the hadron total spin. Consequently, in the laboratory frame there can be
additional to Lorenz boost contribution to correlations between $p_N$ and
$p_H$. This contribution is smearing with growth of statistics and can be also
neglected if typical $\gamma$-factor of hadrons is large,
$\gamma_H=E_H/M_H\gg1$. Hence, in the laboratory frame, 
the distribution of heavy leptons over spatial momentum is given by 
\begin{equation}
\label{Sec3:3+}
\begin{split}
\frac{dN_N}{dp_{N,L}dp^2_{N,T}}&=\sum_H \tau_H\cdot \int 
\frac{dB_H\l H\to N+\dots\r}{dE_N}\cdot dE_N\\ &\times\int 
d^3 {\bf n}_\gamma \cdot 
\delta\l {\bf p_N} - {\bf p_H} - {\bf n}_\gamma \cdot \sqrt{E_N^2-M_N^2} \r
 \cdot
\frac{dN_H}{dp_{H,L}dp^2_{H,T}}\;,
\end{split}
\end{equation}
where we integrate over unit sphere boosted to laboratory frame and sum up all
contributions from all relevant hadrons; $dB_H\l H\to N+\dots\r$ is a
differential inclusive branching ratio of hadron $H$ into heavy neutrino.
These branching ratios can be straightforwardly obtained for each hadron with
help of the standard technique used to calculate weak decays in the framework
of the SM.  Indeed, in both models (MSM and $\nu$MSM) neutrinos are produced
mostly via virtual $W$-boson (charged current): the only difference is that in
$\nu$MSM neutrinos are massive. For heavy neutrinos this
results\footnote{Also, in models with heavy neutrinos values of hadronic form
  factors governing semileptonic width are changed in accordance with shift in
  virtuality of $W$-boson.} in enhancement of pure leptonic decay modes which
are strongly suppressed in the SM by charged lepton masses.
 
The heavier the quark the lower its production rate; hence, a class of the
lightest kinematically allowed hadrons saturates heavy neutrino production. 
As we explained in Section~\ref{Sec:exp-limits}, neutrinos in phenomenologically
viable $\nu$MSM are likely to be heavier than pion. If neutrino $N_I$ is lighter than
kaon, the dominant source of neutrinos is decaying kaons,
\begin{align}
\label{Sec3:4*}
K^\pm&\to l_\alpha^\pm N_I ~,\\
\label{Sec3:4**}
K_L&\to \pi^\mp l_\alpha^\pm N_I ~.
\end{align}
The two-body decays \eqref{Sec3:4*} have been already studied in
literature (see, e.g., Refs.~\cite{Shrock:1981wq,Gronau:1984ct}). For
convenience, the differential branching ratio is presented in
Appendix~\ref{Appendix-meson-decays}. Contribution of three-body
decays \eqref{Sec3:4**} to neutrino production is suppressed by phase
volume factor; as the largest impact they give a few per cent at
$M_N\simeq M_\pi$; the corresponding differential branching ratio is
presented in Appendix~\ref{Appendix-meson-decays}.

 In  models with neutrino heavier than kaon but lighter than charmed hadrons,
decays of those latter dominate neutrino production. The largest partial width
to heavy neutrinos is exhibited by $D_s$-meson which leptonic decays $D_s\to
l_\alpha N_I$ are not suppressed by CKM mixing angles as compared to similar
decays of $D$-mesons,  $D^\pm\to l^\pm_\alpha N_I$. Semileptonic three-body
decay modes
\begin{align}
\label{Sec3:5*}
D_s\to\eta^{(')}l_\alpha N_I\;,~~~& D\to K l_\alpha N_I~,
\\
D_s\to\phi l_\alpha N_I\;,~~~& D\to K^* l_\alpha N_I
\label{Sec3:5**}
\end{align}
are unsuppressed by CKM-mixing as well, and are sub-dominant in
general. For sufficiently light neutrinos, $M_K\lesssim M_N\lesssim
700$~MeV, $D$-meson semileptonic decay modes give contribution comparable to
$D_s\to l_\alpha N_I$ at $M_K\lesssim M_N\lesssim 700$~MeV, because
the $D$-meson total production dominates over $D_s$-production in
hadronic collisions.  Differential branching ratios of the leptonic
decays and the semileptonic decays \eqref{Sec3:5*} of charmed mesons are 
provided by general formulae in
Appendix~\ref{Appendix-meson-decays}, where the expression of 
differential branching ratio to vector mesons $V=\phi,K^*$ 
\eqref{Sec3:5**} is also presented. 
Both the rest of kinematically allowed three-body decay modes and four-body
decay modes, e.g. $D\to K\pi l_\alpha N_I$, are strongly suppressed by either
CKM mixing or phase volume factor and can be neglected.  The largest
contribution from charmed baryons comes from the decay $\Lambda_c\to\Lambda
l_\alpha N$ and is negligibly small for heavy neutrino production.

In models where neutrino masses are within the range 2~GeV~$\lesssim
M_{N}\lesssim$~5~GeV, neutrinos are produced mostly in decays of beauty
mesons.  These are also mostly leptonic and semileptonic decays, which
branching ratios are described by general formulae presented in
Appendix~\ref{Appendix-meson-decays}. As compared to $D$-meson decays,
$B$-meson decays into heavy neutrinos are strongly suppressed by off-diagonal
entries of CKM matrix. For neutrinos lighter than about
2.5~GeV semileptonic modes to charm mesons, e.g. $B\to D^{(*)}lN_I$, 
dominate over leptonic mode $B\to lN_I$ because of both larger
CKM-mixing, $\left|V_{bc}\right|\gg\left|V_{bu}\right|$, and larger
values of hadronic form factors,\footnote{This is a consequence of
strong overlapping between quark wave functions in the meson required
to produce virtual $W$-boson in case of leptonic decay.}  $f_B/M_B\ll
f_+,f_0$. $B_c\to l N_I$ is more promising, but $B_c$-production in
hadron collisions is suppressed. For heavier neutrinos leptonic modes
dominate.  The baryon contribution is subdominant at any $M_N$.

Note, that additional, but always subdominant, contribution to heavy neutrino
production comes from decays of $\tau$-leptons (if
kinematically allowed), which emerge
as results of decays of $D_s$- and $B$-mesons. 

The total number of produced heavy leptons $N_N$ is given by the
integration of eq.~\eqref{Sec3:3+} over ${\bf n}_\gamma$ and $E_N$.
For order-of-magnitude estimates one can use the following simple
approximation,
\[
N_N=\sum_H N_H\cdot \Br\l H\to N\dots\r\;,
\]
with $N_H$ being a total number of produced hadrons $H$, which in turn 
can be estimated as 
\[
N_H=N_Q\cdot \Br\l Q\to H \r\;,
\]
where $N_Q$ is a total number of produced heavy quarks $Q$ and $\Br\l Q\to H
\r$ is a relative weight of the channel $Q\to H$ in $Q$-quark hadronization. 
For strange meson the reasonable estimate is 
$\Br\l s\to K^-\r=\Br\l s\to K_L\r$. Following Ref.~\cite{Lourenco:2006vw} we
set $\Br\l c\to D^+\r=0.4\cdot\Br\l c\to D_0\r$ and assuming 
$\Br\l c\to D_s\r=\Br\l c\to \Lambda_c \r$ obtain for relevant hadrons 
\[
\Br\l c\to D^+\r=0.2\;,~~~\Br\l c\to D_0\r=0.5\;,~~~ \Br\l c\to D_s\r=0.15\;.
\]
For beauty mesons we use~\cite{PDG}
\[
\Br\l b\to B^+\r=\Br\l b\to B^0\r=0.4\;,~~~\Br\l b\to B_s\r=0.1\;. 
\]

\begin{figure}[!htb]
\centerline{
\includegraphics[width=0.33\textwidth]{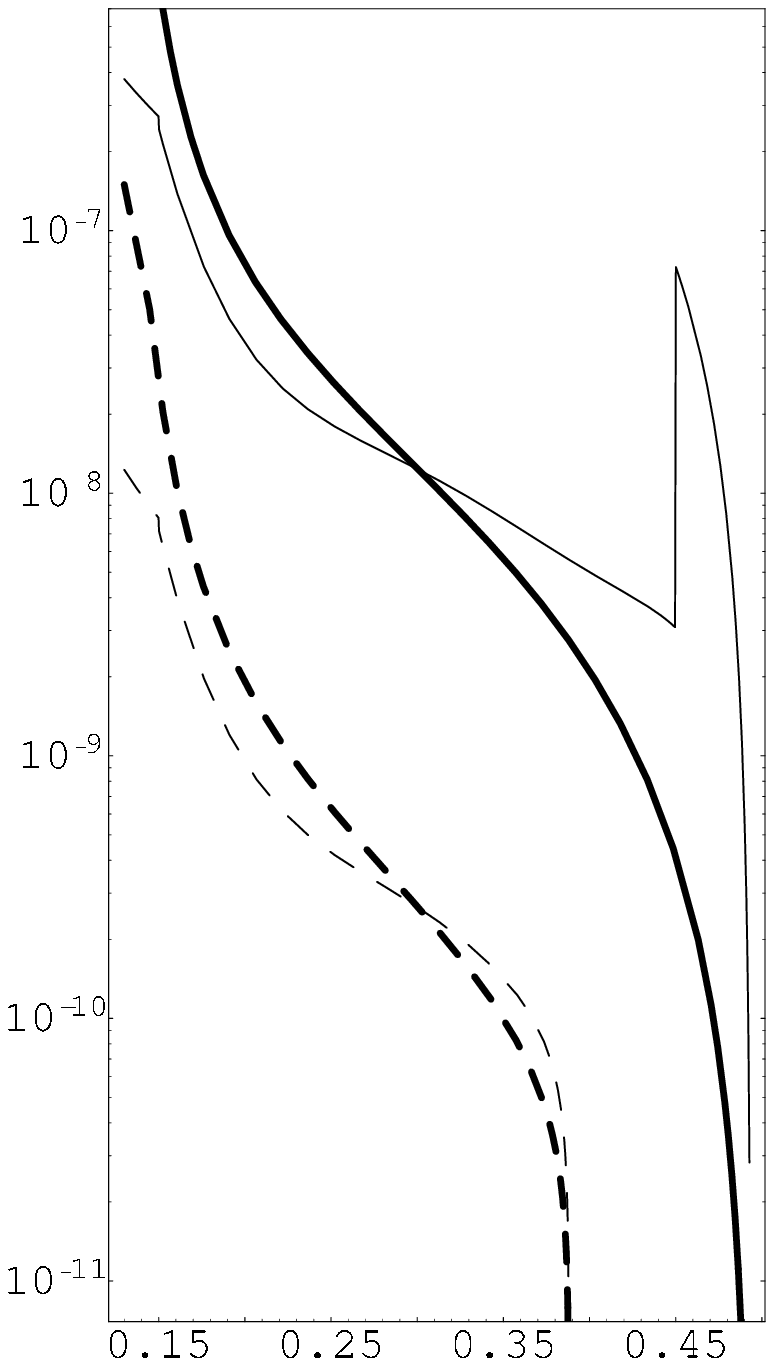}
\includegraphics[width=0.33\textwidth]{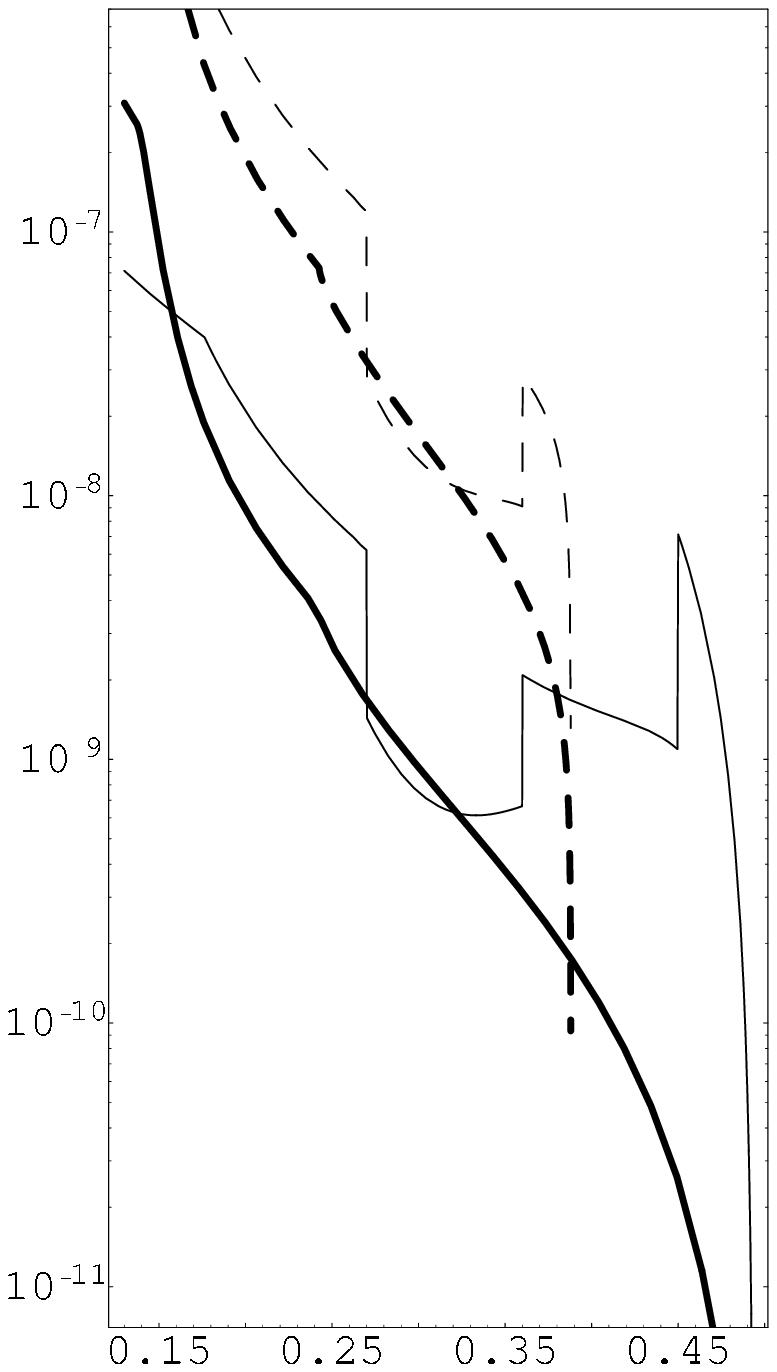}
\includegraphics[width=0.33\textwidth]{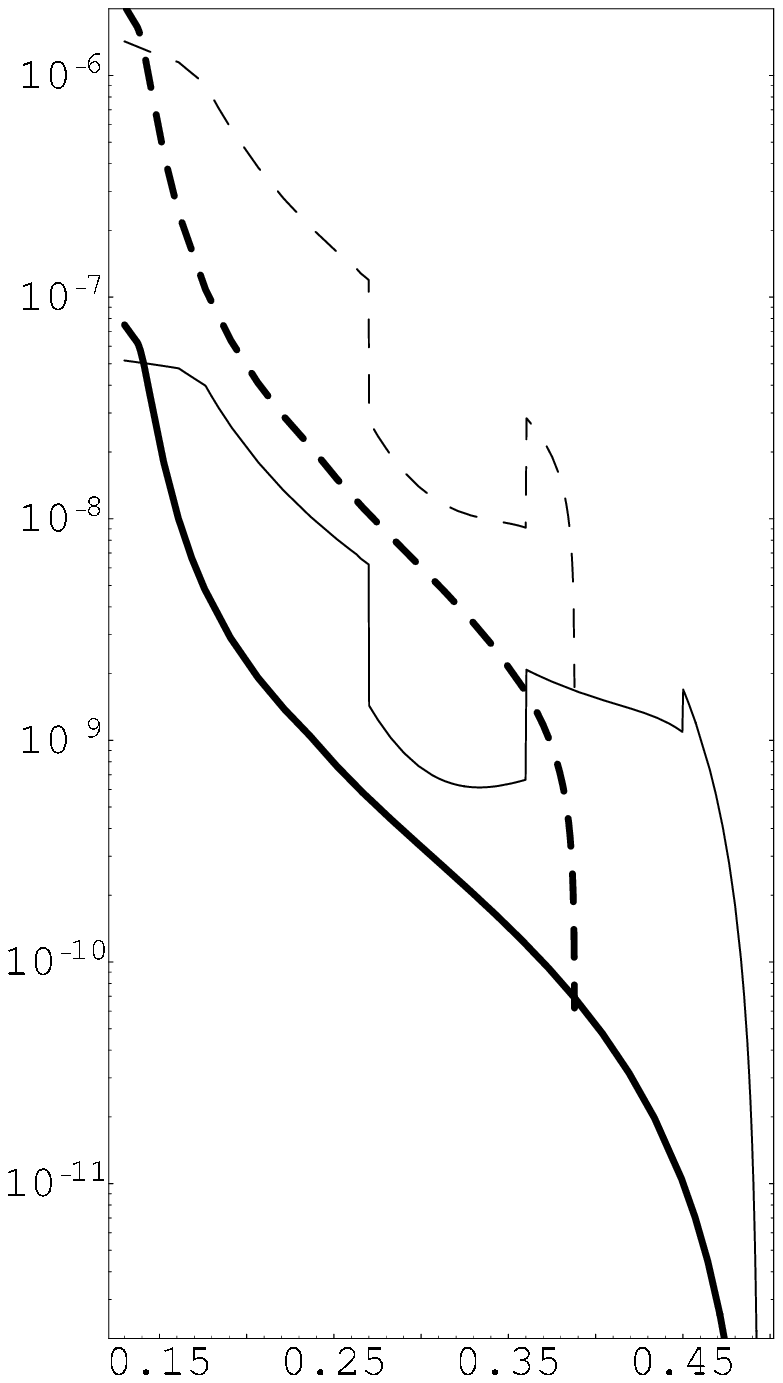}}
\begin{picture}(0,0)(0,0)
\put(-160,40){{\small$M_N$, GeV}}
\put(-10,40){{\small$M_N$, GeV}}
\put(155,40){{\small$M_N$, GeV}}
\put(-105,280){{\small a)}}
\put(50,280){{\small b)}}
\put(210,280){{\small c)}}
\end{picture}
\vskip -1.5cm
\caption{Branching ratios of decays $K\to e N_I$ (solid lines)
and $K\to \mu N_I$ (dashed lines) as functions of heavy
neutrino mass $M_N$ in models: a) I, b) II, c) III. In a
phenomenologically viable model and heavy neutrino mass within
$M_\pi\lesssim M_N\lesssim M_K$, the branching ratios are confined
between corresponding thin and thick lines which show upper and lower
limits on $U^2$ from Fig.~\ref{Nu-lifetime}b, respectively.
\label{leptonic-widths-kaons}
}
\end{figure}

\begin{figure}[!htb]
\centerline{
\includegraphics[width=0.33\textwidth]{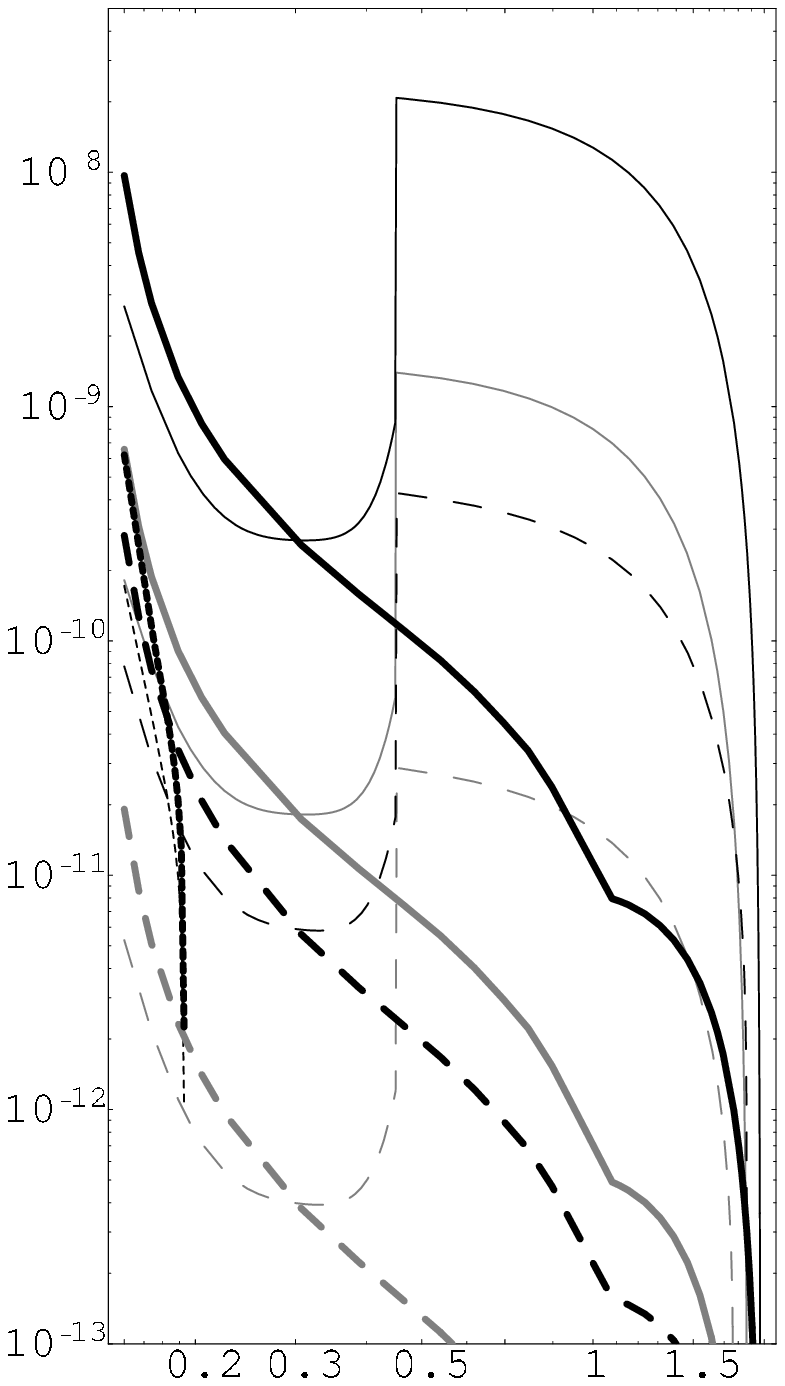}
\includegraphics[width=0.33\textwidth]{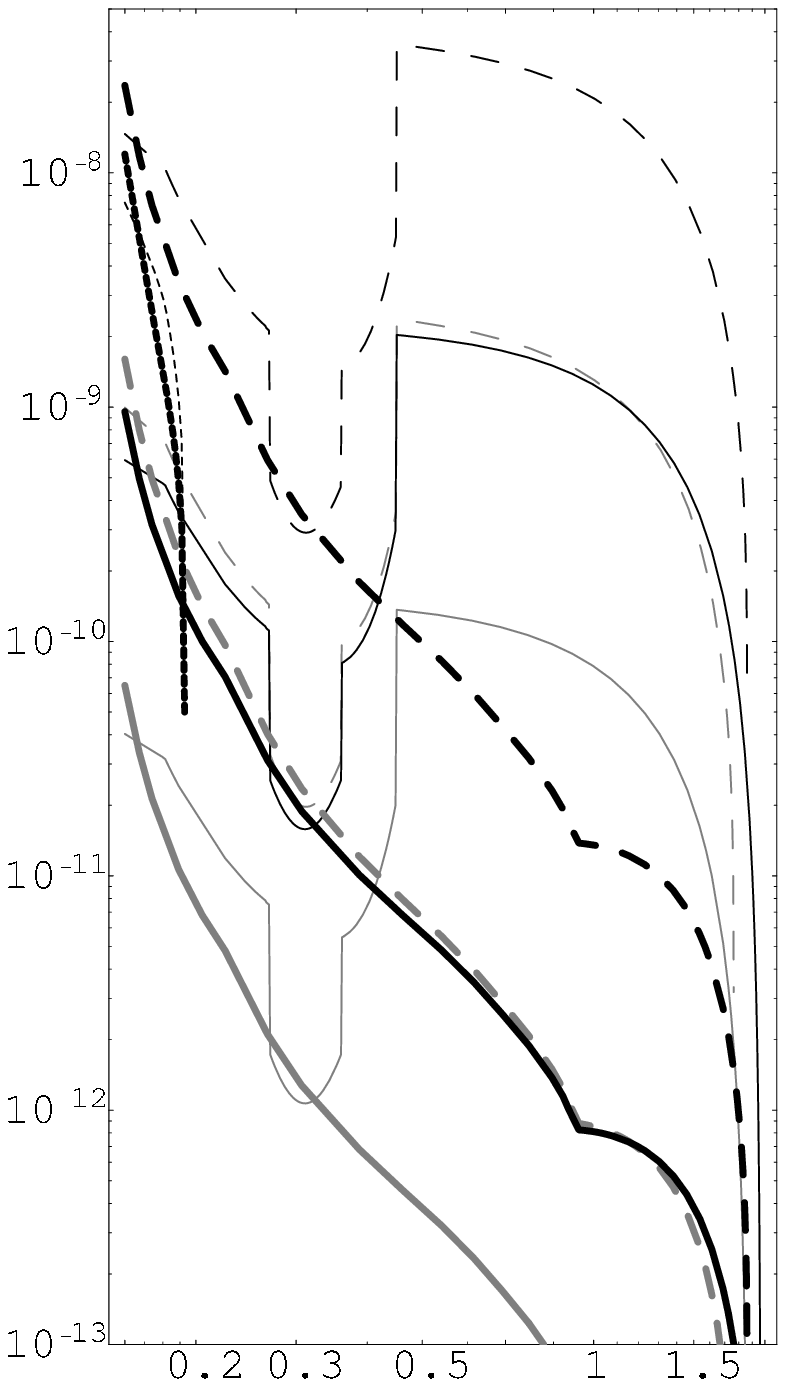}
\includegraphics[width=0.33\textwidth]{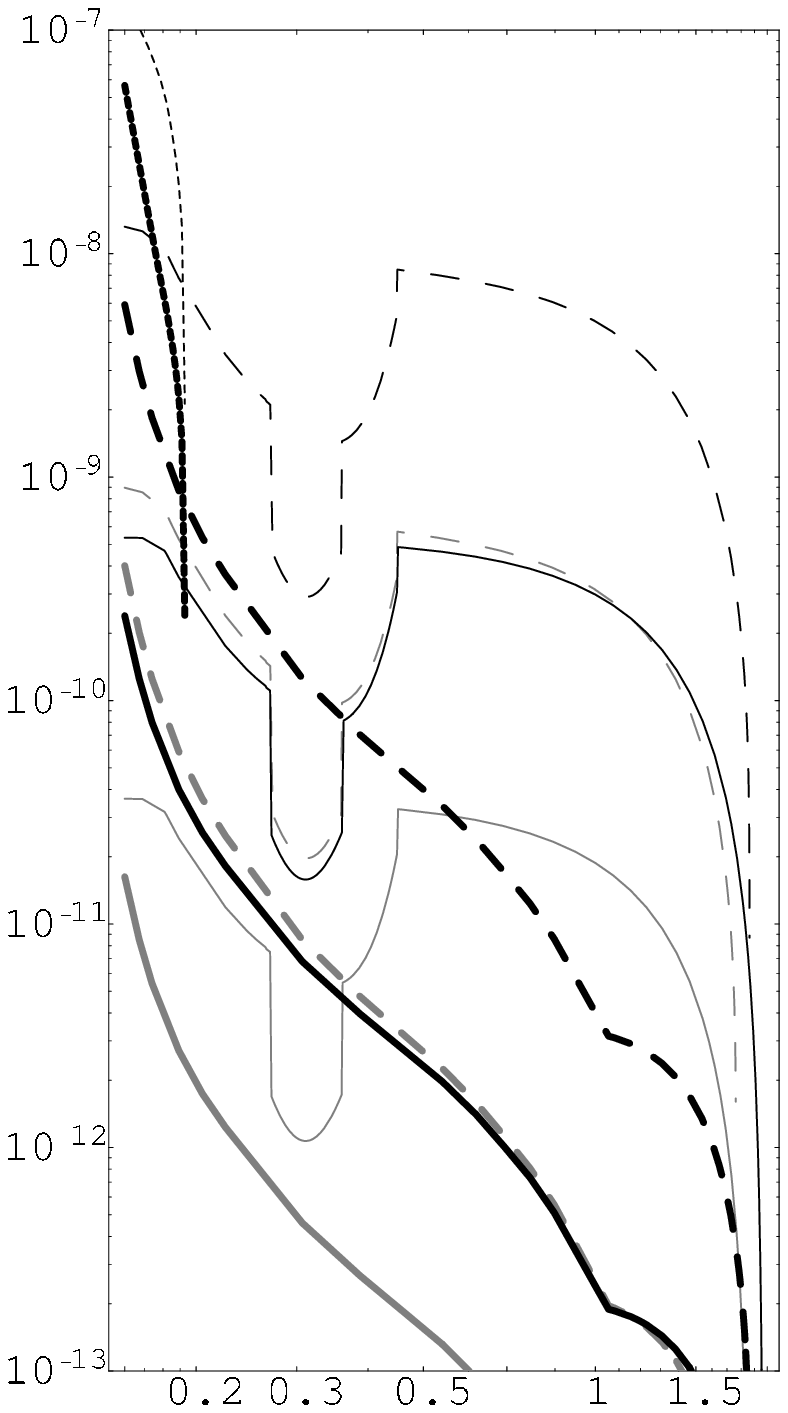}}
\begin{picture}(0,0)(0,0)
\put(-165,40){{\small$M_N$, GeV}}
\put(-10,40){{\small$M_N$, GeV}}
\put(150,40){{\small$M_N$, GeV}}
\put(-100,280){{\small a)}}
\put(60,280){{\small b)}}
\put(220,280){{\small c)}}
\end{picture}
\vskip -1.5cm
\caption{Branching ratios of decays $D\to e N_I$ (gray solid
lines), $D\to \mu N_I$ (gray long-dashed lines), $D_s\to e
N_I$ (black solid lines), $D_s\to \mu N_I$ (black long-dashed lines) and
$D_s\to \tau N_I$ (black short-dashed lines) as functions of heavy
neutrino mass $M_N$ in models: a) I, b) II, c) III.  In a
phenomenologically viable model and heavy neutrino mass within
$M_\pi\lesssim M_N\lesssim M_D$, the branching ratios are confined
between corresponding thin and thick lines which show upper and lower
limits on $U^2$ from Fig.~\ref{Nu-lifetime}b, respectively.
\label{leptonic-widths-charm}
}
\end{figure}

\begin{figure}[!htb]
\centerline{
\includegraphics[width=0.33\textwidth]{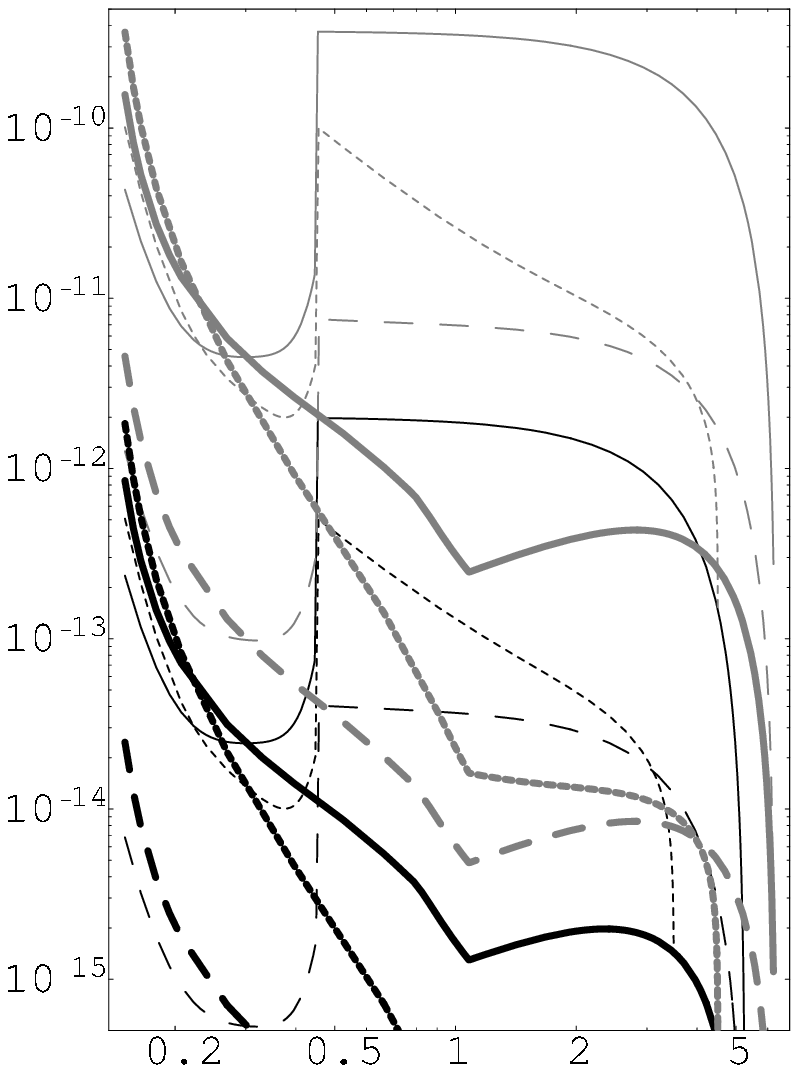}
\includegraphics[width=0.33\textwidth]{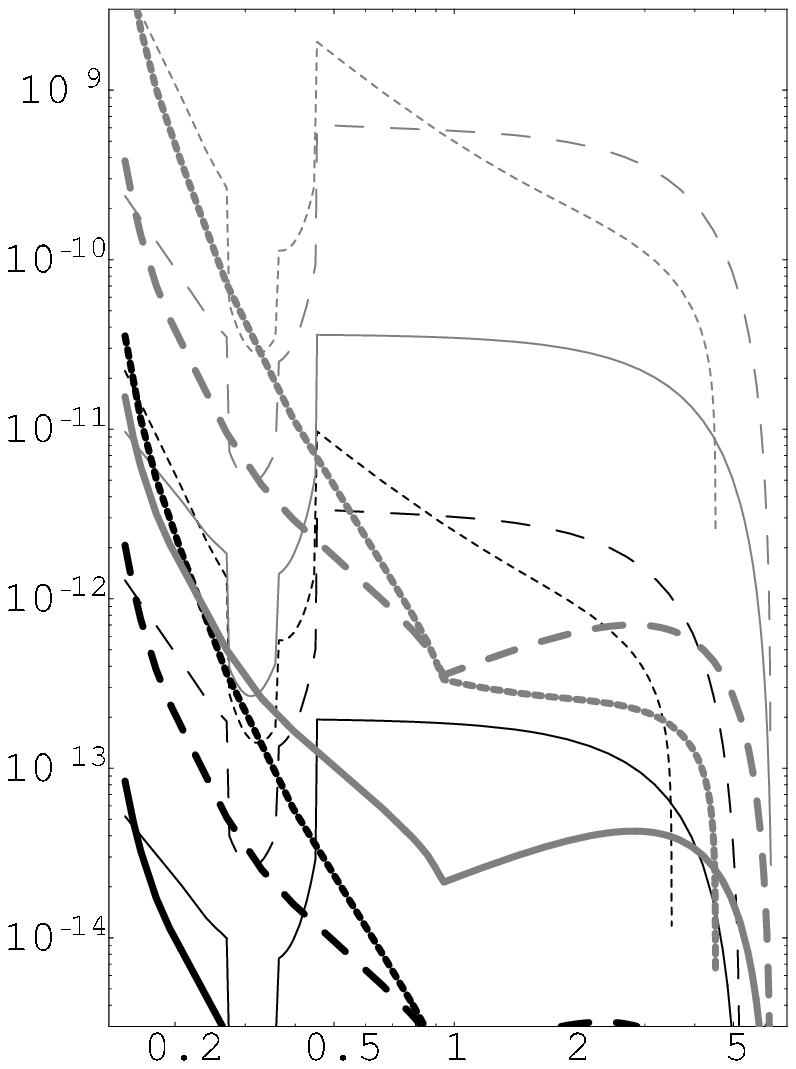}
\includegraphics[width=0.33\textwidth]{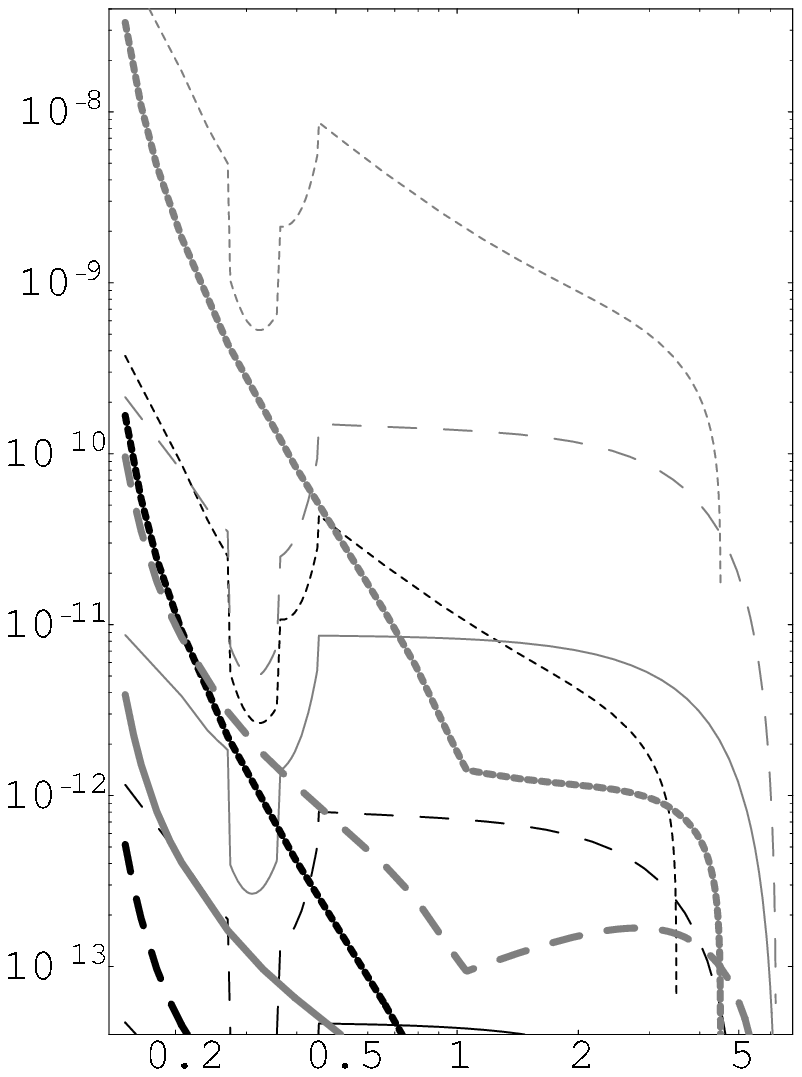}}
\begin{picture}(0,0)(0,0)
\put(-170,30){{\small$M_N$, GeV}}
\put(-10,30){{\small$M_N$, GeV}}
\put(155,30){{\small$M_N$, GeV}}
\put(-95,220){{\small a)}}
\put(60,220){{\small b)}}
\put(220,220){{\small c)}}
\end{picture}
\vskip -1.cm
\caption{Branching ratios of decays $B\to e N_I$ (black solid
lines), $B\to \mu N_I$ (black long-dashed lines), $B\to
\tau N_I$ (black short-dashed lines), $B_c\to e
N_I$ (gray solid lines), $B_c\to \mu N_I$ (gray long-dashed lines) and
$B_c\to \tau N_I$ (gray short-dashed lines) as functions of heavy
neutrino mass $M_N$ in models: a) I, b) II, c) III. In a
phenomenologically viable model and heavy neutrino mass within
$M_\pi\lesssim M_N\lesssim M_B$, the branching ratios are confined
between corresponding thin and thick lines which show upper and lower
limits on $U^2$ from Fig.~\ref{Nu-lifetime}b, respectively.
\label{leptonic-widths-beauty}
}
\end{figure}

\begin{figure}[!htb]
\centerline{
\includegraphics[width=0.33\textwidth]{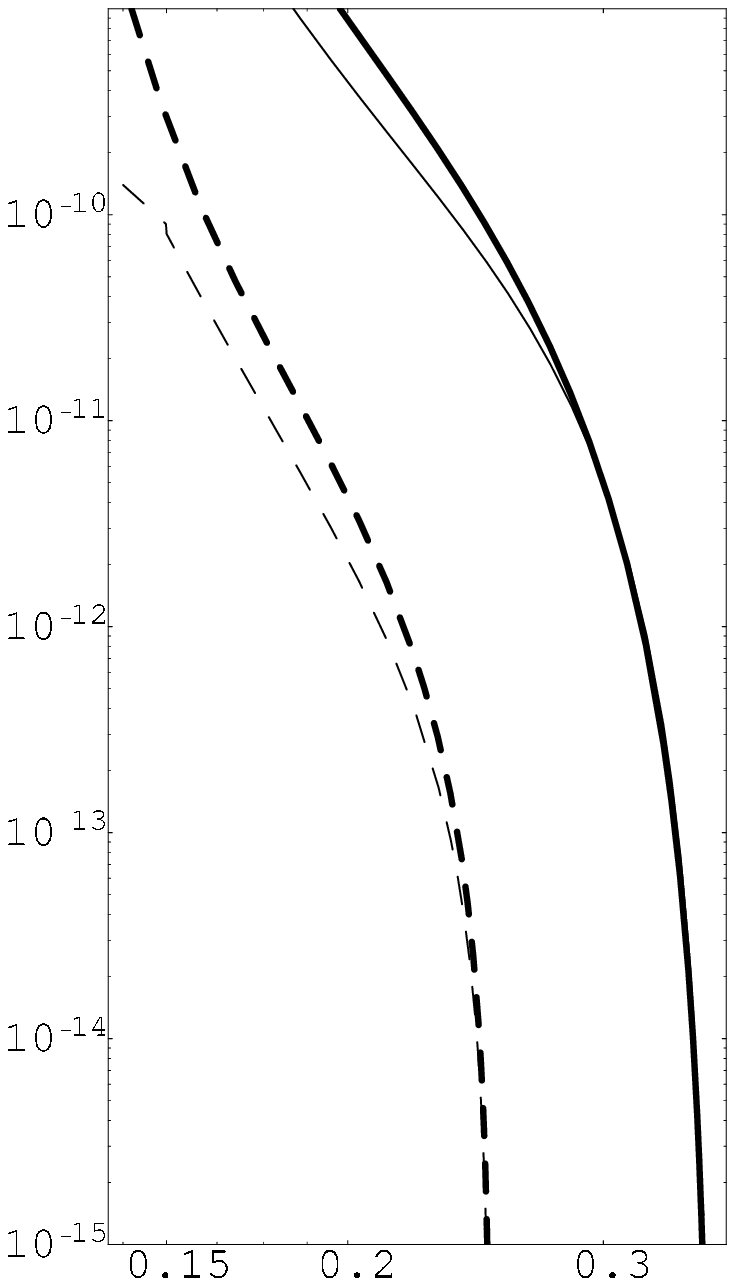}
\includegraphics[width=0.33\textwidth]{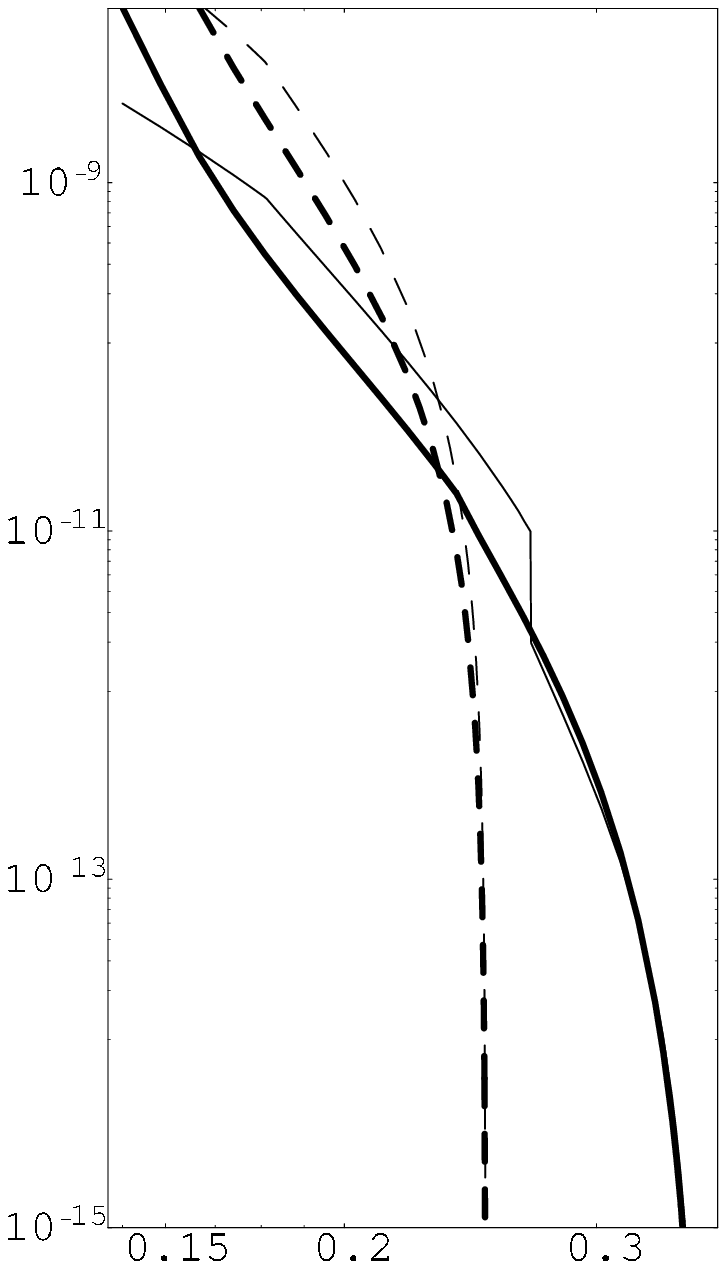}
\includegraphics[width=0.33\textwidth]{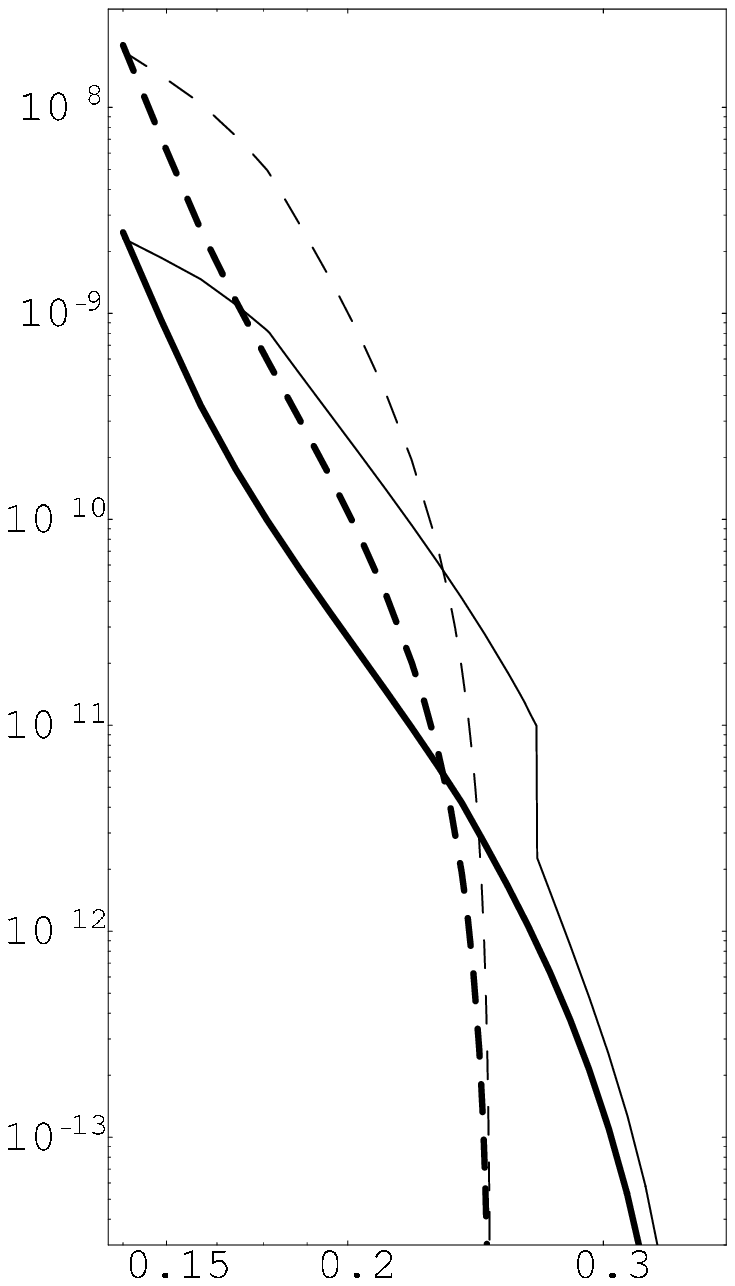}}
\begin{picture}(0,0)(0,0)
\put(-160,40){{\small$M_N$, GeV}}
\put(0,40){{\small$M_N$, GeV}}
\put(155,40){{\small$M_N$, GeV}}
\put(-110,280){{\small a)}}
\put(50,280){{\small b)}}
\put(210,280){{\small c)}}
\end{picture}
\vskip -1.5cm
\caption{Branching ratios of semileptonic decays $K\to\pi e N_I$ (solid
lines) and $K\to\pi \mu N_I$ (dashed lines) as functions of heavy neutrino
mass $M_N$ in models: a) I, b) II, c) III. In a
phenomenologically viable model and heavy neutrino mass within
$M_\pi\lesssim M_N\lesssim M_K$, the branching ratios are confined
between corresponding thin and thick lines which show upper and lower
limits on $U^2$ from Fig.~\ref{Nu-lifetime}b, respectively; 
form factors are taken from Refs.~\cite{PDG}.
\label{semileptonic-widths-K} 
}
\end{figure}

\begin{figure}[!htb]
\centerline{
\includegraphics[width=0.33\textwidth]{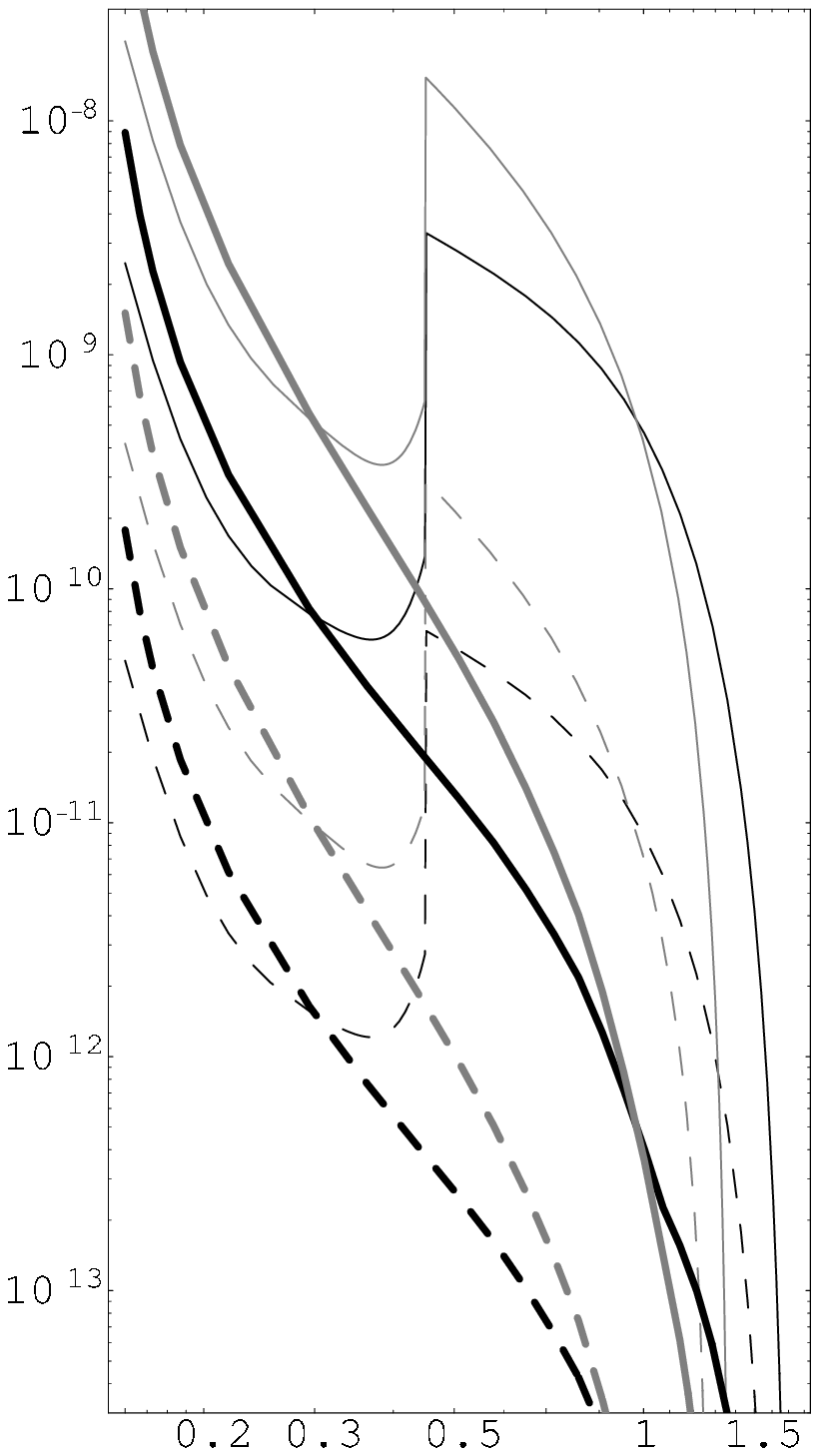}
\includegraphics[width=0.33\textwidth]{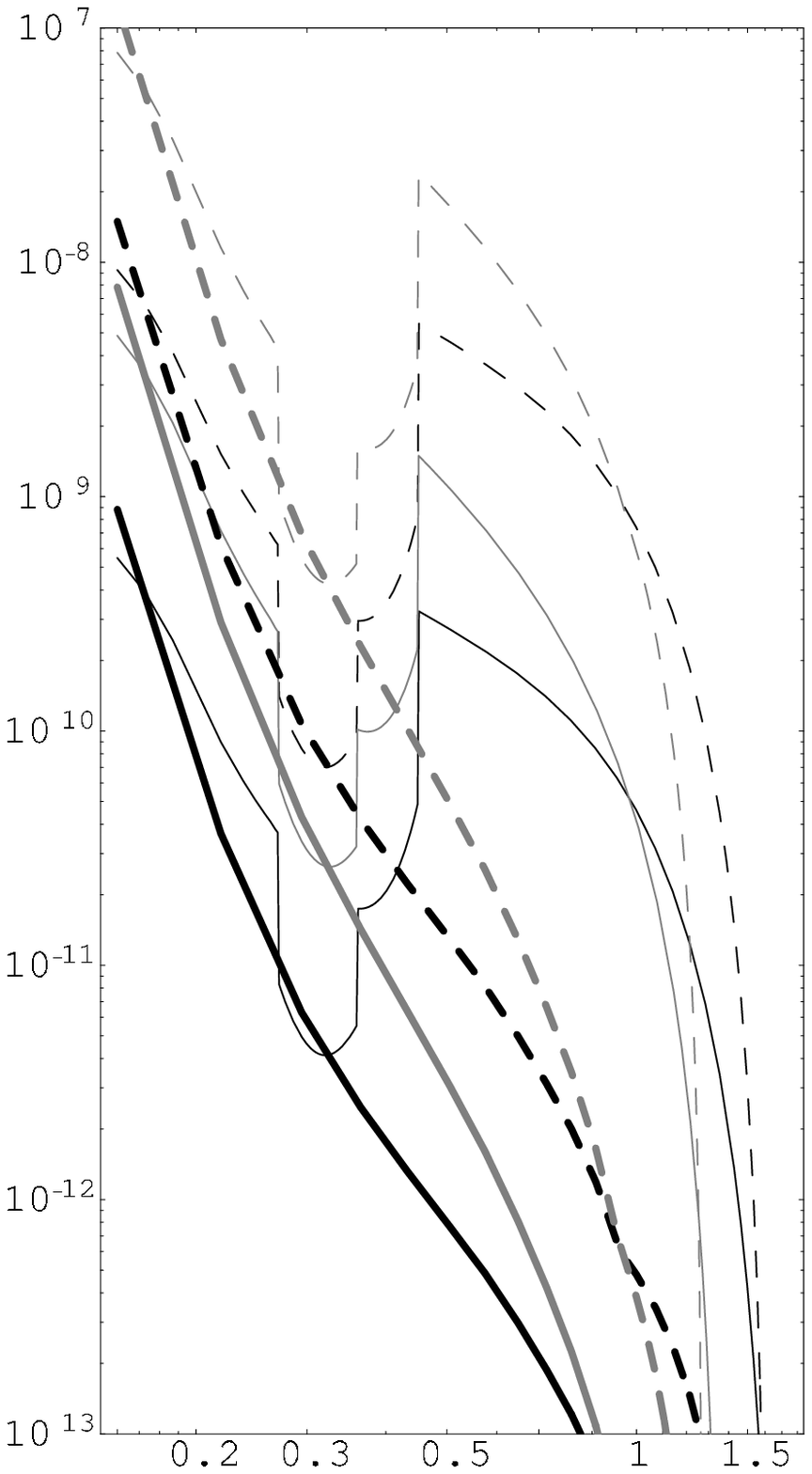}
\includegraphics[width=0.33\textwidth]{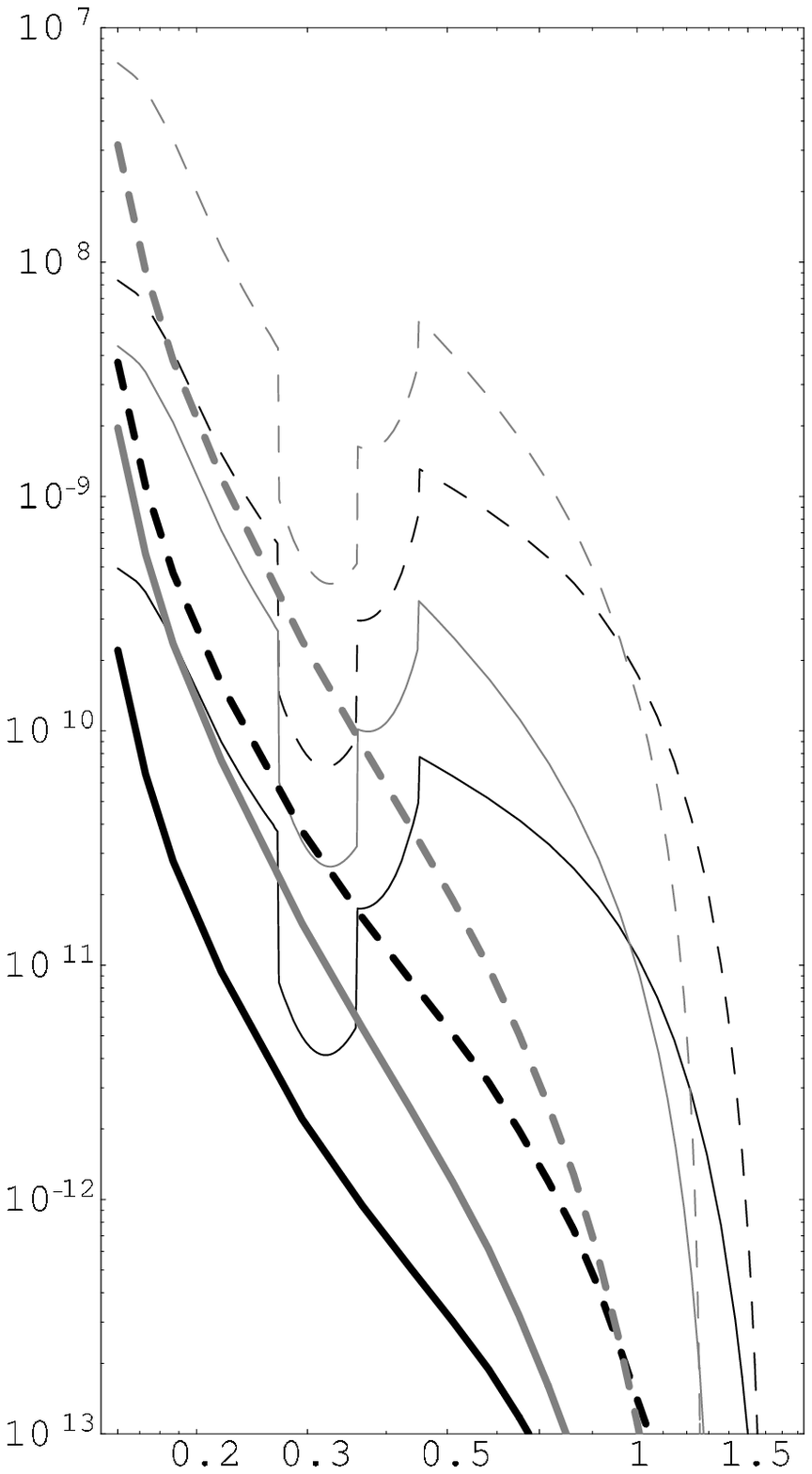}}
\begin{picture}(0,0)(0,0)
\put(-160,35){{\small$M_N$, GeV}}
\put(0,35){{\small$M_N$, GeV}}
\put(155,35){{\small$M_N$, GeV}}
\put(-110,280){{\small a)}}
\put(50,280){{\small b)}}
\put(210,280){{\small c)}}
\end{picture}
\vskip -1.3cm
\caption{Branching ratios of semileptonic decays $D\to\pi e N_I$ (black solid
lines), $D\to\pi \mu N_I$ (black dashed lines), $D\to K e N_I$ (gray solid
lines) and $D\to K \mu N_I$ (gray dashed lines) as functions of heavy neutrino
mass $M_N$ in models: a) I, b) II, c) III. In a
phenomenologically viable model and heavy neutrino mass within
$M_\pi\lesssim M_N\lesssim M_D$, the branching ratios are confined
between corresponding thin and thick lines which show upper and lower
limits on $U^2$ from Fig.~\ref{Nu-lifetime}b, respectively; 
form factors are taken from Refs.~\cite{Shipsey:2006gf}. 
\label{semileptonic-widths-D}    
}
\end{figure}

\begin{figure}[!htb]
\centerline{
\includegraphics[width=0.33\textwidth]{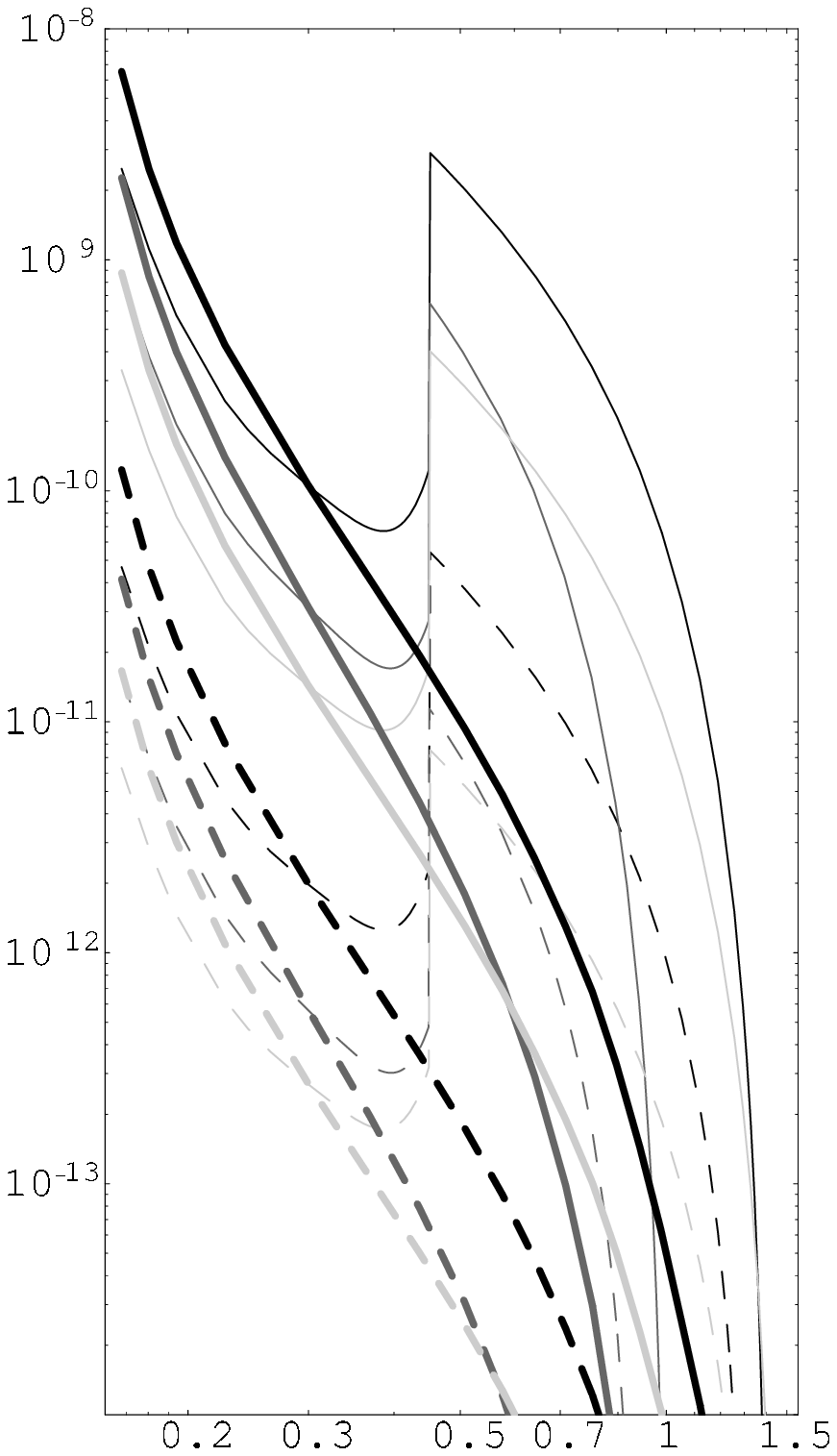}
\includegraphics[width=0.33\textwidth]{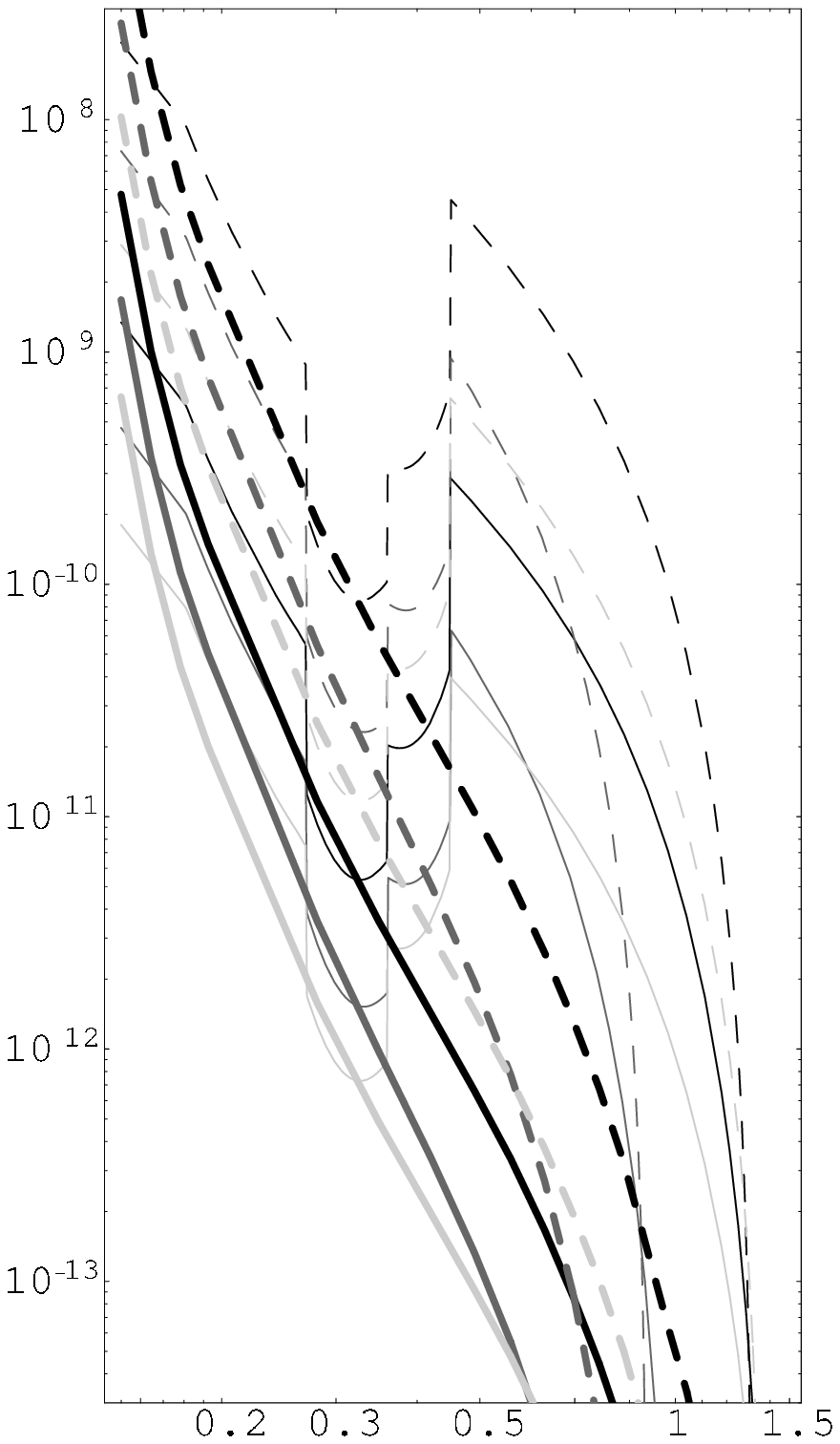}
\includegraphics[width=0.33\textwidth]{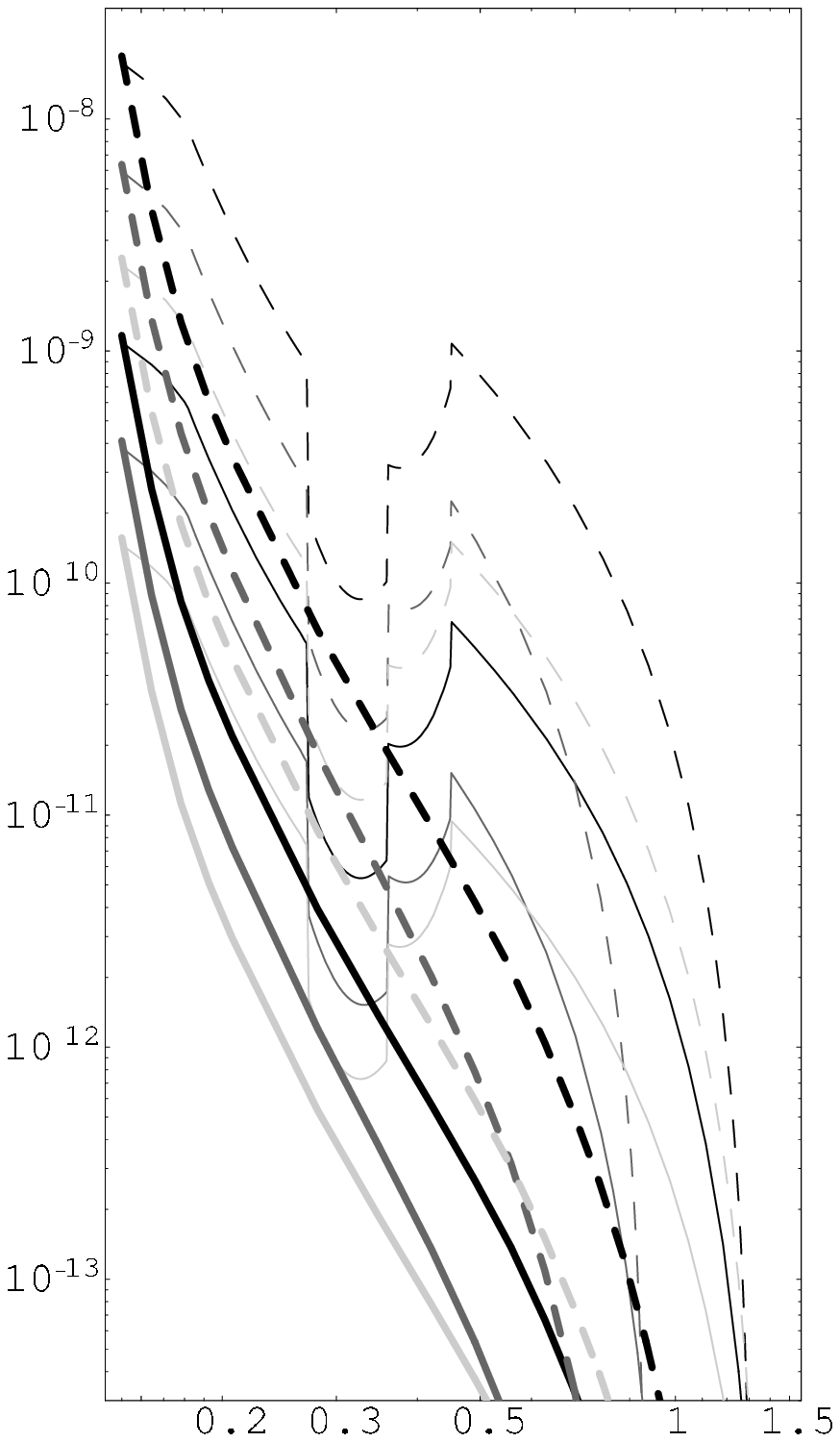}}
\begin{picture}(0,0)(0,0)
\put(-160,40){{\small$M_N$, GeV}}
\put(0,40){{\small$M_N$, GeV}}
\put(155,40){{\small$M_N$, GeV}}
\put(-110,280){{\small a)}}
\put(50,280){{\small b)}}
\put(210,280){{\small c)}}
\end{picture}
\vskip -1.5cm
\caption{Branching ratios of semileptonic decays $D_s\to X l N_I$,
$X=\eta,\eta',K$ (black, dark gray, light gray lines), 
$l=e,\mu$ (solid and dashed lines),  
as functions of heavy neutrino mass $M_N$ in models: a) I, b) II, c)
III. In a
phenomenologically viable model and heavy neutrino mass within
$M_\pi\lesssim M_N\lesssim M_D$, the branching ratios are confined
between corresponding thin and thick lines which show upper and lower
limits on $U^2$ from Fig.~\ref{Nu-lifetime}b, respectively; form factors
are taken from Ref.~\cite{Melikhov:2000yu}.
\label{semileptonic-widths-Ds}    
}
\end{figure}

\begin{figure}[!htb]
\centerline{
\includegraphics[width=0.33\textwidth]{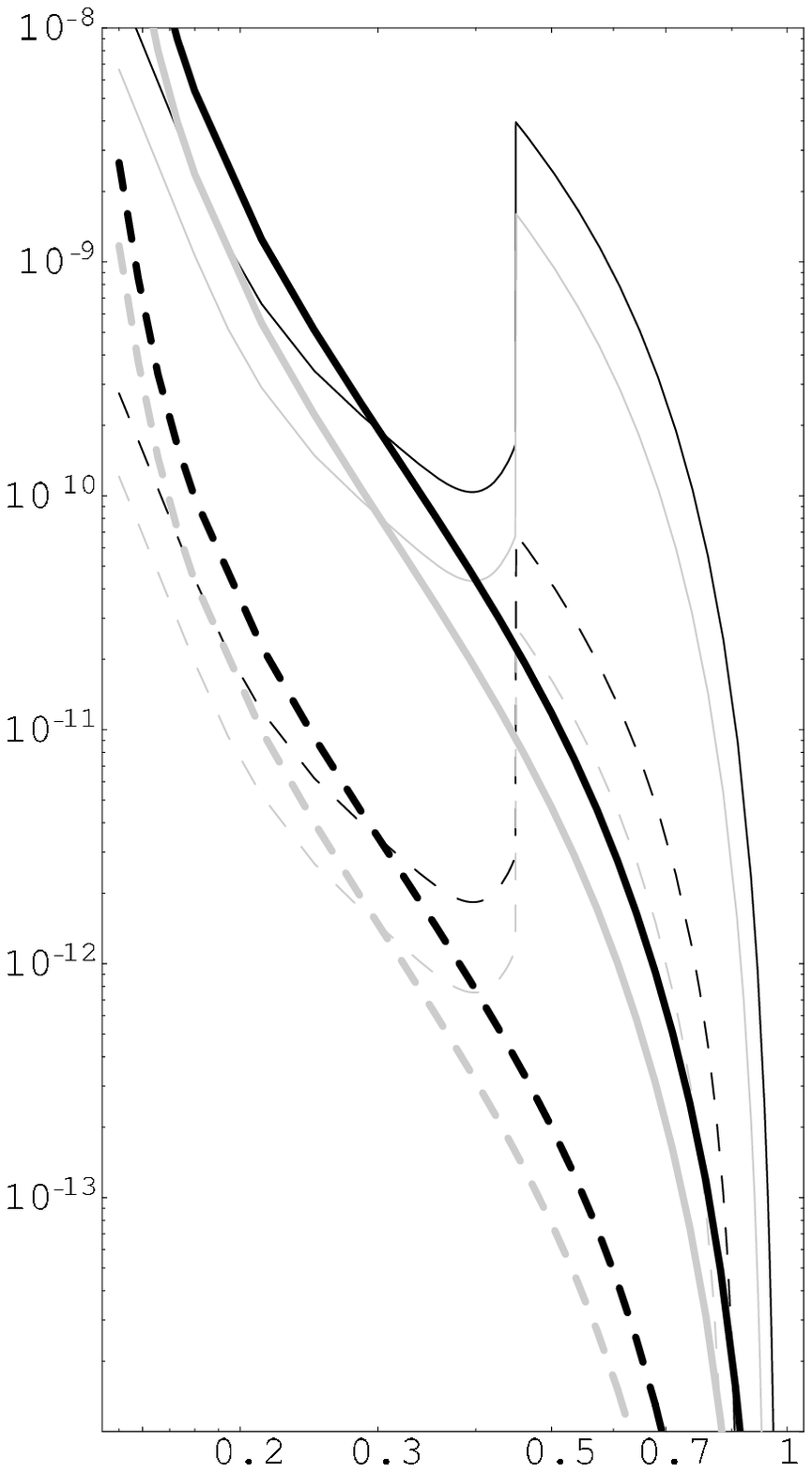}
\includegraphics[width=0.33\textwidth]{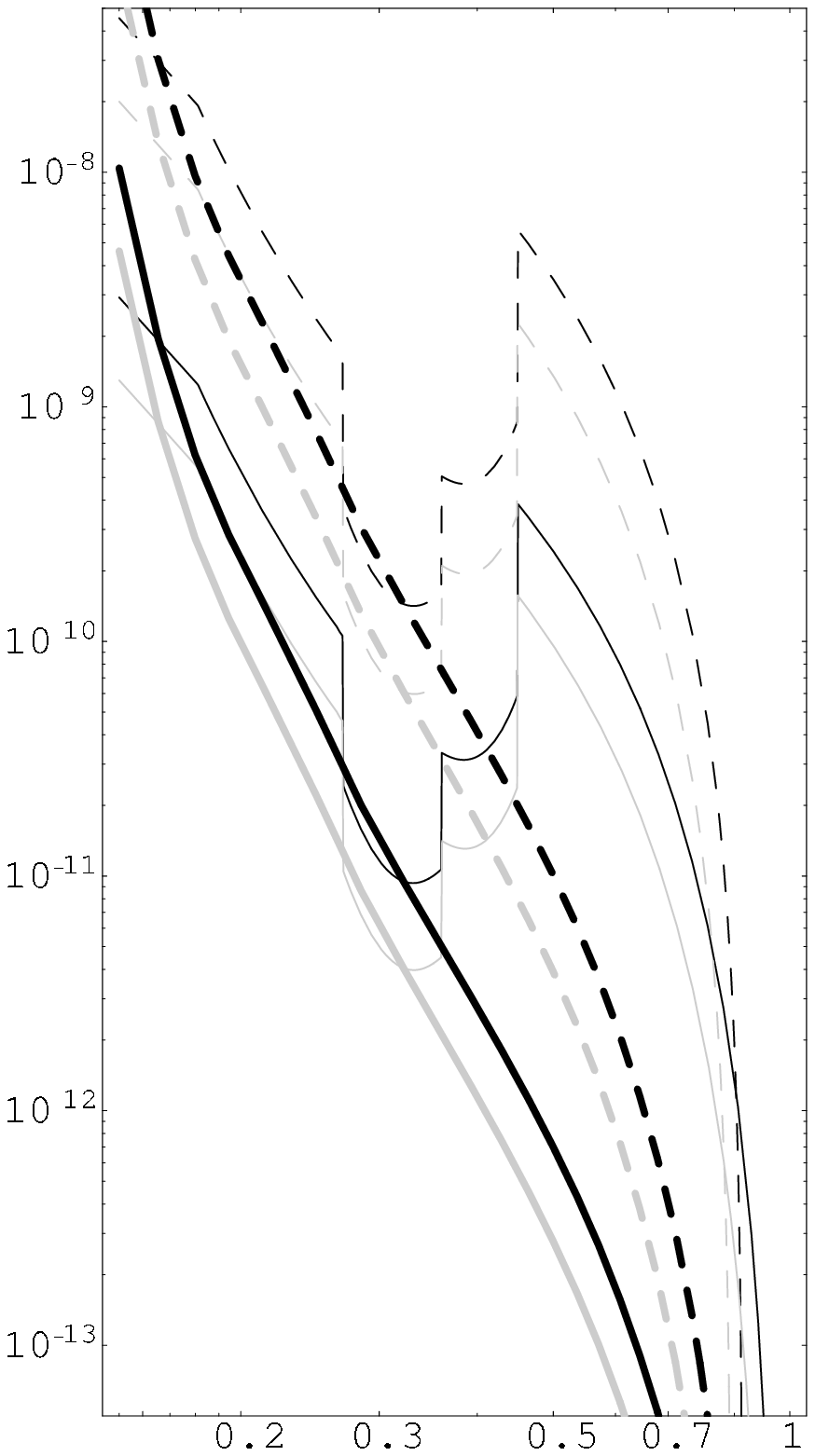}
\includegraphics[width=0.33\textwidth]{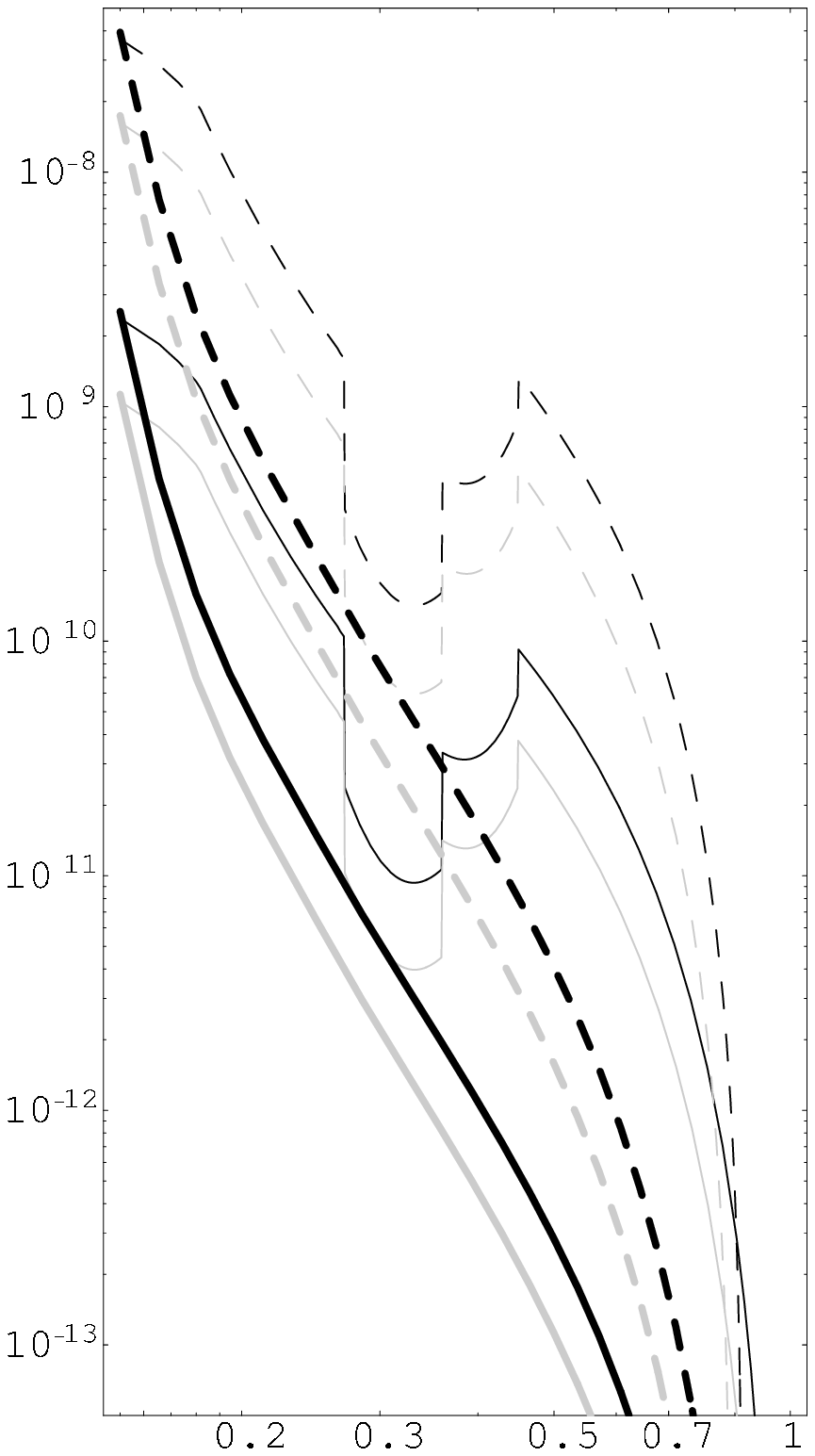}}
\begin{picture}(0,0)(0,0)
\put(-160,30){{\small$M_N$, GeV}}
\put(0,30){{\small$M_N$, GeV}}
\put(155,30){{\small$M_N$, GeV}}
\put(-110,280){{\small a)}}
\put(50,280){{\small b)}}
\put(210,280){{\small c)}}
\end{picture}
\vskip -1cm
\caption{Branching ratios of semileptonic decays  
$D\to K^* e N_I$ (black solid lines), $D\to K^* \mu N_I$ (black dashed 
 lines), $D_s\to \phi e N_I$ (gray solid lines) and 
$D_s\to \phi \mu N_I$ (gray dashed lines) 
as functions of heavy neutrino mass $M_N$ in models: a) I, b) II, c)
III. In a
phenomenologically viable model and heavy neutrino mass within
$M_\pi\lesssim M_N\lesssim M_D$, the branching ratios are confined
between corresponding thin and thick lines which show upper and lower
limits on $U^2$ from Fig.~\ref{Nu-lifetime}b, respectively; form factors
are taken from
Refs.~\cite{Cheng:2003sm,Melikhov:2000yu}.
\label{semileptonic-widths-D-to-vectors}
}
\end{figure}

\begin{figure}[!htb]
\centerline{
\includegraphics[width=0.33\textwidth]{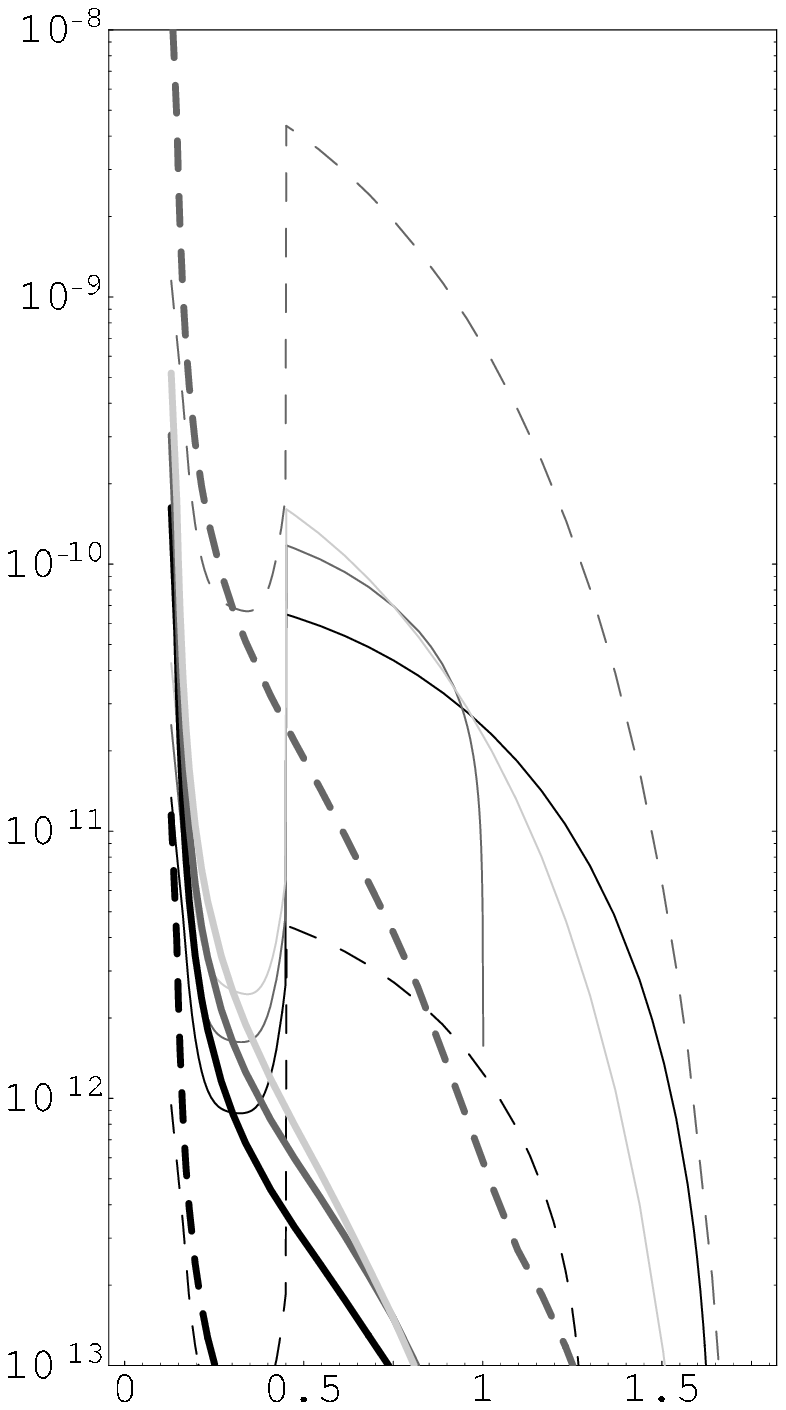}
\includegraphics[width=0.33\textwidth]{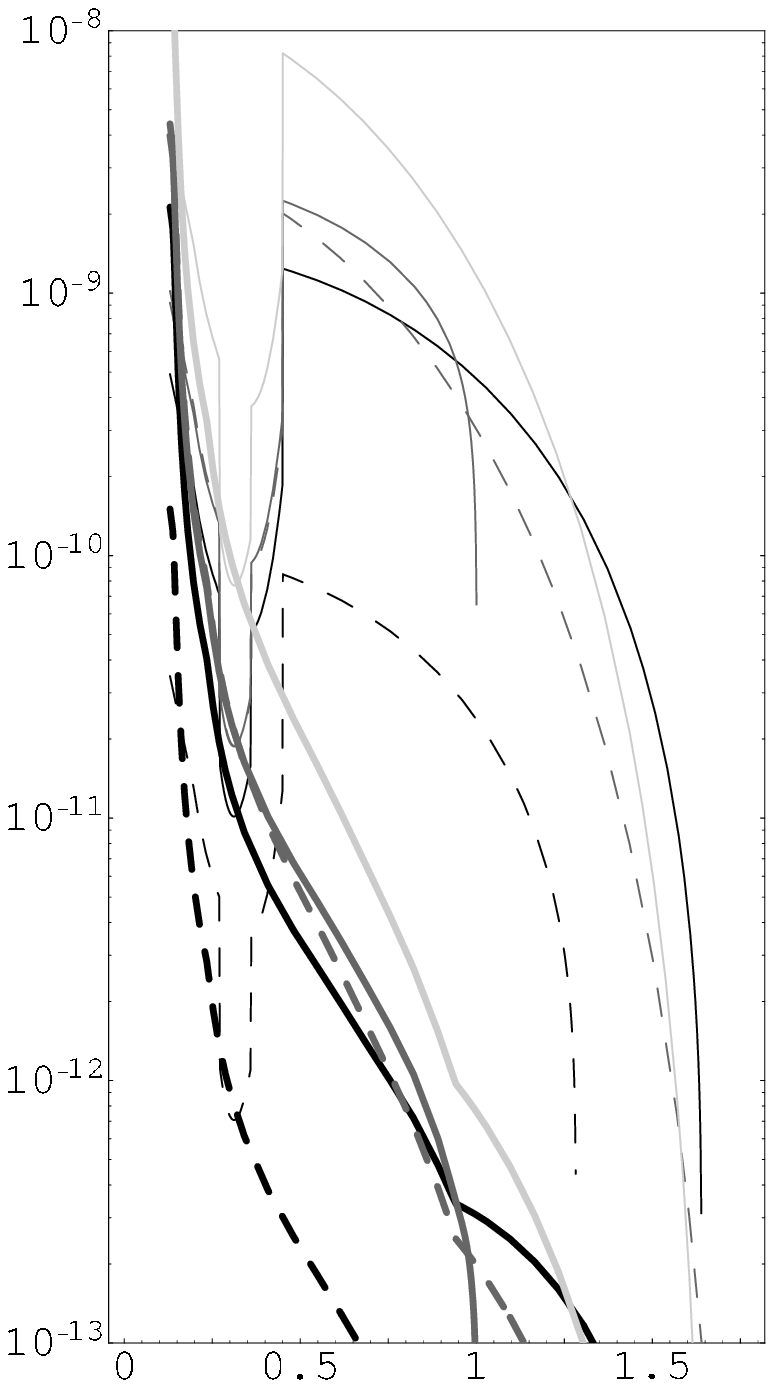}
\includegraphics[width=0.33\textwidth]{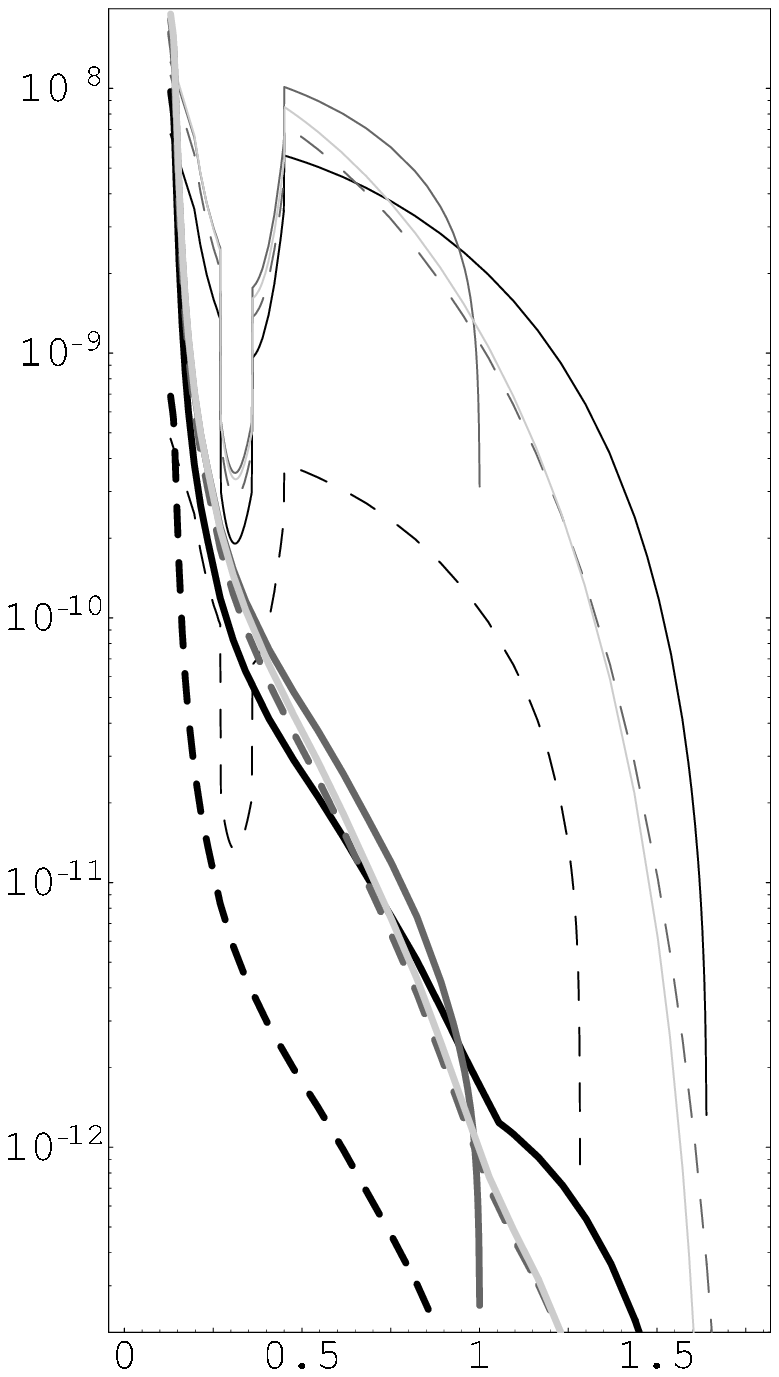}}
\begin{picture}(0,0)(0,0)
\put(-160,40){{\small$M_N$, GeV}}
\put(0,40){{\small$M_N$, GeV}}
\put(155,40){{\small$M_N$, GeV}}
\put(-110,280){{\small a)}}
\put(50,280){{\small b)}}
\put(210,280){{\small c)}}
\end{picture}
\vskip -1.5cm
\caption{Branching ratios of decays $\tau\to\pi N$ (black solid lines), 
$\tau\to K N$ (black dashed lines), $\tau\to \rho N$ (dark gray solid
  lines), 
$\tau\to \nu e N_I$ (sum over all
active neutrino species, dark gray dashed lines), $\tau\to \nu \mu N_I$
(sum over all active neutrino species, light gray solid lines) as
functions of heavy neutrino mass $M_N$ in models: a) I, b) II, c)
III. In a phenomenologically viable model and heavy neutrino mass
within $M_\pi\lesssim M_N\lesssim M_\tau$, the branching ratios are
confined between corresponding thin and thick lines which show upper
and lower limits on $U^2$ from Fig.~\ref{Nu-lifetime}b, respectively.
\label{leptonic-widths-tau}
}
\end{figure}

\begin{figure}[!htb]
\centerline{
\includegraphics[width=0.33\textwidth]{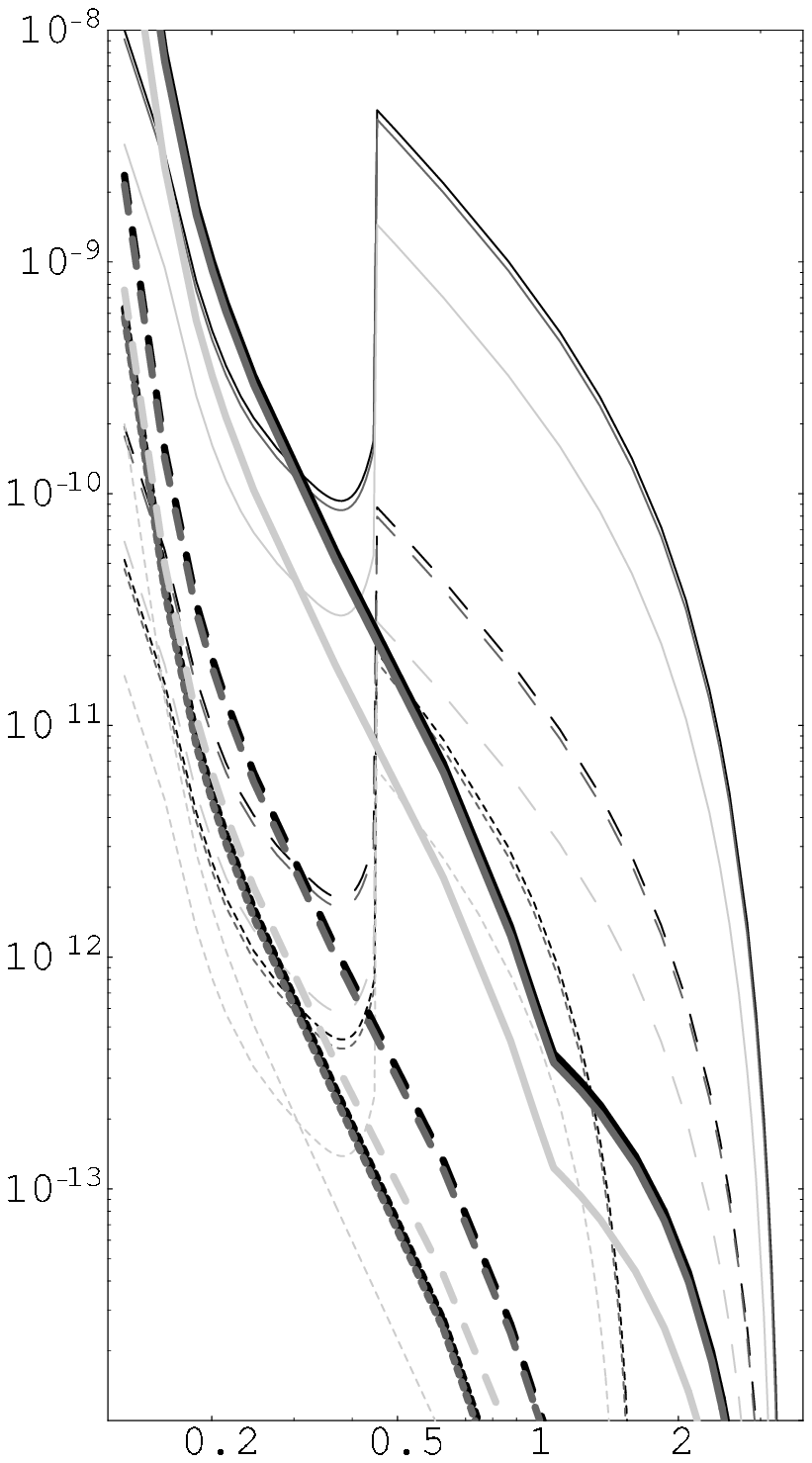}
\includegraphics[width=0.33\textwidth]{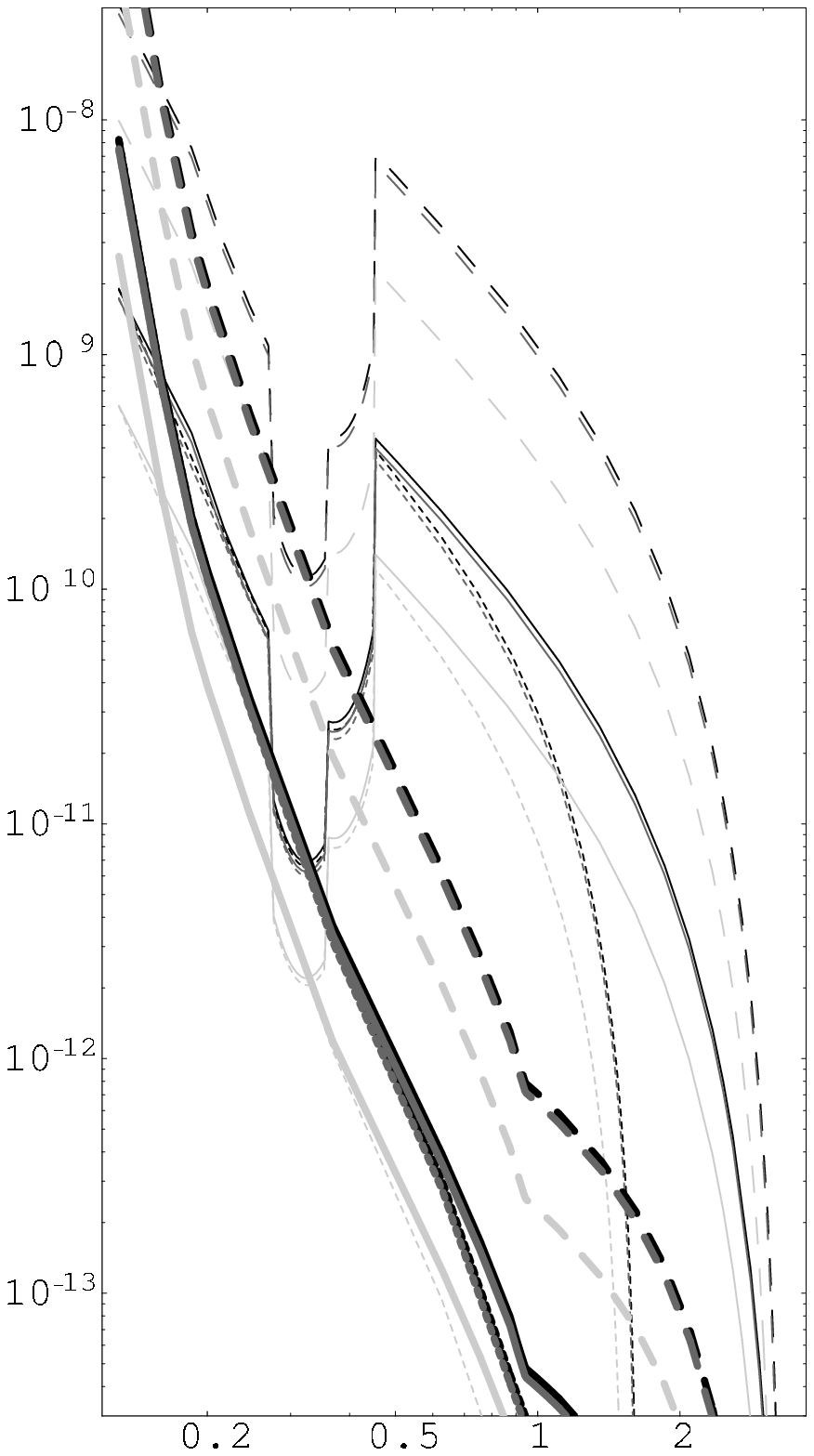}
\includegraphics[width=0.33\textwidth]{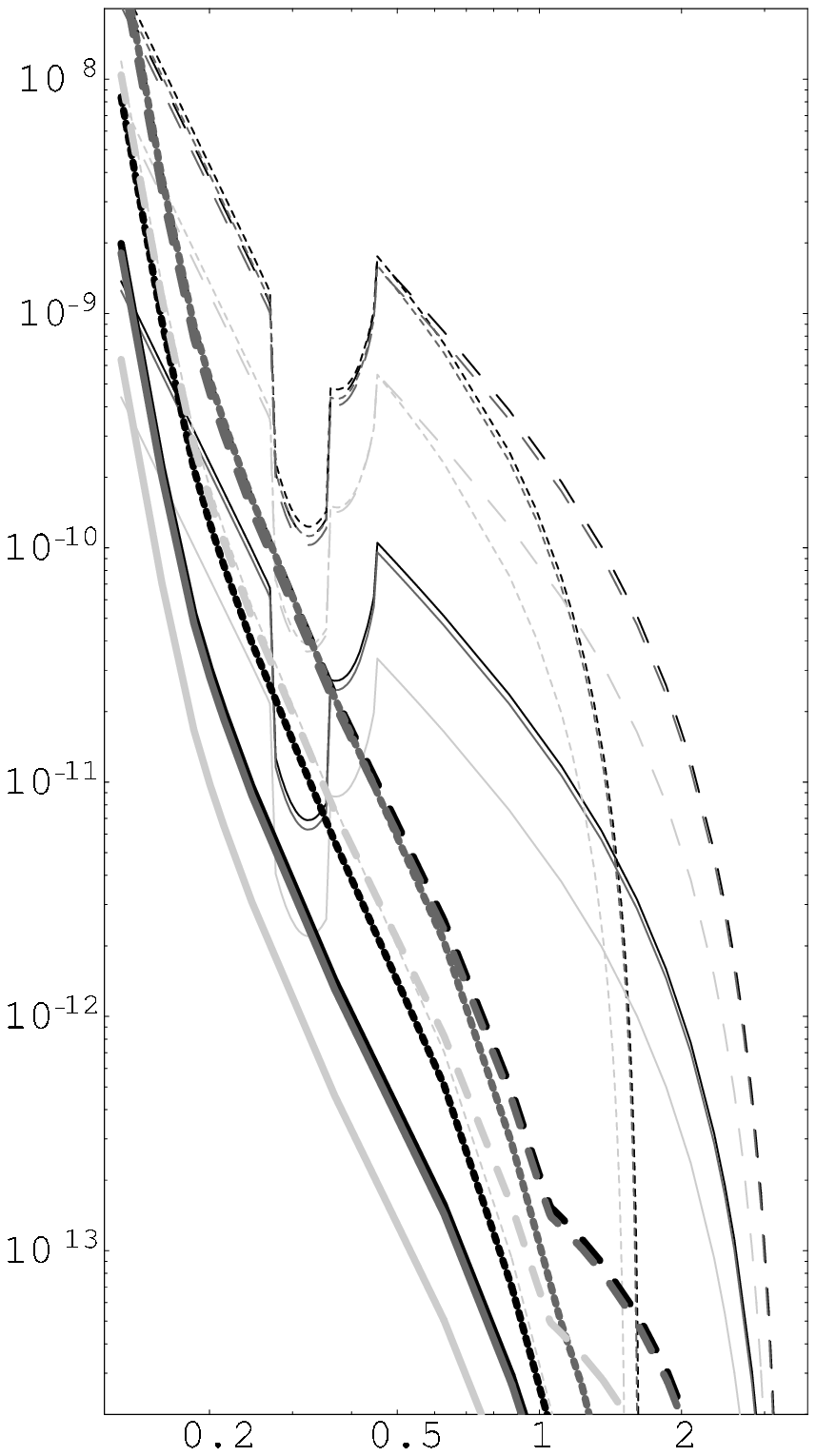}}
\begin{picture}(0,0)(0,0)
\put(-160,35){{\small$M_N$, GeV}}
\put(0,35){{\small$M_N$, GeV}}
\put(155,35){{\small$M_N$, GeV}}
\put(-110,280){{\small a)}}
\put(50,280){{\small b)}}
\put(210,280){{\small c)}}
\end{picture}
\vskip -1.3cm
\caption{Branching ratios of semileptonic decays $B\to D l N_I$ (black
lines), $B_s\to D_s l N_I$ (dark gray lines) and $B_c\to \eta_c l N_I$
(light gray lines), $l=e,\mu,\tau$ (solid, long dashed and short
dashed lines) as functions of heavy neutrino mass $M_N$ in models: a)
I, b) II, c) III.  In a phenomenologically viable model and heavy
neutrino mass within $M_\pi\lesssim M_N\lesssim M_B$, the branching
ratios are confined between corresponding thin and thick lines which
show upper and lower limits on $U^2$ from Fig.~\ref{Nu-lifetime}b,
respectively; form factors for $B_s$-meson decays are taken to be
equal to the form factors for $B$-meson decays from
Ref.~\cite{Cheng:2003sm}, for $B_c$-meson decays we adopted form
factors from Ref.~\cite{Ebert:2003cn}.
\label{semileptonic-widths-beauty-a}
}
\end{figure}

\begin{figure}[!htb]
\centerline{
\includegraphics[width=0.33\textwidth]{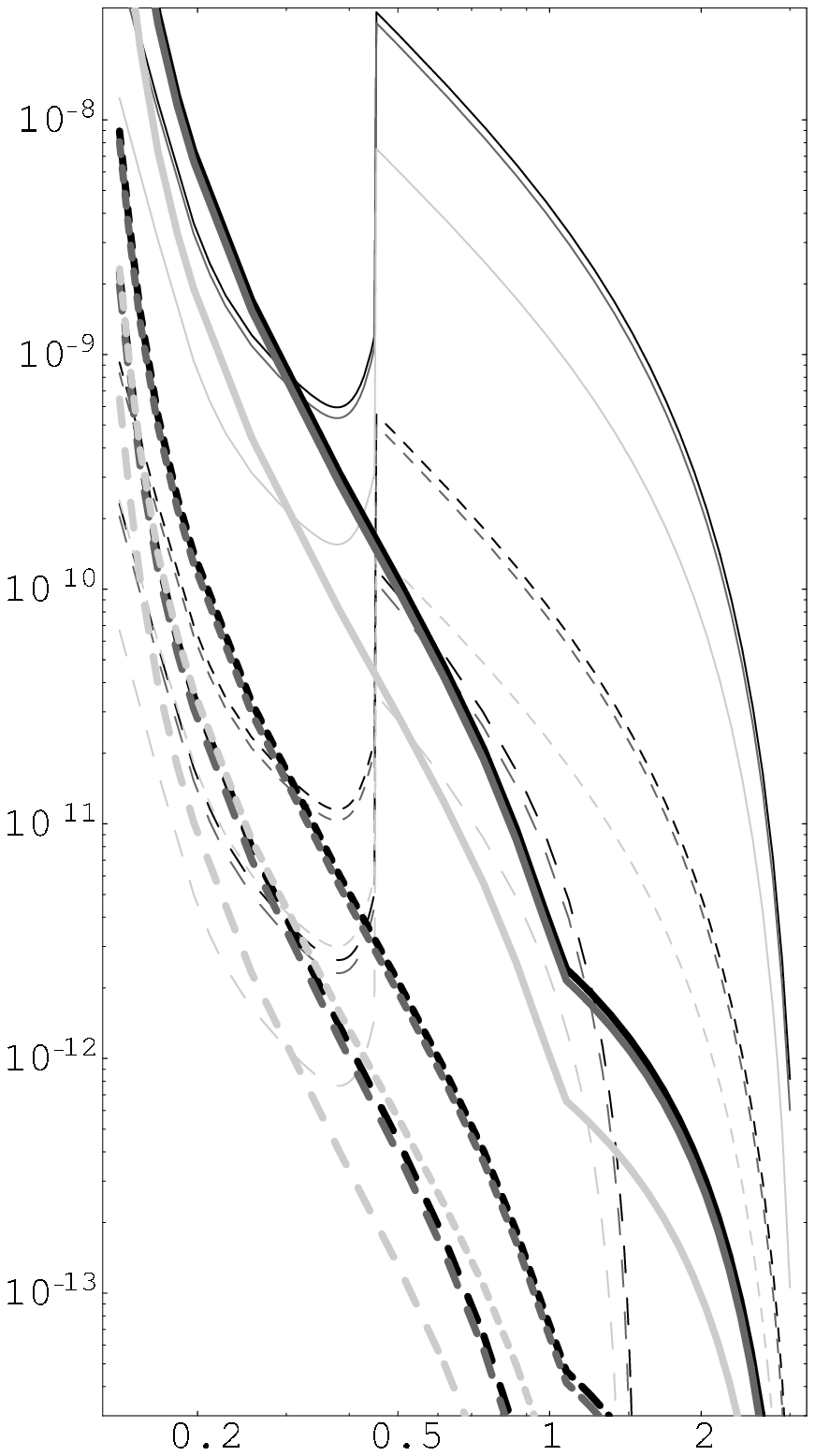}
\includegraphics[width=0.33\textwidth]{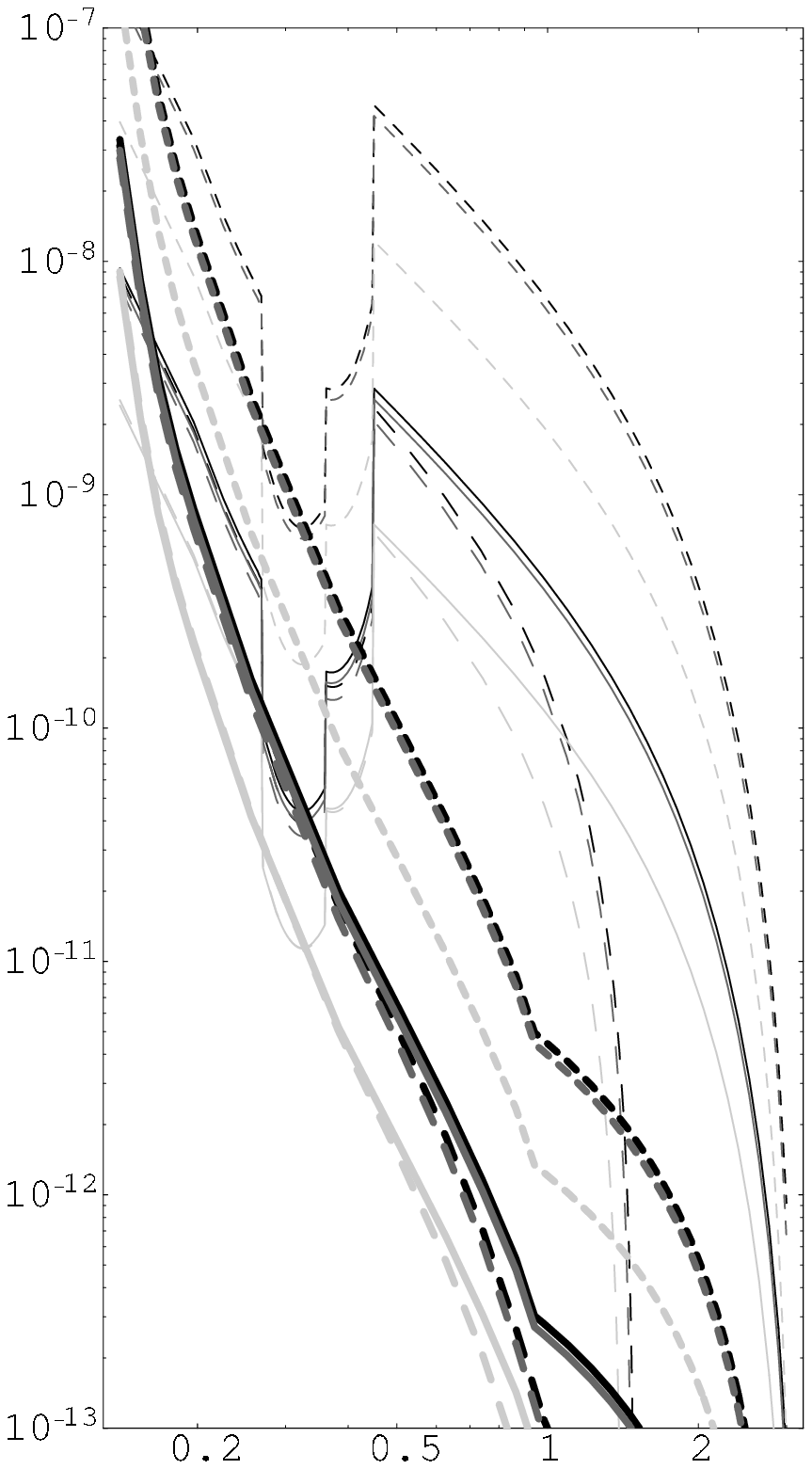}
\includegraphics[width=0.33\textwidth]{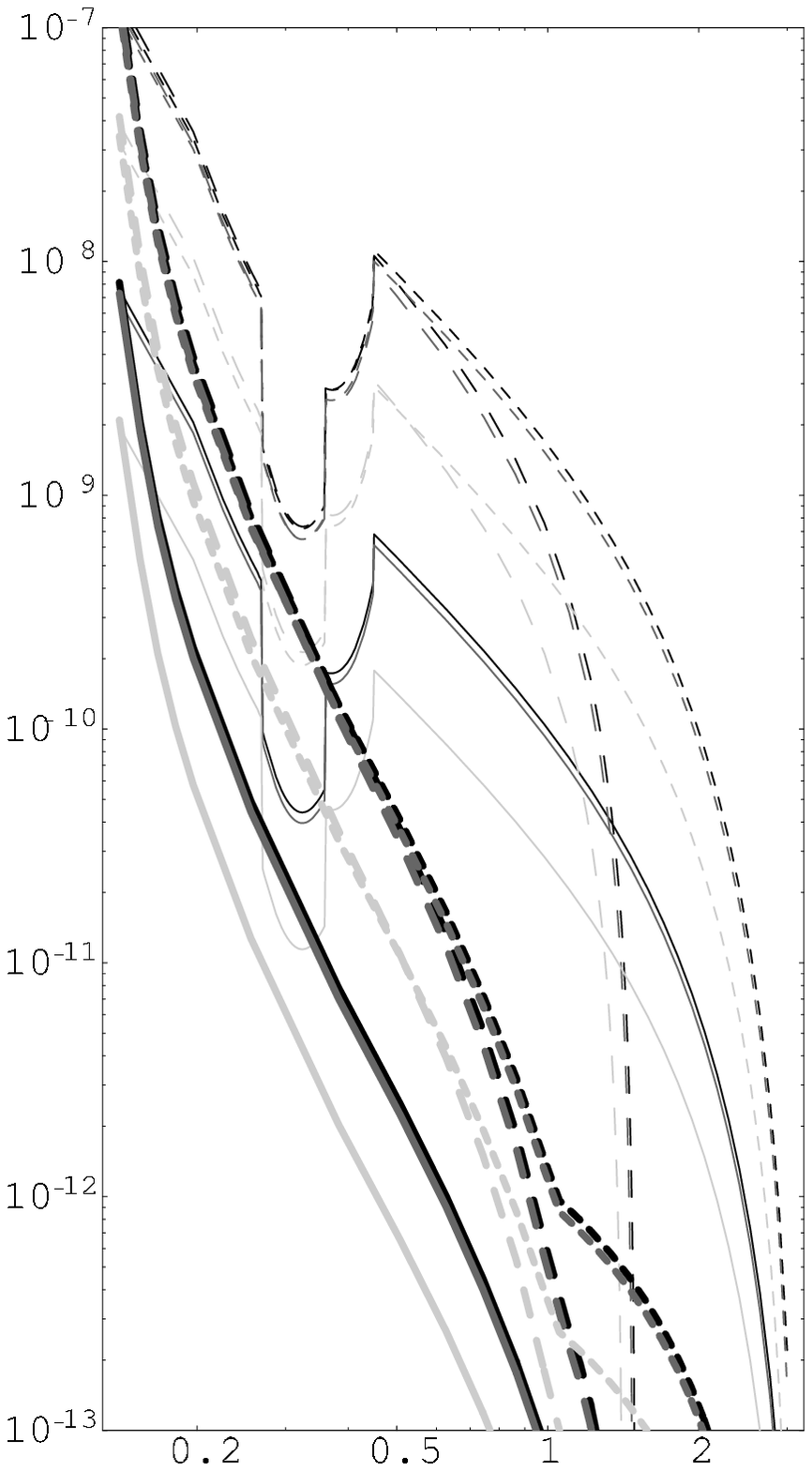}}
\begin{picture}(0,0)(0,0)
\put(-160,30){{\small$M_N$, GeV}}
\put(0,30){{\small$M_N$, GeV}}
\put(155,30){{\small$M_N$, GeV}}
\put(-110,280){{\small a)}}
\put(50,280){{\small b)}}
\put(210,280){{\small c)}}
\end{picture}
\vskip -1cm
\caption{Branching ratios of semileptonic decays 
$B\to D^* l N_I$ (black lines), $B_s\to D^*_s l
N_I$ (dark gray lines) and   
$B_c\to J/\psi l N_I$ (light gray lines),  
$l=e,\mu,\tau$ (solid, short dashed and long dashed lines)
as functions of heavy neutrino mass $M_N$ in models: a) I, b) II, c)
III. In a
phenomenologically viable model and heavy neutrino mass within
$M_\pi\lesssim M_N\lesssim M_D$, the branching ratios are confined
between corresponding thin and thick lines which show upper and lower
limits on $U^2$ from Fig.~\ref{Nu-lifetime}b, respectively; 
form factors for $B_s$-meson decays 
are taken to be equal to the form factors for $B$-meson decays from Ref.~\cite{Cheng:2003sm}, 
for $B_c$-meson decays we adopted form factors from Ref.~\cite{Ebert:2003cn}.
\label{semileptonic-widths-beauty-b}
}
\end{figure}

\begin{figure}[!htb]
\centerline{
\includegraphics[width=0.33\textwidth]{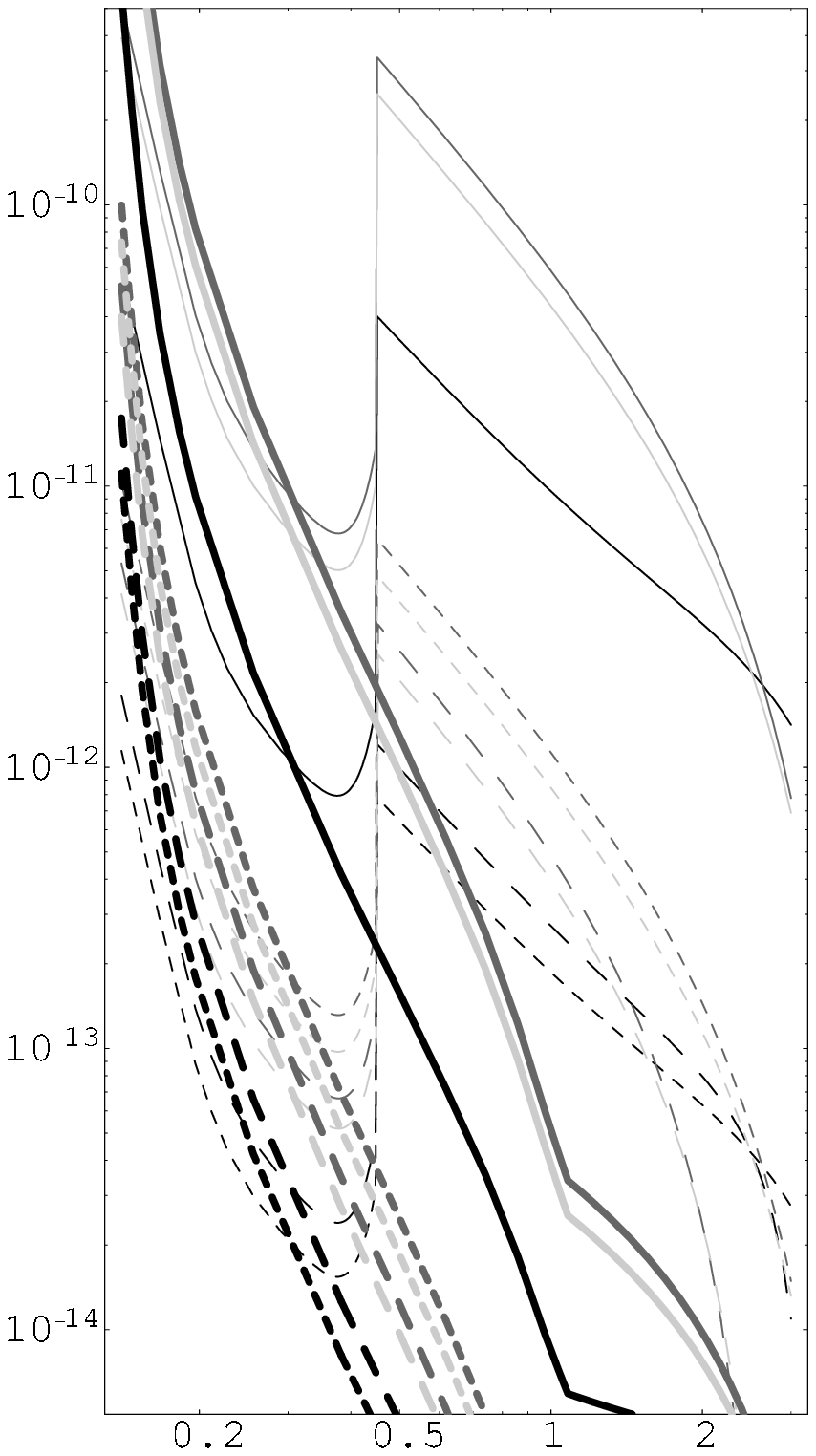}
\includegraphics[width=0.33\textwidth]{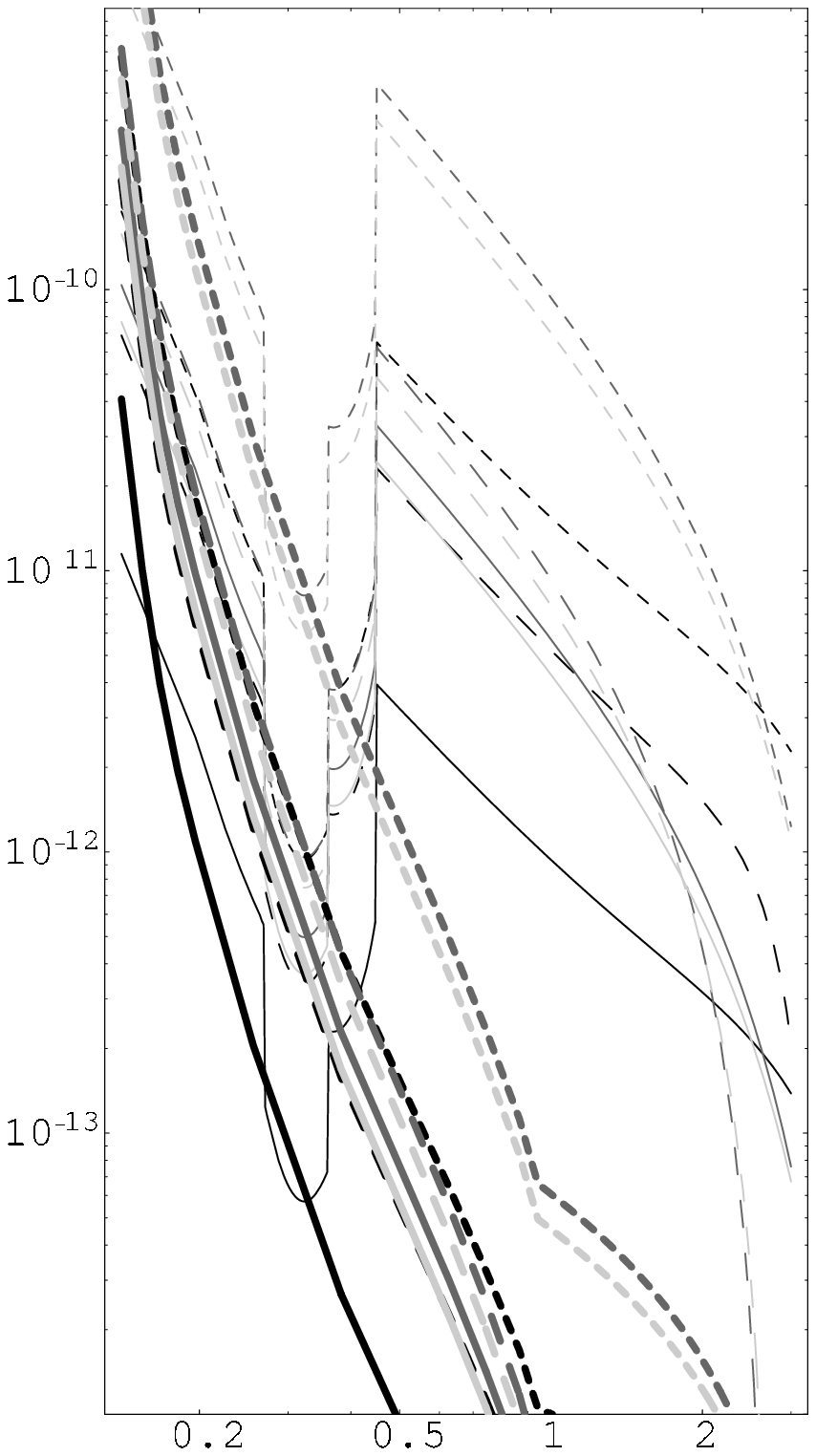}
\includegraphics[width=0.33\textwidth]{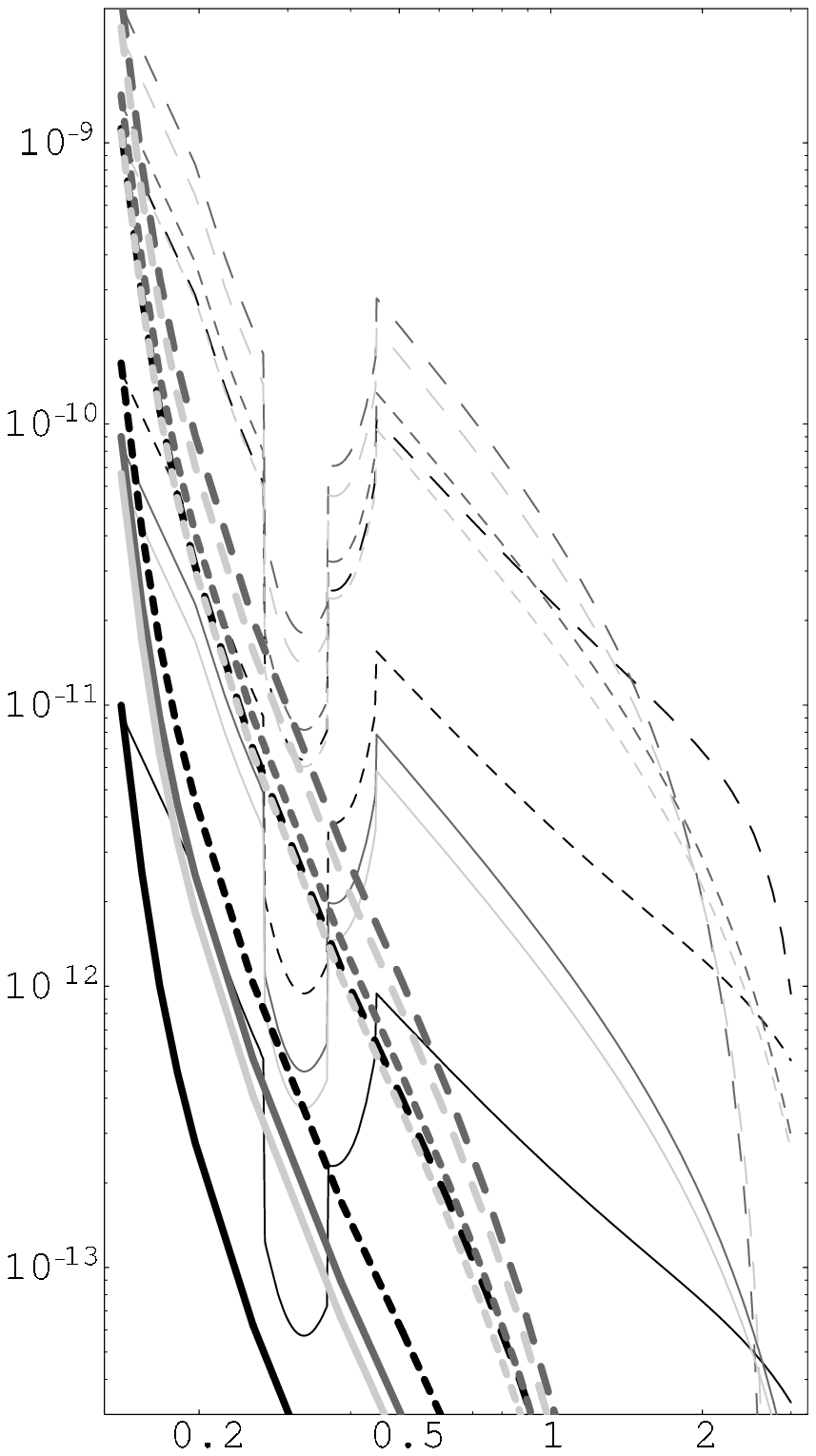}}
\begin{picture}(0,0)(0,0)
\put(-160,35){{\small$M_N$, GeV}}
\put(0,35){{\small$M_N$, GeV}}
\put(155,35){{\small$M_N$, GeV}}
\put(-110,280){{\small a)}}
\put(50,280){{\small b)}}
\put(210,280){{\small c)}}
\end{picture}
\vskip -1.3cm
\caption{Branching ratios of semileptonic decays $B\to \pi l N_I$
(black lines), $B\to \rho l N_I$ (dark gray lines) and $B_s\to K^* l
N_I$ (light gray lines), $l=e,\mu, \tau$ (solid, long dashed and short
dashed lines) as functions of heavy neutrino mass $M_N$ in models: a)
I, b) II, c) III.In a phenomenologically viable model and heavy
neutrino mass within $M_\pi\lesssim M_N\lesssim M_D$, the branching
ratios are confined between corresponding thin and thick lines which
show upper and lower limits on $U^2$ from Fig.~\ref{Nu-lifetime}b,
respectively; form factors are taken from
Refs.~\cite{Cheng:2003sm,Melikhov:2000yu}.
\label{semileptonic-widths-beauty-c}
}
\end{figure}

For each heavy quark $Q$ the dominant contribution to heavy neutral
lepton production comes from leptonic and semileptonic decays of
mesons.  The limits on branching ratios for relevant decays are
plotted in 
Figs.~\ref{leptonic-widths-kaons}-\ref{semileptonic-widths-beauty-c}
as function of neutrino mass for three benchmark models.   
Within $\nu$MSM the interesting branching ratios 
are confined between corresponding thin (upper limit) 
and thick (lower limit) lines: inside these
regions all limits on $U^2$ plotted in Fig.~\ref{Nu-lifetime} 
are fulfilled, in a given model the neutrino mass region, where the
corresponding thin line is below the corresponding thick line, is
disfavoured.  Rate doubling due to heavy neutrino degeneracy is taken
into account.   

The two-body decays can be searched for to
probe $\nu$MSM: produced charged leptons are monochromatic with
spatial momenta 
\[
|{\bf p_l}|=\sqrt{ \l \frac{M_H^2+M_N^2-M_l^2}{2M_H}\r^2 -M_N^2}\;.
\] 
The positions of these  peaks in charged lepton spectra and their
heights are correlated obviously  for different modes and mesons.
These features is a very clean signature of heavy leptons. From the
plots in  Fig. \ref{leptonic-widths-charm} one concludes that 
statistics of billions charmed hadrons is needed to probe $\nu$MSM
with  neutrino of masses 0.5~GeV~$\lesssim M_N\lesssim 2$~GeV. In
models with lighter neutrinos kaon decays are important and required
statistics is smaller. Contrary, in models with heavier neutrinos
statistics has to be larger and it is a challenging task for future
$B$-factories.  Note that the set of phenomenologically interesting
models where neutrinos are produced in kaon decays can be examined
completely, as it requires billions of kaons and collected World
statistics is much larger. 

Semileptonic decays also contribute to heavy lepton production, but spectra 
of outgoing leptons and mesons are not monoenergetic, making this process be
less promising probe of $\nu$MSM heavy neutrinos.  

\begin{figure}[!htb]
\centerline{
\includegraphics[width=0.33\textwidth]{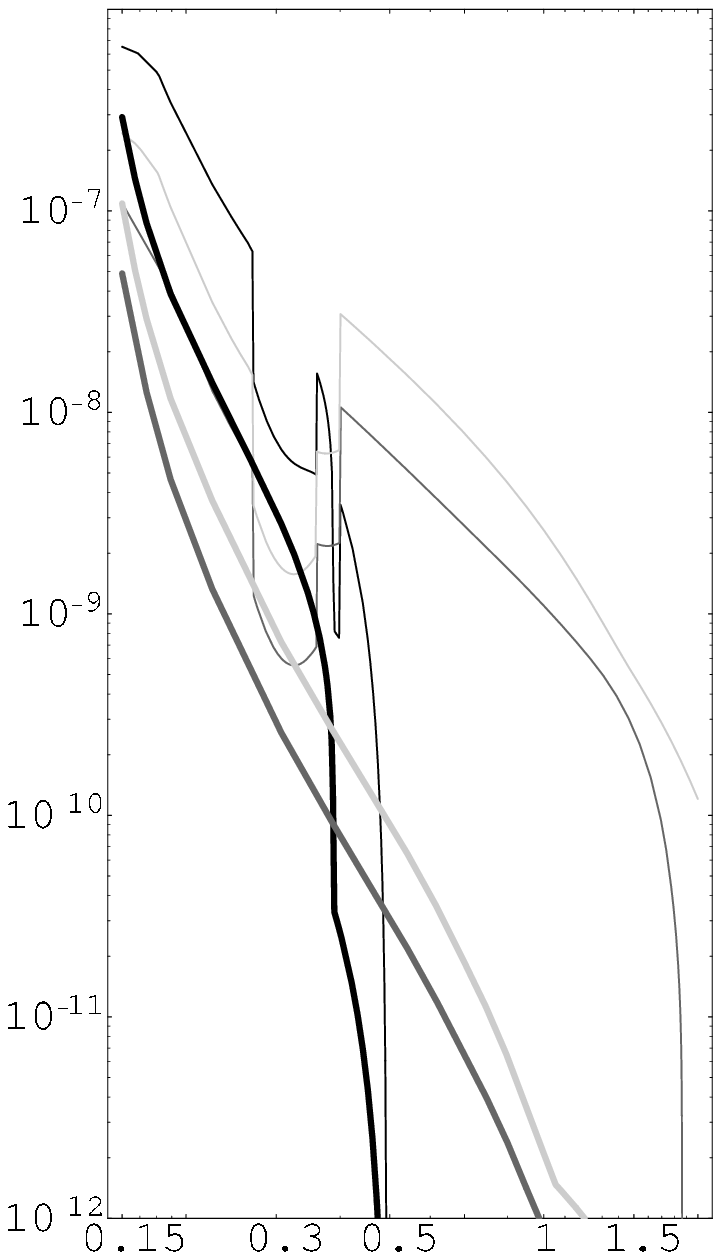}
\includegraphics[width=0.33\textwidth]{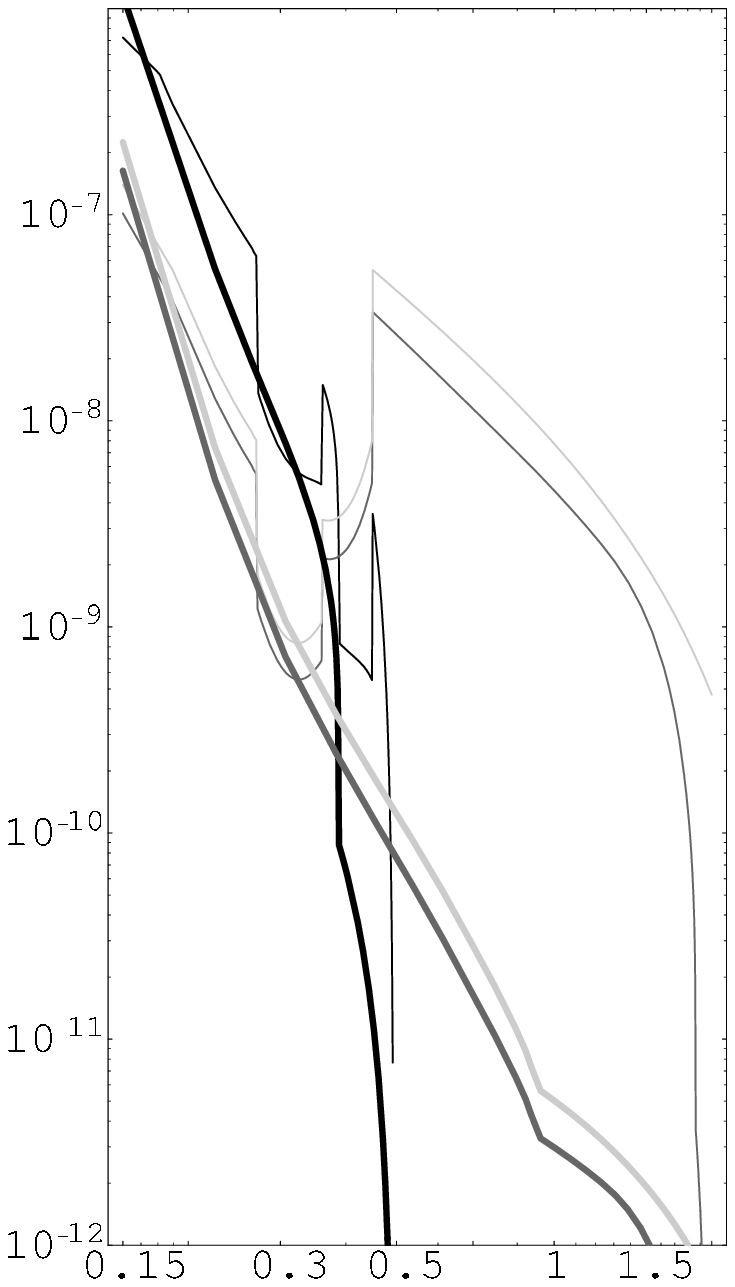}
\includegraphics[width=0.33\textwidth]{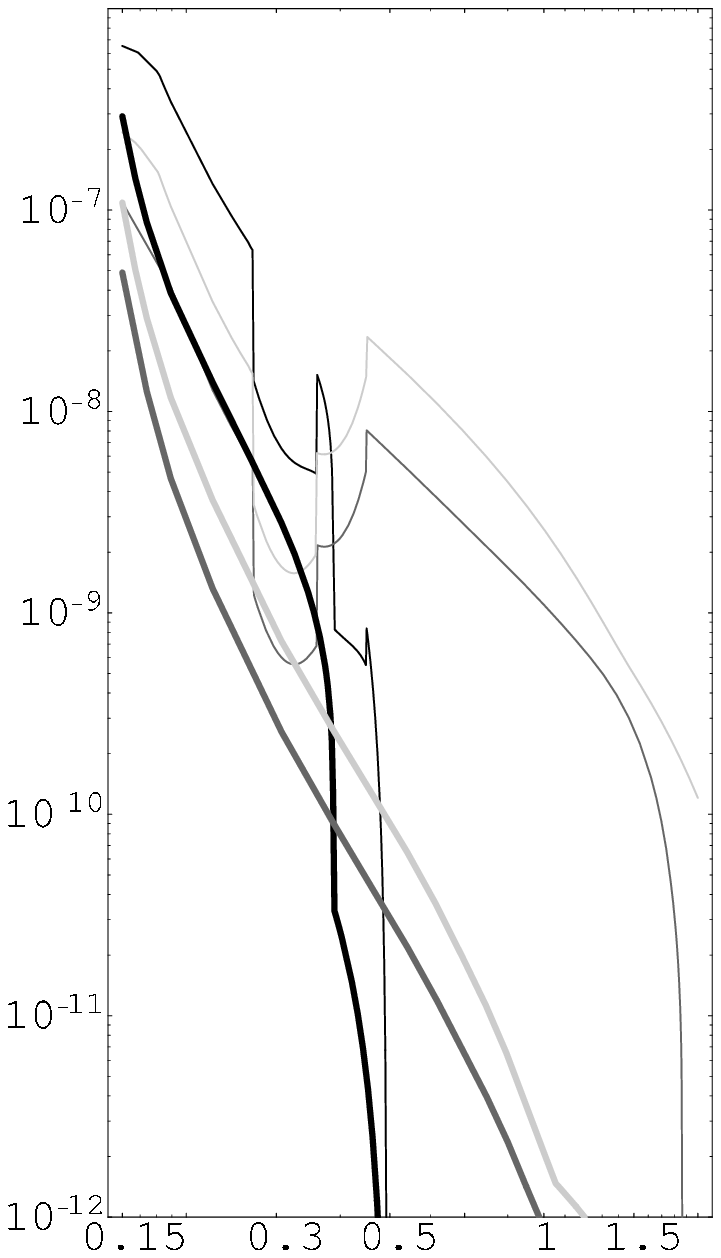}}
\begin{picture}(0,0)(0,0)
\put(-160,40){{\small$M_N$, GeV}}
\put(0,40){{\small$M_N$, GeV}}
\put(155,40){{\small$M_N$, GeV}}
\put(-110,280){{\small a)}}
\put(50,280){{\small b)}}
\put(210,280){{\small c)}}
\end{picture}
\vskip -1.5cm
\caption{Inclusive heavy lepton production by strange (black 
lines), charm (dark gray lines) and beauty (light gray lines) hadrons in models:
a) I, b) II and c) III; within $\nu$MSM the interesting rates are
between corresponding thin and thick lines which
show upper and lower limits on $U^2$ from Fig.~\ref{Nu-lifetime}b,
respectively.
\label{quark-contribution}
}
\end{figure}

To illustrate
the relative weight of different  mesons in total neutrino production we plot
in Fig.~\ref{quark-contribution} 
the quantity 
\[
\xi_Q\equiv \sum_{H}\xi_{Q,H}\;,~~~~~ \xi_{Q,H}\equiv \Br\l Q\to
H\r\cdot \Br\l H\to N\dots\r
\]
(where all considered above leptonic and semileptonic decays of strange,  
charmed  and beauty mesons are taken into account, $Q=s,c,b$) 
within relevant ranges of neutrino masses $M_N$.    

With a reasonable estimate of strange, charm and beauty cross sections at
large energies~\cite{Lourenco:2006vw}  
\[
\sigma_{pp\to s}\sim 1/7\cdot \sigma_{pp}^{total}\;,~~~~
\sigma_{pp\to c}\sim 10^{-3}\cdot \sigma_{pp}^{total}\;,~~~~
\sigma_{pp\to b}\sim 10^{-5}\cdot \sigma_{pp}^{total}\;,~~~~
\]
one concludes that to produce a few neutrinos lighter than kaon, 
$10^7$-$10^{10}$ collisions is required, while for heavier neutrinos the
statistics should be four orders of magnitude (0.5~GeV~$\lesssim
M_N\lesssim$~2~GeV) or even eight orders of magnitude (2~GeV~$\lesssim
M_N\lesssim$~4~GeV) larger.  

Note in passing that in our considerations baryon decays as well as
decays with more than three particles in a final state have been
neglected. These additional contributions to neutrino production are
expected to be insignificant.

\section{Prospects for future experiments}

Generally, there are two types of processes where heavy neutrinos can
be searched for: neutrino production hadron decays and neutrino decays
into SM particles.

In Section~\ref{production} we presented plots with hadron branching
ratios to neutrinos in the frameworks of the three benchmark models. 
>From these plots one can conclude that statistics expected at
proposed Super B-factories give a chance to explore $\nu$MSM with
neutrinos lighter than about 1~GeV and probe some part of parameter
space, if neutrino masses are in 1-2~GeV range. For heavier neutrinos
typical branching ratios become too small, so even with large number
of available hadrons actually small uncertainties in prediction of 
background can make any searches insensitive.

For heavier neutrinos the most promising experiments are beam-target
experiments with high intensity of a beam and high energy of incident
protons. Heavy neutrinos from decays of numerous secondary hadrons
will travel some distance and then decay into SM particles with
branching ratios discussed in Section~\ref{decays}. With lifetime in
the range $10^{-1}\div10^{-5}$~s neutrino covers a distance in exceed
of one kilometer, so a detector aimed at searches for neutrino decay
signatures should be placed  at an appropriate small distance from
the  target to avoid decrease of statistics due to neutrino beam
divergence.   In what follows we consider the experimental setup
with appropriately thin target, assuming that produced in beam-target
collision hadrons decay freely without further interaction inside the
target.  So, this is not a classical beam-dump setup. For classical
beam-dump experiment secondary kaons interact in material before
decay, that change their contribution to production of neutrinos with
$M_N<M_K$, which estimate requires additional study. Heavier neutrinos
are produced mostly by $D$- and $B$-mesons, which even in beam-dump
setup decay before interaction. {\it Hence, for $M_N>M_K$ our results
obtained below are valid for beam-dump experiment as well. } 

The total number of neutrinos produced by $N_{POT}$ incident upon a
target protons with energy $E$ is given by 
\[
N_N(E)=\sum_{Q=u,d,s,\dots}\xi_Q \cdot \frac{\sigma_{pA\to
Q}(E)}{\sigma^{total}_{pA}(E)}\cdot N_{POT}(E)\cdot M_{pp}(E)\;,
\]
where $A$ refers to the target material and $M_{pp}(E)$ is a total
multiplicity (average number of secondary particles in proton-proton
collision).  Here we suppose that all beam protons interact once
  in the target; the account of finite thickness of the target is
  straightforward and results in effective decrease in $N_{POT}$.  
Assuming as a reasonable approximation at large $E$
\[
\frac{\sigma_{pA\to Q}(E)}{\sigma^{total}_{pA}(E)}\approx
\frac{\sigma_{pp\to Q}(E)}{\sigma^{total}_{pp}(E)} \equiv 
\chi_Q(E)
\;,
\] 
we arrived at 
\[
N_N(E)=\sum_{Q=s,c,b}\xi_Q \cdot 
\chi_Q(E)\cdot N_{POT}(E)\cdot M_{pp}(E)\;.
\]
Below we present the
numerical estimates for four high energy beams available today or will
be available in the nearest future: CNGS, NuMi, JPARC (T2K setup) and
TeVatron (NuTeV setup). The relevant parameters of these beams are
presented in Table~\ref{beams-details}.
\begin{table}[!htb]
\begin{tabular}{|c||c|c|c|c|c|c|c|c|c|}
\hline 
Experiment & $E$, GeV & $N_{POT}$, $10^{19}$ & $M_{pp}$~\cite{PDG}
& $\chi_s$~\cite{Fragmentation-in-PYTHIA} &
$\chi_c$~\cite{Lourenco:2006vw} & $\chi_b$~\cite{Lourenco:2006vw} &
$\la p^K_L\ra $, GeV & $\la p^D_L\ra $, GeV & $\la p^B_L\ra $, GeV
\\\hline CNGS~\cite{CNGS} & 400 & 4.5 & 13 & $1/7$ & $0.45\cdot
10^{-3}$ & $3\cdot 10^{-8}$ & 44 & 58 & 58 \\ NuMi~\cite{NuMi} & 120 &
5 & 11 & $1/7$ & $1\cdot 10^{-4}$ & $10^{-10}$ & 24 & 24 & 24 \\ T2K~\cite{T2K} &
50 & 100 & 7 & $1/7$ & $1\cdot 10^{-5}$& $10^{-12}$ & 8.5 & 10 & 10 \\
NuTeV~\cite{NuTeV} & 800 & 1 & 15 & $1/7$ & $1\cdot 10^{-3}$ & $2\cdot
10^{-7}$ & 68 & 82 & 82 \\\hline
\end{tabular}
\caption{
Adopted values of relevant for heavy neutrino production parameters of
several experiments.
\label{beams-details}
}
\end{table}  
To estimate the mean longitudinal momenta $\la p_{H,L}\ra $ of $D$- and
$B$-mesons we make use of the parameterization
\[
\frac{d\sigma}{dx_F}\propto \l 1 -
x_F\r^c\;,~~~x_F\equiv\frac{p_{H,L}}{p_{H,L}^{max}}\;,
\] 
with $c=7.7$ for $E=800$~GeV~\cite{Ammar:1988ta,Kodama:1991jk}, $c=4.9$
for $E=400$~GeV~\cite{Aguilar-Benitez:1987rc} and $c=3$ for
$E=120$~GeV and $E=50$~GeV as an factor-of-two estimate. In case of
kaons we use the estimate 
\[
\la p_{K,L}\ra = \frac{1}{2} \l \la p_{D,L}\ra+\frac{E}{M_{pp}}\r\;,
\]

As we show in Section~\ref{production}, the dominant
contribution to the total neutrino production in collisions come
mostly from two-body hadron decays. Thus, with neutrino longitudinal
momentum uniformly distributed in hadron rest frame one gets for
average neutrino momentum in laboratory frame 
\[
\la p_{N,L}\ra_H=\frac{1}{2} \la p_{H,L} \ra \cdot \l 1
+\frac{M_N^2}{M_H^2}\r\;.
\] 
If neutrino decay length exceeds detector length $\Delta l$, 
the total number of neutrino decays inside the fiducial
volume is 
\[
N_N^{decays}=N_N(E)\cdot \frac{\Delta l}{\tau_N}\cdot 
\sum_{H}\frac{M_N}{\la p_{N,L}\ra_H}\cdot\epsilon_{N}^H\;
\] 
with $\epsilon_{N}^H$ being a relative contribution of a given hadron
$H$ to total neutrino production, 
\[
\epsilon_{N}^H=\frac{N_H(E)\cdot {\rm Br} \l H\to N\dots\r}{N_N(E)}\;,
\]
where the number of produced hadrons of a type $H$ is estimated as 
\[
N_H(E)=N_{POT}(E)\cdot M_{pp}(E)\cdot\chi_Q(E)\cdot{\rm Br} \l Q\to H\r\;.
\] 
Finally we obtain for the total
number of neutrino decays inside the detector 
\[
N_N^{decays}=N_{POT}\cdot M_{pp}\cdot \frac{\Delta l}{\tau_N}\cdot 
\sum_{Q,H}\chi_Q\cdot\xi_{Q,H}
\cdot \frac{M_N}{\la p_L^N\ra_H}\;.
\]

\begin{figure}[!htb]
\centerline{
\includegraphics[width=0.5\textwidth]{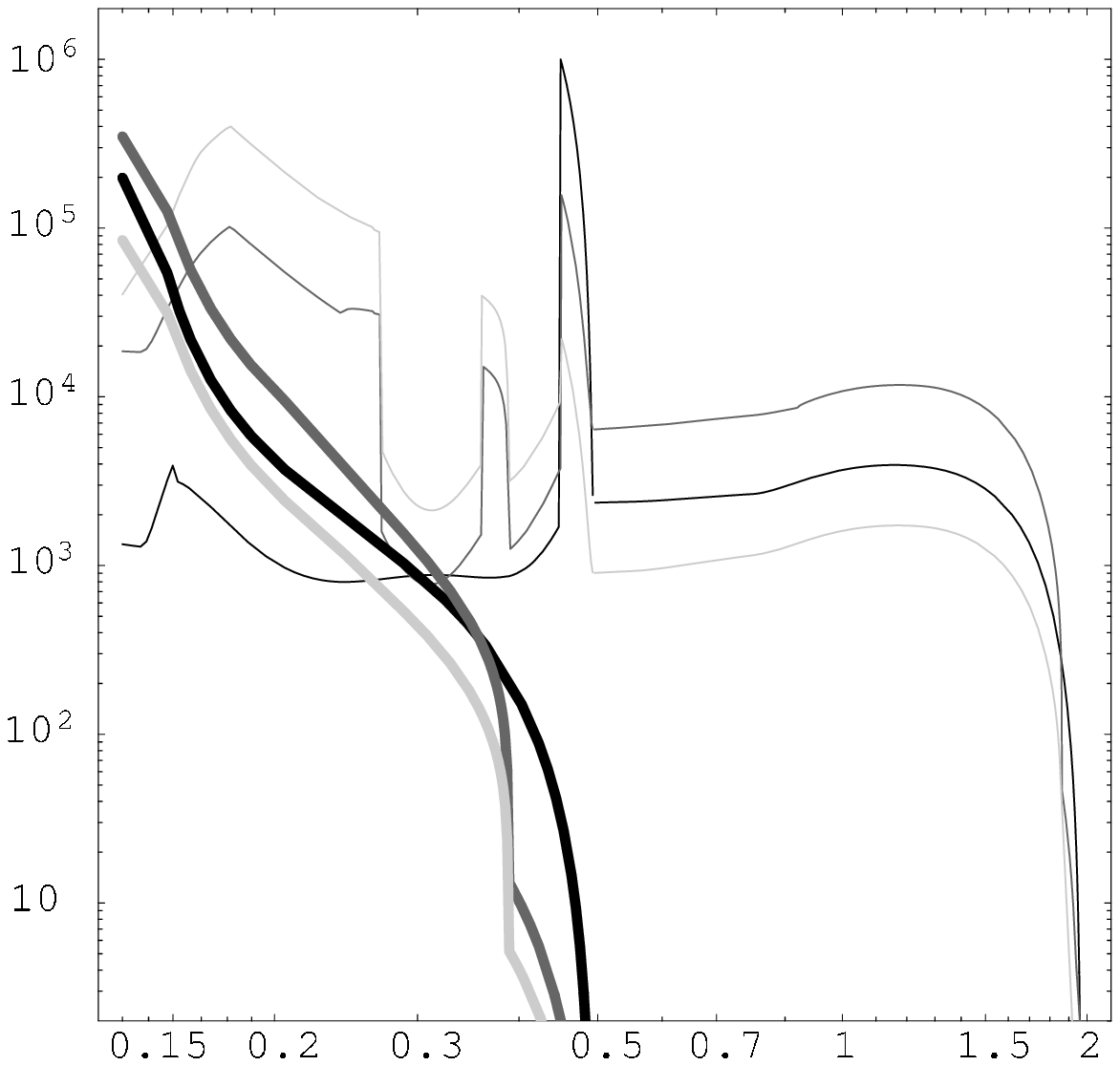}
\includegraphics[width=0.5\textwidth]{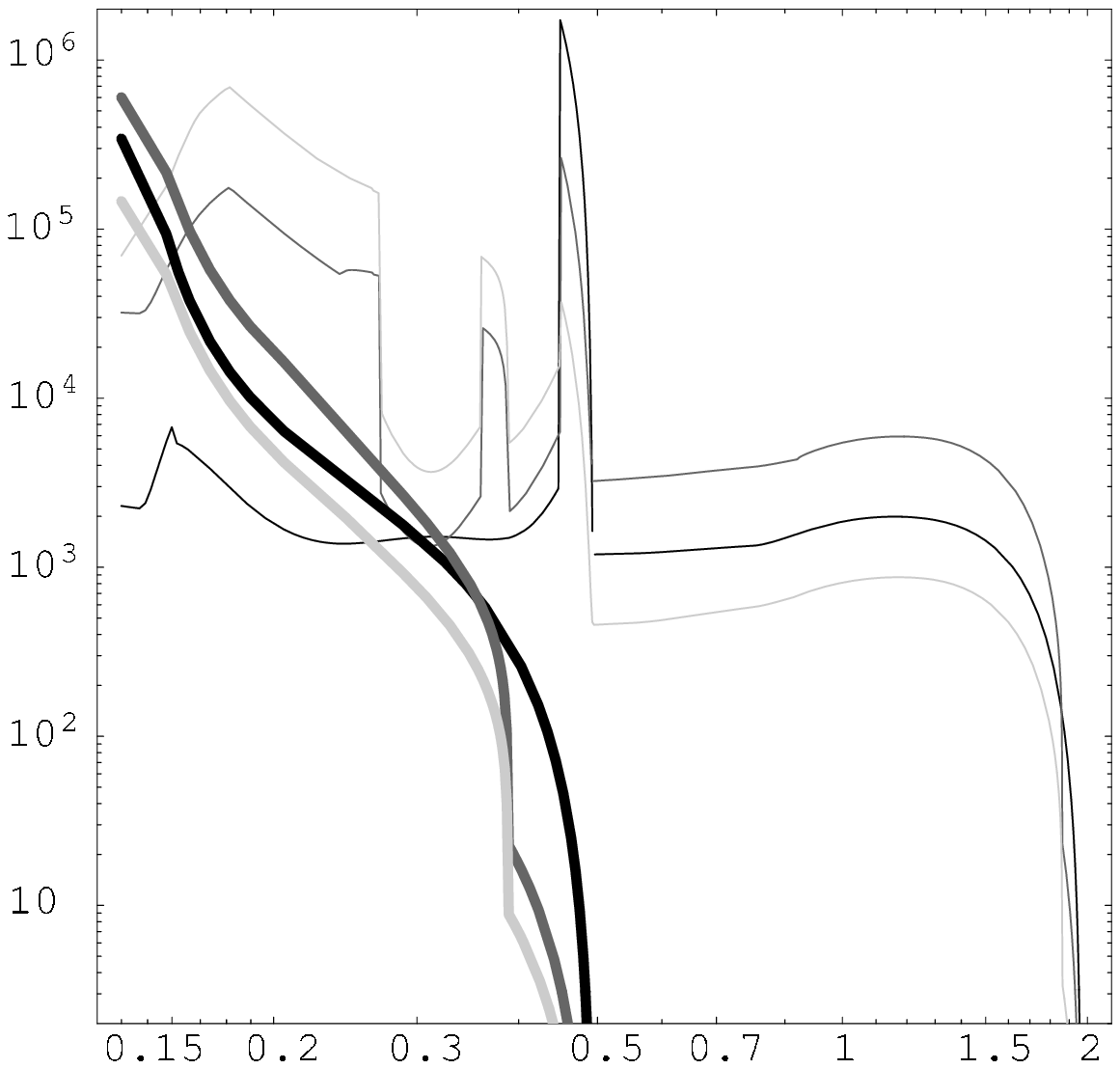}
}
\centerline{
\includegraphics[width=0.5\textwidth]{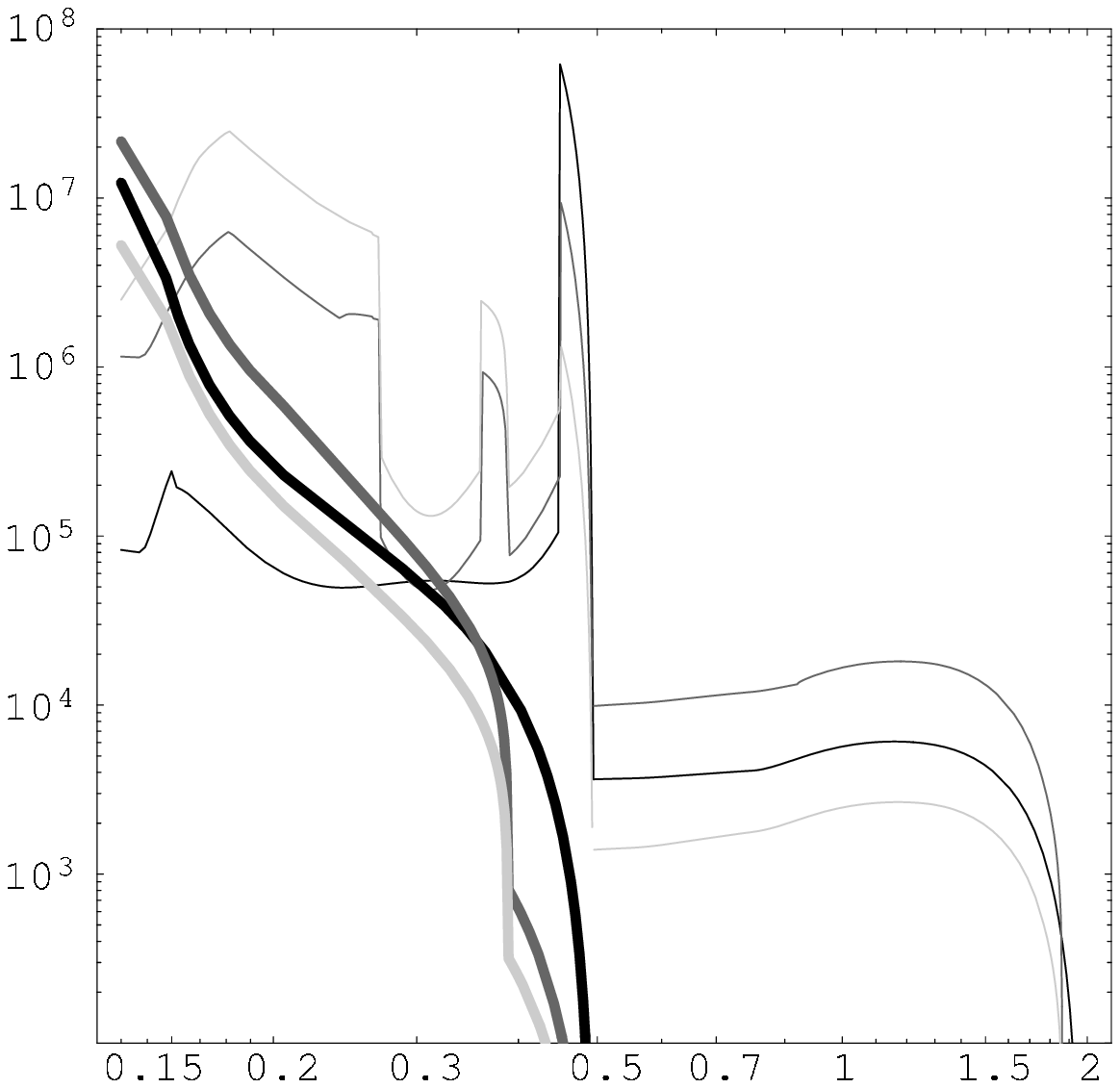}
\includegraphics[width=0.5\textwidth]{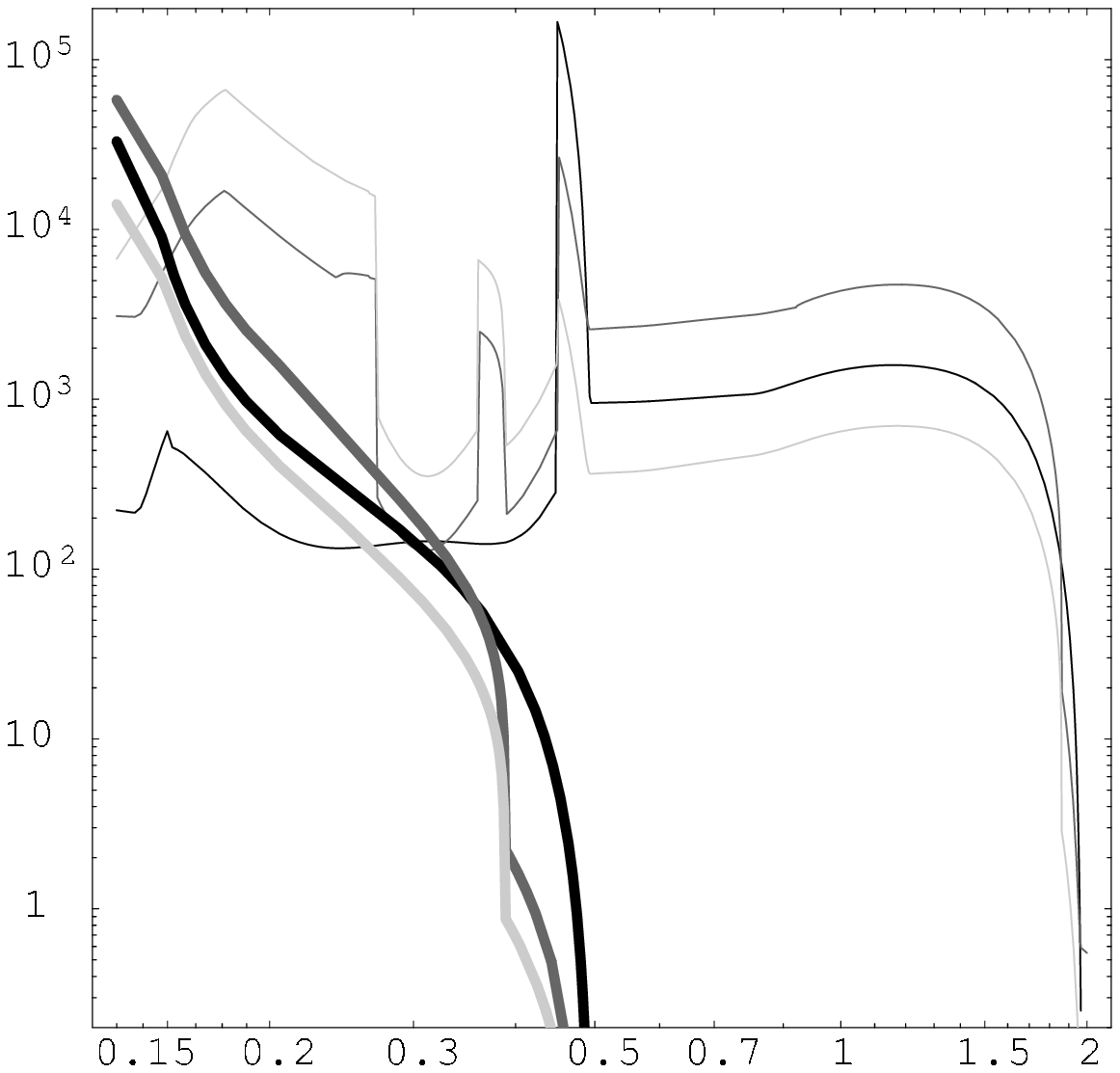}
}
\begin{picture}(0,0)(0,0)
\put(-170,300){a)}
\put(70,300){b)}
\put(-170,70){c)}
\put(70,70){d)}
\put(120,255){$M_N$, GeV}
\put(-115,255){$M_N$, GeV}
\put(-115,20){$M_N$, GeV}
\put(120,20){$M_N$, GeV}
\end{picture}
\vskip -0.8cm
\caption{Number of sterile neutrino decays within 5m-length fiducial
  volume for a) CNGS, b) NuMI, c) T2K, d) NuTeV beams as a function of
  sterile neutrino mass $M_N$. Black, dark gray and light gray lines refer to
  benchmark models I, II and III, respectively; in phenomenologically
  viable models the number of decay events are confined by corresponding
  thin (upper limits) and thick (lower limits) lines.
\label{beam-dump-signal}
}
\end{figure}

For the four available beams with parameters presented in
Table~\ref{beams-details} and $\Delta l=5$m the quantitative
predictions are given in Fig.~\ref{beam-dump-signal} 
for the three benchmark models {\it with account of all experimental and
theoretical constraints; } both lower and upper bounds scale with mixing
as $U^4$. One has to multiply these numbers by the
value of the corresponding branching ratio (see plots in
Section~\ref{decays}), if interested in a particular neutrino decay
mode.
 
Note in passing that in these estimates we neglected neutrino beam
spreading due to nonzero average transverse momentum $\la
p_{N,T}\ra$. With a detector of width $\Delta l_T$  
placed at a distance $l$ from a target this is justified if 
\[
\zeta\equiv \frac{\la p_T^N\ra_H}{\la p_L^N\ra_H} 
\cdot \frac{l}{\Delta l_T}\sim \frac{M_H}{\la p_L^N\ra_H}
\cdot\frac{l}{\Delta l_T}\lesssim 1\;.
\]
Otherwise the predictions for neutrino signal {\it have to be
corrected by a suppression factor } of order $1/\zeta^2$. Similar
suppression factor should be accounted for in case of off-axis
detector. Likewise, as the next approximation to $N_N^{decays}$ one
has to consider the realistic distributions of hadron and neutrino
momenta instead of their average values. Finally, in case of real
neutrino experiment, additional suppression can emerge due to focusing
system. For a given experiment with fixed geometry and detector not
tuned to searches for heavy neutrinos, these
suppression factors can be obtained after dedicated studies;  
they can be as large as one-two orders of magnitude, 
depending on the neutrino mass. This studies are 
beyond the scope of this work.  

\section{Conclusions}
A pair of relatively light and almost degenerate Majorana leptons is
an  essential ingredient of the $\nu$MSM. In this model these
particles are responsible for observed pattern of neutrino masses and
mixings and for baryon asymmetry of the Universe. In the present
paper we analysed  the properties of the singlet fermions of the
$\nu$MSM, which can be used for their experimental search. In
particular, we discussed their production in decays of different
mesons and in $pp$ collisions. We studied the decays of singlet
fermions in a wide range of their masses and other parameters,
consistent with cosmological considerations.

In the $\nu$MSM the strength of the coupling of singlet fermions to
ordinary leptons is bounded both from {\em above} by cosmology
(baryon asymmetry of the Universe) and from {\em below} by neutrino
oscillation experiments. In addition, a lower bound on the strength
of interaction comes from Big Bang Nucleosynthesis. We analysed the
latter constraint in some detail and showed that it is stronger than
the one coming from neutrino experiments for masses smaller than  
1~GeV.

The only particle physics searches for neutral fermions that were
able to enter into the cosmologically interesting region of masses
and couplings of neutral fermions, described in this paper, is the
CERN PS191 experiment \cite{Bernardi:1985ny,Bernardi:1987ek}. It was
performed some 20 years ago and since then no improvements of the
bounds were made. This experiment, together with the BBN
considerations, indicates that the masses of neutral leptons should
be larger than the pion mass, though definite exclusion of the region
below $\pi$-meson would require more theoretical (BBN analysis) and
experimental work. Quite interestingly, with an order of magnitude
improvement of the bounds on the strength of interaction the whole
region of masses below kaon mass can be scanned and the neutral
leptons could be either ruled out or found in this mass range. It
looks likely that the significant improvement of the results of
\cite{Bernardi:1985ny,Bernardi:1987ek} can be achieved by the
reanalysis of the {\em existing} data of KLOE and of E949
experiments, and that the third stage of NA48 would allow to settle
down the question.

The search for the neutral fermions that are heavier than 
K-mesons would require dedicated experiments similar to the CERN
PS191 experiment, but with higher intensity and higher energy proton
beams. We argued that the use of the CNGS, NuMI, T2K or NuTeV beams
can allow to touch the interesting range of mixings for the lepton
mass below charm, whereas for going above charm much more intensive
accelerators would be necessary.  

In conclusion, it is quite possible that the already existing machines can be
used for the search of the physics beyond the Minimal Standard Model,
responsible for neutrino oscillations, dark matter and baryon
asymmetry of the Universe. Clearly, to study the CP-violation in
interactions of neutral leptons more statistics would be required,
calling for the intensity (rather than energy) increase of the proton
beams.

{\bf Acknowledgements.}

We thank A. Dolgov, S. Eidelman, Yu. Kudenko, P. Pakhlov and
F. Vannucci for useful discussions.  This work was supported in part
by the Swiss National Science Foundation, by the Russian Foundation of
Basic Research grants 05-02-17363 (DG), by the grants of the President
of the Russian Federation NS-7293.2006.2 (Government contract
02.445.11.7370) and MK-2974.2006.2 (DG).

\newpage
\appendix
\section{Sterile neutrino decays}
\label{Appendix-neutrino-decays}

Formulae for most decay rates presented here can be obtained
straightforwardly by making use of the formulae for Dirac neutrinos
from Ref.~\cite{Johnson:1997cj}; we checked them and found to be
correct. Formulae for decay rates into fixed final states are
identical in Dirac and Majorana cases.  Decays $N\to\pi^0\nu$ and
$N\to\rho\nu$ have not been considered there. Our result for decay
$N\to\pi^0\nu$ differs from the estimate used in
Ref.\cite{Dolgov:2000jw} by an additional phase volume factor and by a
factor $1/2$.  Hence, the neutrino life-time used in
Ref.~\cite{Dolgov:2000jw} to get BBN limits should be multiplied by
$2/(1-M_\pi^2/M_N^2)$, and corresponding limit on neutrino mixing
angle should be divided by this factor.

Two-body decay modes are
{\footnotesize
\begin{align*}
\Gamma\l N\to \pi^0\nu_\alpha \r&=\frac{|U_\alpha|^2}{32\pi}G_F^2f_\pi^2
M_N^3\cdot \l 1-\frac{M_\pi^2}{M_N^2}\r^2\;,\\
\Gamma\l N\to H^+ l_\alpha^- \r&=\frac{|U_\alpha|^2}{16\pi}G_F^2|V_{H}|^2f_H^2
M_N^3\cdot\l \l 1-\frac{M_l^2}{M_N^2}\r^2-\frac{M_H^2}{M_N^2}
\l 1+ \frac{M_l^2}{M_N^2}\r \r\\&\times 
\sqrt{\l 1-\frac{\l M_H -M_l\r^2}{M_N^2} 
\r \l 1-\frac{\l M_H +M_l\r^2}{M_N^2} \r}\;,\\
\Gamma\l N\to \eta\nu_\alpha \r&=\frac{|U_\alpha|^2}{32\pi}G_F^2f_\eta^2
M_N^3\cdot \l 1-\frac{M_\eta^2}{M_N^2}\r^2\;,\\
\Gamma\l N\to \eta'\nu_\alpha \r&=\frac{|U_\alpha|^2}{32\pi}G_F^2f_{\eta'}^2
M_N^3\cdot \l 1-\frac{M_{\eta'}^2}{M_N^2}\r^2\;,\\
\Gamma\l N\to \rho^+ l_\alpha^- \r&=
\frac{|U_\alpha|^2}{8\pi}\frac{g^2_\rho}{M^2_\rho}G_F^2|V_{ud}|^2
M_N^3\cdot\l \l 1-\frac{M_l^2}{M_N^2}\r^2+\frac{M_\rho^2}{M_N^2}
\l 1+ \frac{M_l^2-2M_\rho^2}{M_N^2}\r \r\\&\times 
\sqrt{\l 1-\frac{\l M_\rho -M_l\r^2}{M_N^2} \r 
\l 1-\frac{\l M_\rho +M_l\r^2}{M_N^2} \r}\;,\\
\Gamma\l N\to \rho^0 \nu_\alpha \r&=
\frac{|U_\alpha|^2}{16\pi}\frac{g^2_\rho}{M^2_\rho}G_F^2
M_N^3\cdot\l 1+ 2\frac{M_\rho^2}{M_N^2}\r 
\cdot\l 1-\frac{M_\rho^2}{M_N^2}\r^2\;,
\end{align*}
} where $G_F$ is Fermi coupling constant, $f_\eta=1.2f_\pi$,
$f_{\eta'}=-0.45f_\pi$, $g_\rho=0.102$~GeV$^2$~\cite{PDG}; hereafter
and for CKM matrix elements we use values from Ref.~\cite{PDG}, while
for meson decay constants we used most recent values from
Refs.~\cite{PDG,Stone:2006cc}. 
\begin{table}[!htb]
\begin{tabular}{|c|c|c|c|c|c|c|c|}
\hline 
$H$ & $\pi^+$ & $K^+$ & $D^+$ & $D_s$ & $B^+$ & $B_s$ & $B_c$ \\ 
\hline
$f_H$, MeV & 130 & 159.8 & 222.6 & 280.1 & 190 & 230 & 480 \\ 
\hline
$V_H$ & $V_{ud}$ & $V_{us}$ & $V_{cd}$ & $V_{cs}$ 
& $V_{ub}$ & $V_{us}$ & $V_{cb}$\\ 
\hline
\end{tabular}
\end{table}

Three body decay modes read
{\footnotesize
\begin{align*}
\nonumber
\Gamma\l N\to \sum_{\alpha,\beta}\nu_\alpha\bar\nu_\beta \nu_\beta \r&=
\frac{G_F^2M_N^5}{192\pi^3}\cdot\sum_\alpha |U_\alpha|^2\;,\\
\Gamma\l N\to l^-_{\alpha\neq\beta}l^+_\beta  \nu_\beta \r&=
\frac{G_F^2M_N^5}{192\pi^3}\cdot |U_\alpha|^2 \l
1-8x_l^2+8x_l^6-x_l^8-12x_l^4 \log x_l^2 \r\;,~~~x_l=\frac{{\rm max }\left[
  M_{l_\alpha},\;M_{l_\beta}\right]}{M_N}\;,  
\\
\Gamma\l N\to \nu_\alpha l_\beta^+l_\beta^- \r&=
\frac{G_F^2M_N^5}{192\pi^3}\cdot |U_\alpha|^2 \cdot 
\Biggl[\l
  C_1\cdot(1-\delta_{\alpha\beta})+C_3\cdot\delta_{\alpha\beta}\r  
\biggl(
\l 1-14x_l^2-2x_l^4-12x_l^6\r\sqrt{1-4x_l^2}\\&
+12 x_l^4 \l x_l^4-1\r L \biggr) 
+ 4 
\l C_2\cdot(1-\delta_{\alpha\beta}) +C_4\cdot\delta_{\alpha\beta}  \r 
\biggl( x_l^2 \l 2+10 x_l^2-12 x_l^4\r \sqrt{1-4x_l^2} \\\nonumber
&+6x_l^4\l 1-2 x_l^2+2x_l^4\r L\biggl)\Biggr]\;,
\end{align*}
}
with 
{\footnotesize
\[
L=\log\left[
\frac{1-3x_l^2-\l 1-x_l^2 \r\sqrt{1-4x_l^2}}
{x_l^2\l 1+\sqrt{1-4 x_l^2}\r}\right]\;,~~~~x_l\equiv\frac{M_l}{M_N}\;,
\]
}
and
{\footnotesize
\begin{align*}
C_1&=\frac{1}{4}\l 1-4\sin^2\theta_w+8\sin^4\theta_w\r\;,
&C_2=\frac{1}{2}\sin^2\theta_w\l 2\sin^2\theta_w-1\r\;,\\
C_3&=\frac{1}{4}\l 1+4\sin^2\theta_w+8\sin^4\theta_w\r\;,
&C_4=\frac{1}{2}\sin^2\theta_w\l 2\sin^2\theta_w+1\r\;.
\end{align*}
}

The Majorana neutrino total decay rate is a sum of all rates presented
above multiplied by a factor of 2, which accounts for charge-conjugated
decay modes.

\section{Decays into sterile neutrino}
\label{Appendix-meson-decays}

Differential branching ratio of pseudoscalar meson leptonic 
decays into sterile neutrinos reads 
{\footnotesize
\begin{align}
\nonumber
\frac{d\Br\l H^+\to l_\alpha^+ N\r}{dE_N}
&=\tau_H\cdot 
\frac{G_F^2f_H^2M_HM_N^2}{8\pi}
|V_H|^2 |U_\alpha|^2 \cdot 
\l 
1 - \frac{M_N^2}{M_H^2} + 2 \frac{M_l^2}{M_H^2} + 
\frac{M_l^2}{M_N^2} \l 1 - \frac{M_l^2}{M_H^2} \r
\r\\\label{Sec3:5++}
&\times
\sqrt{\l 1 + \frac{M_{N}^2}{M_H^2} - \frac{M_l^2}{M_H^2} \r^2 - 
4\frac{M_{N}^2}{M_H^2}}\cdot \delta\l E_N-\frac{M_H^2-M_l^2+M_N^2}{2M_H}\r
\;,
\end{align}
}
where $\tau_H$ is the meson life-time~\cite{PDG}. 

For differential branching ratios of pseudoscalar meson semileptonic decays
one has 
{\footnotesize
\begin{equation}
\label{Sec3:5+}
\begin{split}
&\frac{d\Br\l H\to H' l_\alpha^+ N\r}{dE_N} =\tau_{H}\cdot
|U_\alpha|^2\cdot\frac{|V_{HH'}|^2G_F^2}{64\pi^3M_H^2}
\times \int dq^2 \Biggl( f^2_-(q^2)\cdot \l q^2\l M_N^2+M_l^2 \r -\l
M_N^2-M_l^2\r^2 \r \\ 
&
+2f_+(q^2)f_-(q^2) \l M_N^2\l
2M_H^2-2M_{H'}^2-4E_N M_H- M_l^2+M_N^2+q^2 \r + M_l^2 \l 4E_N M_H +
M_l^2-M_N^2-q^2 \r \r\\ 
& 
f_+^2(q^2) \biggl( \l 4E_N M_K +
M_l^2-M_N^2-q^2\r \l 2M_K^2-2M_\pi^2-4E_N M_K- M_l^2+M_N^2+q^2 \r \\ 
&
- \l 2M_K^2+2 M_\pi^2 -q^2\r\l q^2-M_N^2-M_l^2\r \biggr) \Biggr)\;,
\end{split}
\end{equation}
}
where $q^2=(p_l+p_N)^2$ is momentum of leptonic pair, $V_{HH'}$ is
corresponding entry of CKM matrix  and $f_+(q^2)$, $f_-(q^2)$
are dimensionless hadronic form factors~\cite{PDG} can be found in literature.

For three-body decays into vector mesons $V$ one obtains 
{\footnotesize
\begin{equation}
\label{Sec3:5+++}
\begin{split}
\frac{d\Br \l H\to V l_\alpha N\r}{dE_N}
=\tau_{H}\cdot |U_\alpha|^2\cdot
\frac{|V_{HV}|^2G_F^2}{32\pi^3M_{H}} \times\int dq^2 \biggl(
\frac{f_2^2}{2}\l q^2-M_N^2 -M_l^2 +\omega^2\frac{\Omega^2-\omega^2}{M_V^2}\r
\\ 
+ \frac{f_5^2}{2} \l M_N^2+M_l^2\r \l q^2-M_N^2+M_l^2\r\l \frac{\Omega^4}{4M_V^2}-q^2\r
+ 2f_3^2 M_V^2 \l \frac{\Omega^4}{4M_V^2}-q^2\r \l
M_N^2+M_l^2-q^2+\omega^2\frac{\Omega^2-\omega^2}{M_V^2}\r \\
+2 f_3 f_5  \l M_N^2 \omega^2 + \l \Omega^2-\omega^2\r M_l^2\r \l \frac{\Omega^4}{4M_V^2}-q^2\r 
+2 f_1 f_2 \l q^2\l2\omega^2-\Omega^2\r+\Omega^2\l M_N^2-M_l^2\r \r \\
+\frac{f_2f_5}{2} \l \omega^2 
\frac{\Omega^2}{M_V^2}\l M_N^2-M_l^2\r +\frac{\Omega^4}{M_V^2}M_l^2 +2
\l M_N^2-M_l^2\r^2 - 2 q^2\l M_N^2+M_l^2\r 
\r
\\+ f_2f_3 
\l 
\Omega^2\omega^2\frac{\Omega^2-\omega^2}{M_V^2}
+2\omega^2 \l M_l^2-M_N^2\r +\Omega^2\l M_N^2-M_l^2-q^2\r\r
\\
+f_1^2
\l
\Omega^4\l q^2 -M_N^2+M_l^2\r-2M_V^2\l q^4-\l M_N^2-M_l^2\r^2\r 
+2 \omega^2\Omega^2\l M_N^2-q^2-M_l^2 \r 
+2 \omega^4  q^2
\r\biggr)\;,
\end{split}
\end{equation}
}
where 
$\omega^2=M_{H}^2-M_V^2+M_N^2-M_l^2-2M_{H}E_N$ and 
$\Omega^2=M_{H}^2-M_V^2-q^2$; form factors $f_i(q^2)$ can be  
expressed via standard 
axial form factors $A_0(q^2),A_1(q^2),A_2(q^2)$ and vector form
factor $V(q^2)$ as  
{\footnotesize
\begin{align*}
&f_1=\frac{V}{M_{H}+M_V}\;,~~~f_2=\l M_{H}+M_V\r \cdot
    A_1\;,~~~
f_3=-\frac{A_2}{M_{H}+M_V}\;,\\
&f_4=\l M_V \l 2A_0-A_1-A_2\r+M_{H}\l A_2-A_1\r
    \r\cdot\frac{1}{q^2}\;,~~~
f_5=f_3+f_4\;,
\end{align*}
}
which can be found in literature. 

For two-body decays of $\tau$-lepton into heavy neutrino and meson we obtain 
{\footnotesize
\begin{align*}
\frac{d\Br \l \tau\to H N\r}{dE_N}&=\tau_\tau\cdot 
\frac{|U_\tau|^2}{16\pi}G_F^2|V_{H}|^2f_H^2
M_\tau^3\cdot\l \l 1-\frac{M_N^2}{M_\tau^2}\r^2-\frac{M_H^2}{M_\tau^2}
\l 1+ \frac{M_N^2}{M_\tau^2}\r \r\\&\times 
\sqrt{\l 1-\frac{\l M_H -M_N\r^2}{M_\tau^2} 
\r \l 1-\frac{\l M_H +M_N\r^2}{M_\tau^2} \r}\cdot \delta\l
E_N-\frac{M_\tau^2-M_H^2+M_N^2}{2M_\tau}\r  \;,\\
\frac{d\Br \l \tau\to \rho N\r}{dE_N}&=\tau_\tau\cdot 
\frac{|U_\tau|^2}{8\pi}\frac{g^2_\rho}{M^2_\rho}G_F^2|V_{ud}|^2
M_\tau^3\cdot\l \l 1-\frac{M_N^2}{M_\tau^2}\r^2+\frac{M_\rho^2}{M_\tau^2}
\l 1+ \frac{M_N^2-2M_\rho^2}{M_\tau^2}\r \r\\&\times 
\sqrt{\l 1-\frac{\l M_\rho -M_N\r^2}{M_\tau^2} \r 
\l 1-\frac{\l M_\rho +M_N\r^2}{M_\tau^2} \r}\cdot \delta\l
E_N-\frac{M_\tau^2-M_\rho^2+M_N^2}{2M_\tau}\r
\;,
\end{align*}
}
where $\tau_\tau$ is $\tau$-lepton life-time. 
For three-body decays of $\tau$-lepton one has  
{\footnotesize
\begin{align*}
\frac{d\Br \l \tau\to \nu_\tau l_\alpha N\r}{dE_N}&=\tau_\tau\cdot 
\frac{|U_\alpha|^2}{2\pi^3}G_F^2M_\tau^2\cdot E_N 
\l 1+\frac{M_N^2-M_l^2}{M_\tau^2}-2\frac{E_N}{M_\tau} 
\r 
\l 1-\frac{M_l^2}{M_\tau^2+M_N^2-2E_NM_\tau}\r
\sqrt{E_N^2-M_N^2}
\;,\\
\frac{d\Br \l \tau\to \bar\nu_\alpha l_\alpha N\r}{dE_N}&= \tau_\tau\cdot 
\frac{|U_\tau|^2}{4\pi^3}G_F^2M_\tau^2 
\l 1-\frac{M_l^2}{M_\tau^2+M_N^2-2E_NM_\tau}\r^2  \sqrt{E_N^2-M_N^2} 
\\
&\times
\l
\l M_\tau-E_N \r \l 1-\frac{M_N^2+M_l^2}{M_\tau^2} \r - 
\l 1-\frac{M_l^2}{M_\tau^2+M_N^2-2E_NM_\tau}\r 
\l \frac{\l M_\tau-E_N\r^2}{M_\tau}+\frac{E_N^2-M_N^2}{3M_\tau} \r
\r
\;.
\end{align*}
}

Note that omitted here charge-conjugated processes also contribute
to Majorana neutrino production.


 \end{document}